\pdfoutput=1
\documentclass[twocolumn,trackchanges]{aastex631}

\makeatletter
\let\frontmatter@title@above\relax
\makeatother

\usepackage{amsmath,amssymb,amstext}

\usepackage[all]{hypcap}
\hypersetup{hypertexnames=false}
\usepackage[T1]{fontenc} 
\usepackage[caption=false]{subfig}
\usepackage{booktabs}

\usepackage{xcolor}

\usepackage{amsthm}

\shortauthors{Giron, Krief \& Steinberg}
\numberwithin{equation}{section}

\begin{document}

\title{Multigroup Radiation Diffusion with Compton Scattering on a Moving Mesh: Implementation in \textsc{rich} and Application to Tidal Disruption Events}
\shorttitle{Multigroup Radiation Diffusion on a Moving Mesh. II}

\author[0000-0003-2234-1319]{Itamar Giron}
\affiliation{Racah Institute of Physics, The Hebrew University, 9190401 Jerusalem, Israel}
\email{itamar.giron@mail.huji.ac.il}

\author[0000-0002-7093-6467]{Menahem Krief}
\email{menahem.krief@mail.huji.ac.il}
\affiliation{Racah Institute of Physics, The Hebrew University, 9190401 Jerusalem, Israel}

\author[0000-0003-0053-0696]{Elad Steinberg}
\email{elad.steinberg@mail.huji.ac.il}
\affiliation{Racah Institute of Physics, The Hebrew University, 9190401 Jerusalem, Israel}

\begin{abstract}
Compton scattering exchanges energy between radiation and hot gas and changes the spectrum in high-energy astrophysical systems such as tidal disruption events (TDEs). Because it couples all photon energies and can act faster than the hydrodynamic time step, many radiation-hydrodynamics (RHD) simulations omit its spectral redistribution or treat its energy exchange in the gray approximation. We extend the multigroup flux-limited diffusion module of the moving-mesh RHD code \textsc{rich} to include Compton scattering with the exact thermal Compton kernel. In our semi-implicit scheme for the thermal radiative transfer equation, the temperature dependence of the kernel enters a generalized Fleck factor, and the group Compton operator conserves photon number and satisfies detailed balance. In equilibration tests, the scheme agrees with independent solutions, relaxes without absorption and emission to a Bose--Einstein spectrum, and remains stable for time steps that do not resolve the Compton equilibration timescale. We apply it to a three-dimensional simulation of the disruption of a $0.5\,M_\odot$ star by a $10^6\,M_\odot$ black hole. We find that during the X-ray flash, inverse Compton scattering dominates the coupling between the gas and the radiation near the nozzle shock, where it adds X-ray energy much faster than thermal emission and may affect the dynamics. Outside the X-ray thermalization radii, its instantaneous X-ray production rate during the flash is comparable to or larger than the escaping X-ray luminosity, so it may contribute substantially to the X-ray emission from TDEs.
\end{abstract}

\section{Introduction}
\label{sec:introduction}

Radiation hydrodynamics (RHD) determines the evolution and the observed emission of many astrophysical systems, from star-forming regions \citep{Krumholz+07a,Grudic+21} and accretion flows around black holes \citep{Jiang+19} to stellar collisions and tidal disruption events \citep{Brutman+24,steinberg_streamdisk_2024,Huang+25b}. Because the opacity varies strongly over the wide range of photon energies, a growing number of codes use multigroup methods, including \textsc{castro} \citep{Zhang_2013}, \textsc{athena}++ \citep{Stone2020,Jiang22}, SuperNu \citep{Wollaeger_2013}, \textsc{quokka} \citep{Krumholz_2024}, and \textsc{rich}\footnote{\url{https://gitlab.com/eladtan/RICH}} \citep{Giron_2026}. These codes solve for the radiation field with flux-limited diffusion (FLD) \citep{Pomeraning-Levermore,Levermore84,Zhang_2013,Giron_2026}, a two-moment method in the mixed frame with the closure of \citet{Levermore84} \citep{Krumholz_2024}, discrete ordinates \citep{Jiang22}, or implicit Monte Carlo (IMC) \citep{fleck_implicit_1971,Wollaeger_2013}.

Compton scattering is the inelastic scattering of photons by free electrons \citep{Compton1923,Klein1929}. In a thermal plasma, repeated scatterings change the radiation spectrum and exchange energy between the radiation and the gas \citep{Kompaneets1957,RybickiLightman1979,Pozdnyakov1983}. Hot electrons upscatter soft photons (inverse Compton scattering), and electron recoil makes photons with energies well above the thermal energy of the electrons lose energy. Comptonization shapes the X-ray spectra of accretion flows, their coronae, and neutron-star atmospheres \citep{Madej2017}, as well as the spectra of active galactic nuclei and gamma-ray bursts \citep{Canfield1987,Peer2005}. It also produces the thermal Sunyaev--Zeldovich effect \citep{BIRKINSHAW199997}.
Compton scattering is difficult to include in RHD for two reasons. First, the thermal Compton kernel couples all photon energies and depends on the gas temperature. When the energy exchange per scattering is small, the kernel can be replaced by the Kompaneets (Fokker--Planck) approximation \citep{Kompaneets1957,WINSLOW1995262,Jiang22}. Many RHD simulations instead use a gray energy-exchange term with a single radiation temperature \citep{Jiang+19,Martire+25} or photon number \citep{SadowskiNarayan15}. Full-kernel treatments, needed when individual scatterings produce large energy shifts, have been developed for deterministic transport \citep{TILL2020108990,mcgraw_till_warsa_2021,MCGRAW2023111980} and for Monte Carlo transport \citep{Keady2017,Ryan+15}. They require accurate group-to-group Compton coefficients, which we compute from the exact thermal Compton kernel with the method of \citet{compton_kernel_integration_paper}, implemented in the \textsc{mace}\footnote{\url{https://github.com/ItamarShmelo/MACE}} code. Second, the Compton equilibration timescale can be much shorter than the hydrodynamic time step, so explicit treatments are subject to a restrictive stability condition \citep{densmore_2007,Densmore2009}. Implicit methods avoid this restriction by linearizing the temperature dependence of the coupling within the time step. The IMC method \citep{fleck_implicit_1971,Wollaber23022016} does this for absorption and emission, and \citet{Densmore2009} apply it to the Compton Fokker--Planck equation. 

In tidal disruption events (TDEs), Compton scattering can affect the emission and the radiation--matter coupling, and thus the dynamics of the debris. A TDE occurs when a star passes close enough to a massive black hole to be torn apart by its tidal field \citep{Hills1975,Rees1988}. TDEs are rare in any individual galaxy \citep{Stone+20}, but time-domain surveys have found dozens in the optical, UV, and X-ray bands \citep{vanVelzen+20,Saxton+20,Gezari21}, and the Vera C. Rubin Observatory and ULTRASAT are expected to discover many more \citep{BricmanGomboc20,Shvartzvald+24}.
The classical picture predicts soft X-ray emission from a rapidly formed accretion disk \citep{Rees1988}, but most optically selected TDEs have lower temperatures and larger inferred radii, and many show little early X-ray emission \citep{vanVelzen+20,Gezari21,Roth+20}. This has been attributed to reprocessing of the disk X-rays by surrounding gas \citep{Guillochon+14}, including optically thick envelopes or outflows \citep{Dai+18,Thomsen2022}, or to shocks in the returning debris when circularization is slow \citep{Piran+15,Shiokawa+15}. The hydrodynamics of disk formation remain uncertain \citep{BonnerotStone21,Price+24,Chan+26}. At pericenter, the vertical compression of the returning stream produces the nozzle shock \citep{BonnerotLu22,Andalman+25}, and the stream collides with itself in a self-crossing shock \citep{BonnerotStone21}. Global RHD simulations of disk formation have only recently become possible \citep{Bonnerot+21,steinberg_streamdisk_2024,Martire+25,Meza+25,Huang+25b}. In \textsc{rich} simulations, shocks near pericenter produce an early X-ray flash \citep{steinberg_streamdisk_2024,Giron_2026}, and X-ray emission during the optical rise has been detected in AT~2022dsb and SRGe~J131014.7+444319 \citep{malyali_transient_2024,Zhang+25}.

Compton scattering can shape the emission of TDEs in several ways \citep{Roth+20}. For example, some TDEs show a hard X-ray tail \citep{Saxton+20,Zabludoff+21}, which in ASASSN-14li can be fit either by Comptonization of disk photons or by a hotter disk component \citep{Kara+18}. \citet{Mummery2021} predict a Compton-scattering corona in TDEs around black holes with masses above about $2\times10^7\,M_\odot$. In the reprocessing layer, repeated scattering in the expanding outflow adiabatically shifts soft X-rays to lower energies \citep{Dai+18,Roth+20}, while in jetted TDEs, Comptonization in the hot jet gas upscatters soft photons from the funnel wall into hard X-rays \citep{CurdNarayan19}. So far, in RHD simulations of TDEs, thermal Compton scattering has either been treated in the gray approximation \citep{Dai+18,CurdNarayan19,Martire+25} or been omitted \citep{Huang+25b,Giron_2026}.

In this paper we extend the multigroup FLD module of \textsc{rich} \citep{Giron_2026} to include Compton scattering with the exact thermal Compton kernel, including induced scattering. We derive a semi-implicit scheme for the thermal radiative transfer (TRT) equation, in which the temperature dependence of the Compton kernel enters a generalized Fleck factor \citep{fleck_implicit_1971}, and reduce it to the multigroup diffusion equation. We verify the scheme in equilibration problems \citep{WINSLOW1995262,MCGRAW2023111980}, including relaxation to a Bose--Einstein spectrum and time steps that do not resolve the Compton equilibration timescale. We then apply it to a three-dimensional simulation of the disruption of a $0.5\,M_\odot$ star by a $10^6\,M_\odot$ black hole. We identify where Compton scattering dominates the energy exchange with the gas, and compare the X-ray energy it produces with the escaping X-ray luminosity. To our knowledge, this is the first implementation of multigroup Compton scattering with the exact thermal Compton kernel in a moving-mesh RHD code. A Monte Carlo implementation of the semi-implicit Compton scheme in \textsc{rich} is also in progress.

The rest of the paper is organized as follows. Section~\ref{sec:transport} derives the semi-implicit TRT scheme, which Sections~\ref{sec:monochromatic_equations} and~\ref{sec:multigroup_equations} reduce to the monochromatic and multigroup diffusion equations. Section~\ref{sec:compton_coefficients} defines the group Compton coefficients, Section~\ref{sec:implementation} describes the numerical scheme in \textsc{rich}, Section~\ref{sec:verification} presents the equilibration tests, and Section~\ref{sec:tde} the application to a TDE. We summarize in Section~\ref{sec:conclusions}.

\section{Thermal Radiative Transfer with Compton Scattering}
\label{sec:transport}

Although this paper focuses on the multigroup diffusion implementation in \textsc{rich}, the semi-implicit treatment of Compton scattering developed here is a general radiation-transport scheme. We therefore first derive it in the most general setting, the Thermal Radiative Transfer (TRT) equation.

\subsection{The Transport Equation}
\label{sec:transport_equation}

We start from the TRT equations with Compton scattering in the form given by \citet{MCGRAW2023111980} for unpolarized radiation in a gas at rest. The gas is in local thermodynamic equilibrium at temperature $T$, so it emits as a blackbody at $T$ and its free electrons follow a relativistic Maxwell--Jüttner distribution at the same temperature \citep{Juttner1911}. Throughout, $\nu$ is the photon energy, so spectral quantities are written per unit photon energy rather than per unit frequency. The specific intensity $I_\nu(\vec{x},\hat{\Omega},t)$ at position $\vec{x}$ and in direction $\hat{\Omega}$ has units of energy per unit area per unit time per unit solid angle per unit photon energy, and satisfies
\begin{multline}
\frac{1}{c}\frac{\partial I_\nu}{\partial t}
+\hat{\Omega}\cdot\nabla I_\nu
+\kappa_\nu I_\nu
=
\frac{c\kappa_\nu B_\nu}{4\pi}
\\
+\int_0^\infty d\nu'\int_{4\pi}d\Omega'
\left(\frac{\nu}{\nu'}\right)
\sigma_s(\nu'\mathbin{\to}\nu,\hat{\Omega}'\cdot\hat{\Omega};\rho,T)
\\
\times I_{\nu'}(\hat{\Omega}')
\left[1+p_\nu I_\nu(\hat{\Omega})\right]
\\
-\int_0^\infty d\nu'\int_{4\pi}d\Omega'\,
\sigma_s(\nu\mathbin{\to}\nu',\hat{\Omega}\cdot\hat{\Omega}';\rho,T)
\\
\times I_\nu(\hat{\Omega})
\left[1+p_{\nu'}I_{\nu'}(\hat{\Omega}')\right]
\label{eq:transport}
\end{multline}
Here $c$ is the speed of light, $\rho$ is the gas density, and $\kappa_\nu$ is the absorption coefficient, per unit length. The Planck spectral energy density, expressed per unit photon energy, is
\begin{equation}
B_\nu(T)=
\frac{8\pi\nu^3}{h^3c^3}
\frac{1}{\exp\!\left(\nu/k_BT\right)-1}
\label{eq:planck_spectral_energy}
\end{equation}
where $h$ is the Planck constant and $k_B$ is the Boltzmann constant.
The macroscopic differential Compton scattering cross section $\sigma_s(\nu'\mathbin{\to}\nu,\hat{\Omega}'\cdot\hat{\Omega};\rho,T)$, per unit photon energy and per unit solid angle, describes scattering from photon energy $\nu'$ and direction $\hat{\Omega}'$ to $\nu$ and $\hat{\Omega}$ in gas of density $\rho$ and temperature $T$. It is obtained by averaging the Klein--Nishina cross section over the Maxwell--Jüttner electron distribution at temperature $T$, then multiplying by the local electron number density \citep{Juttner1911,Klein1929,POMRANING19721047,KERSHAW1986273}. Since the electron distribution is isotropic, the kernel depends on the photon directions only through the scattering-angle cosine $\mu_0=\hat{\Omega}'\cdot\hat{\Omega}$. The induced probability
\begin{equation}
p_\nu=\frac{h^3c^2}{2\nu^3}
\label{eq:induced_probability}
\end{equation}
converts the specific intensity to the photon occupation number, so that $p_\nu I_\nu(\hat{\Omega})$ is the occupation number of the state $(\nu,\hat{\Omega})$.
We write the Compton integrand of Eq.~\eqref{eq:transport} compactly as
\begin{multline}
\mathcal{C}[\sigma;I,\tilde{I}](\nu,\hat{\Omega};\nu',\hat{\Omega}')
\\
\equiv
\left(\frac{\nu}{\nu'}\right)
\sigma(\nu'\mathbin{\to}\nu,\hat{\Omega}'\cdot\hat{\Omega})
I_{\nu'}(\hat{\Omega}')
\left[1+p_\nu \tilde{I}_\nu(\hat{\Omega})\right]
\\
-\sigma(\nu\mathbin{\to}\nu',\hat{\Omega}\cdot\hat{\Omega}')
I_\nu(\hat{\Omega})
\left[1+p_{\nu'}\tilde{I}_{\nu'}(\hat{\Omega}')\right]
\label{eq:compton_integrand}
\end{multline}
where $\sigma$ is the kernel, $I$ is the scattered radiation field, and $\tilde{I}$ is the radiation field in the induced scattering factors. We write the arguments of $\mathcal{C}$ explicitly only when they differ from $(\nu,\hat{\Omega};\nu',\hat{\Omega}')$. The Compton terms of Eq.~\eqref{eq:transport} are then $\int_0^\infty d\nu'\int_{4\pi}d\Omega'\,\mathcal{C}[\sigma_s;I,I]$. The gas internal-energy density $u_m$ satisfies
\begin{multline}
\frac{\partial u_m}{\partial t}
=
c\int_0^\infty\kappa_\nu E_\nu\,d\nu
-c\kappa_PU_r
\\
-\int_0^\infty d\nu\int_{4\pi}d\Omega
\int_0^\infty d\nu'\int_{4\pi}d\Omega'\,
\mathcal{C}[\sigma_s;I,I]
\label{eq:material_energy_transport}
\end{multline}
where
\begin{equation}
E_\nu=\frac{1}{c}\int_{4\pi}I_\nu\,d\Omega
\label{eq:spectral_energy_density}
\end{equation}
is the spectral radiation energy density, with units of energy per unit photon energy per unit volume, $U_r=a_RT^4$ is the blackbody radiation energy density at the gas temperature $T$, with $a_R$ the radiation constant, and
\begin{equation}
\kappa_P=\int_0^\infty\kappa_\nu b_\nu\,d\nu
\label{eq:planck_mean_opacity}
\end{equation}
is the Planck mean opacity, weighted by the normalized Planck spectrum
\begin{equation}
b_\nu=\frac{B_\nu}{U_r},\qquad \int_0^\infty b_\nu\,d\nu=1
\label{eq:b_nu}
\end{equation}

\subsection{Semi-Implicit Treatment of Compton Scattering}
\label{sec:semi_implicit}

As discussed in Section~\ref{sec:introduction}, we treat the temperature-dependent Compton source semi-implicitly. Previous studies have adopted implicit or linearized treatments of Compton energy exchange \citep{TILL2020108990,MCGRAW2023111980}. Here we extend the derivation of the IMC equations \citep{fleck_implicit_1971,Wollaber23022016} to include the temperature dependence of the Compton scattering cross section.

We define the heat capacity ratio
\begin{equation}
\overline{c}_v
\equiv
\frac{\partial u_m}{\partial U_r}
=
\frac{c_v}{4a_RT^3}
\label{eq:cv_bar}
\end{equation}
where $c_v=\partial u_m/\partial T$ is the gas heat capacity and $4a_RT^3=\partial U_r/\partial T$. We consider a time step $t^n\leq t\leq t^{n+1}$ with $\Delta t=t^{n+1}-t^n$, and quantities evaluated at $t^n$ and $t^{n+1}$ carry superscripts $n$ and $n+1$, respectively. We first freeze the temperature-dependent problem data at the beginning of the time step \citep{fleck_implicit_1971,Wollaber23022016}. The quantities $\kappa_\nu$, $b_\nu$, $\kappa_P$, and $\overline{c}_v$ are therefore evaluated at $t^n$, and we omit their superscript.

We integrate Eq.~\eqref{eq:material_energy_transport} over the time step, evaluating $U_r$, the radiation field, and $\sigma_s$ fully implicitly at the end of the step, $t^{n+1}$. The Compton terms are then nonlinear in the material temperature, through $\sigma_s$, and in the radiation field, through the induced scattering. We use the following linearization of the Compton integrand
\begin{multline}
\mathcal{C}[\sigma_s;I,I]^{n+1}
\simeq
\mathcal{C}[\sigma_s^n;I^{n+1},I^n]
\\
+\left(U_r^{n+1}-U_r^n\right)
\mathcal{C}\!\left[\frac{\partial\sigma_s^n}{\partial U_r};I^n,I^n\right]
\label{eq:sigma_linearization}
\end{multline}
The time-discretized gas energy equation is then
\begin{multline}
\overline{c}_v\frac{U_r^{n+1}-U_r^n}{\Delta t}
=
c\int_0^\infty\kappa_\nu E_\nu^{n+1}\,d\nu
-c\kappa_PU_r^{n+1}
\\
-\int_0^\infty d\nu\int_{4\pi}d\Omega
\int_0^\infty d\nu'\int_{4\pi}d\Omega'
\Big\{
\mathcal{C}[\sigma_s^n;I^{n+1},I^n]
\\
+\left(U_r^{n+1}-U_r^n\right)
\mathcal{C}\!\left[\frac{\partial\sigma_s^n}{\partial U_r};I^n,I^n\right]
\Big\}
\label{eq:material_discrete}
\end{multline}
Solving for $U_r^{n+1}$ gives
\begin{multline}
U_r^{n+1}
=
fU_r^n
+\frac{\Delta t}{\overline{c}_v}f
\Bigg[
c\int_0^\infty\kappa_\nu E_\nu^{n+1}\,d\nu
\\
-\int_0^\infty d\nu\int_{4\pi}d\Omega
\int_0^\infty d\nu'\int_{4\pi}d\Omega'
\Big\{
\mathcal{C}[\sigma_s^n;I^{n+1},I^n]
\\
-U_r^n\,
\mathcal{C}\!\left[\frac{\partial\sigma_s^n}{\partial U_r};I^n,I^n\right]
\Big\}
\Bigg]
\label{eq:U_implicit}
\end{multline}
where
\begin{equation}
f=
\left[
1+\frac{c\Delta t}{\overline{c}_v}
\left(\kappa_P+\kappa_C\right)
\right]^{-1}
\label{eq:fleck_continuous}
\end{equation}
is the generalized Fleck factor, and
\begin{multline}
\kappa_C=
\frac{1}{c}
\int_0^\infty d\nu\int_{4\pi}d\Omega
\int_0^\infty d\nu'\int_{4\pi}d\Omega'
\\
\times\mathcal{C}\!\left[\frac{\partial\sigma_s^n}{\partial U_r};I^n,I^n\right]
\label{eq:kappa_C}
\end{multline}
is the Compton coupling coefficient. When $\partial\sigma_s/\partial U_r=0$, $\kappa_C=0$ and Eq.~\eqref{eq:fleck_continuous} reduces to the Fleck factor $f=(1+c\Delta t\,\kappa_P/\overline{c}_v)^{-1}$ of \citet{fleck_implicit_1971}. Equation~\eqref{eq:U_implicit} can be written as
\begin{multline}
U_r^{n+1}-U_r^n
=
\frac{\Delta t}{\overline{c}_v}f
\Bigg[
c\int_0^\infty\kappa_\nu E_\nu^{n+1}\,d\nu
-c\kappa_PU_r^n
\\
-\int_0^\infty d\nu\int_{4\pi}d\Omega
\int_0^\infty d\nu'\int_{4\pi}d\Omega'\,
\mathcal{C}[\sigma_s^n;I^{n+1},I^n]
\Bigg]
\label{eq:U_increment}
\end{multline}

Equation~\eqref{eq:U_increment} gives $U_r^{n+1}$ in terms of the end-of-step radiation field. The transport equation, however, requires the time-continuous $U_r(t)$. We replace the end-of-step quantities by their time-continuous counterparts, $U_r^{n+1}\to U_r(t)$ and $I_\nu^{n+1}\to I_\nu(t)$ in Eq.~\eqref{eq:U_increment} \citep{fleck_implicit_1971,Wollaber23022016}. We apply the same linearization, Eq.~\eqref{eq:sigma_linearization} with $U_r^{n+1}$ and $I^{n+1}$ replaced by $U_r(t)$ and $I(t)$, to the Compton terms of Eq.~\eqref{eq:transport}. Writing $B_\nu=b_\nu U_r^n+b_\nu(U_r-U_r^n)$, substituting $U_r(t)$ into Eq.~\eqref{eq:transport}, and using 
\begin{equation}
    \frac{\Delta t\,f\,c(\kappa_P+\kappa_C)}{\overline{c}_v}=1-f
\end{equation} 
gives the semi-implicit transport equation
\begin{multline}
\frac{1}{c}\frac{\partial I_\nu}{\partial t}
+\hat{\Omega}\cdot\nabla I_\nu
+\kappa_\nu I_\nu
=
\frac{c\kappa_\nu b_\nu U_r^n}{4\pi}
\\
-(1-f)\,c\kappa_PU_r^n\,\phi_\nu(\hat{\Omega})
+\int_0^\infty d\nu'\int_{4\pi}d\Omega'\,
\mathcal{C}[\sigma_s^n;I,I^n]
\\
+(1-f)\,\phi_\nu(\hat{\Omega})
\Bigg[
c\int_0^\infty\kappa_{\nu'}E_{\nu'}\,d\nu'
-\int_0^\infty d\nu'\int_{4\pi}d\Omega'
\\
\times\int_0^\infty d\nu''\int_{4\pi}d\Omega''\,
\mathcal{C}[\sigma_s^n;I,I^n](\nu',\hat{\Omega}';\nu'',\hat{\Omega}'')
\Bigg]
\label{eq:transport_semi_implicit}
\end{multline}
where
\begin{multline}
\phi_\nu(\hat{\Omega})
=
\frac{1}{\kappa_P+\kappa_C}
\Bigg\{
\frac{\kappa_\nu b_\nu}{4\pi}
\\
+\frac{1}{c}\int_0^\infty d\nu'\int_{4\pi}d\Omega'\,
\mathcal{C}\!\left[\frac{\partial\sigma_s^n}{\partial U_r};I^n,I^n\right]
\Bigg\}
\label{eq:reemission_spectrum}
\end{multline}
is the generalized emission spectrum, normalized so that $\int_0^\infty d\nu\int_{4\pi}d\Omega\,\phi_\nu(\hat{\Omega})=1$. 

The terms proportional to $U_r^n$ in Eq.~\eqref{eq:transport_semi_implicit} form the emission source. When $\partial\sigma_s/\partial U_r=0$, $\kappa_C=0$ and $\phi_\nu=\kappa_\nu b_\nu/(4\pi\kappa_P)$, and they reduce to the emission source $fc\kappa_\nu b_\nu U_r^n/(4\pi)$ of \citet{fleck_implicit_1971}. The last term is linear in the radiation field and forms an effective scattering source, in which the fraction $1-f$ of the radiation absorbed by the gas, or given to the gas by Compton scattering, is instantaneously reemitted with the angle-dependent spectrum $\phi_\nu(\hat{\Omega})$ \citep{fleck_implicit_1971}. The remaining fraction $f$ is treated as effective absorption \citep{Wollaber23022016}. The temperature dependence of $\sigma_s$ contributes to $f$ through $\kappa_C$, and to both the emission source and the effective scattering through $\phi_\nu$. For $\kappa_P+\kappa_C>0$, $0<f<1$, and the fraction $1-f$ of effective scattering increases with $\Delta t$. In the limit of elastic scattering, $\sigma_s$ does not depend on the gas temperature, so that $\kappa_C=0$, and the induced scattering terms of Eq.~\eqref{eq:transport} cancel and need not be lagged, the semi-implicit transport equation then reduces to the IMC transport equation of \citet{fleck_implicit_1971}, with elastic scattering added as pure scattering.

Integrating Eq.~\eqref{eq:transport_semi_implicit} over direction and photon energy, and using the normalization of $\phi_\nu$, gives
\begin{multline}
\frac{\partial}{\partial t}\int_0^\infty E_\nu\,d\nu
+\nabla\cdot\vec{F}
=
-f
\Bigg[
c\int_0^\infty\kappa_\nu E_\nu\,d\nu
-c\kappa_PU_r^n
\\
-\int_0^\infty d\nu\int_{4\pi}d\Omega
\int_0^\infty d\nu'\int_{4\pi}d\Omega'\,
\mathcal{C}[\sigma_s^n;I,I^n]
\Bigg]
\label{eq:transport_energy_balance}
\end{multline}
where $\vec{F}=\int_0^\infty\int_{4\pi}\hat{\Omega}I_\nu\,d\Omega\,d\nu$ is the radiation flux. The right-hand side is the net rate of change of the radiation energy density due to the gas, and energy conservation implies an equal and opposite rate of change of the gas internal-energy density \citep{fleck_implicit_1971},
\begin{multline}
\frac{\partial u_m}{\partial t}
=
f
\Bigg[
c\int_0^\infty\kappa_\nu E_\nu\,d\nu
-c\kappa_PU_r^n
\\
-\int_0^\infty d\nu\int_{4\pi}d\Omega
\int_0^\infty d\nu'\int_{4\pi}d\Omega'\,
\mathcal{C}[\sigma_s^n;I,I^n]
\Bigg]
\label{eq:material_semi_implicit}
\end{multline}
The same result follows from substituting $U_r(t)$ into the linearized Eq.~\eqref{eq:material_energy_transport} and collecting terms. When Eqs.~\eqref{eq:transport_energy_balance} and \eqref{eq:material_semi_implicit} are added, the absorption--emission and Compton exchange terms cancel, so the semi-implicit equations satisfy the same energy balance as the TRT equations \citep{Wollaber23022016}. These semi-implicit equations do not depend on how the transport equation is solved, and apply equally to Monte Carlo, discrete-ordinates, or moment-based methods.

\section{The Monochromatic Radiation Diffusion Equation}
\label{sec:monochromatic_equations}

We now reduce the semi-implicit TRT scheme of Section~\ref{sec:transport} to the flux-limited diffusion (FLD) approximation, which is the focus of this paper. We use the comoving-frame, frequency-dependent FLD equations \citep{pomraning1973equations,mihalas1999foundations,Castor_2004,Giron_2026}, which we extend here to include Compton scattering. In contrast to Section~\ref{sec:transport}, we now include bulk material motion, with gas velocity $\vec{v}$. The spectral radiation energy density $E_\nu(\vec{x},t)$ of Eq.~\eqref{eq:spectral_energy_density} satisfies
\begin{multline}
\frac{\partial E_\nu}{\partial t}+\nabla\cdot\left(\vec{v}E_\nu\right)=
-\frac{1-R_{2,\nu}}{2}E_\nu\nabla\cdot\vec{v}
+\nabla\cdot\left(D_\nu\nabla E_\nu\right)\\
+c\kappa_\nu\left(b_\nu U_r-E_\nu\right)
+\left(\nabla\cdot\vec{v}\right)
\frac{\partial}{\partial\nu}
\left[\frac{\nu E_\nu(1-R_{2,\nu})}{2}\right]
\\
+c\int_0^\infty
\left(\frac{\nu}{\nu'}\right)
\sigma_s(\nu'\mathbin{\to}\nu;\rho,T)E_{\nu'}
\left(1+n_\nu\right)d\nu'\\
-c\int_0^\infty
\sigma_s(\nu\mathbin{\to}\nu';\rho,T)E_\nu
\left(1+n_{\nu'}\right)d\nu'
\label{eq:monochromatic}
\end{multline}
Here
\begin{equation}
\sigma_s(\nu'\mathbin{\to}\nu;\rho,T)
=2\pi\int_{-1}^{1}\sigma_s(\nu'\mathbin{\to}\nu,\mu_0;\rho,T)\,d\mu_0
\label{eq:angle_integrated_kernel}
\end{equation}
is the angle-integrated kernel, and
\begin{equation}
n_\nu=\frac{1}{4\pi}\int_{4\pi}p_\nu I_\nu\,d\Omega=\frac{h^3c^3}{8\pi\nu^3}E_\nu
\label{eq:occupation_spectral}
\end{equation}
is the angular mean of the occupation number.
The variable Eddington factor $R_{2,\nu}$, dimensionless radiation gradient $R_\nu$, and Levermore--Pomraning \citep{Pomeraning-Levermore,Levermore84} flux limiter $\lambda_\nu$ are
\begin{equation}
R_{2,\nu}=\lambda_\nu+\lambda_\nu^2R_\nu^2
\label{eq:fld_closure}
\end{equation}
\begin{equation}
    \lambda_{\nu}=\frac{1}{R_{\nu}}\left(\coth R_{\nu}-\frac{1}{R_{\nu}}\right)
    \quad
    R_\nu=\frac{|\nabla E_\nu|}{\kappa_{t,\nu}E_\nu}
\end{equation}
where $\kappa_{t,\nu}$ is the total opacity, the sum of the absorption coefficient and the electron scattering opacity. The diffusion coefficient is
\begin{equation}
D_\nu=\frac{c\lambda_\nu}{\kappa_{t,\nu}}
\label{eq:diffusion_coefficient}
\end{equation}
We refer the reader to \citet{Giron_2026} for a detailed discussion of the non-Compton terms and focus here on Compton scattering.

The Compton terms in Eq.~\eqref{eq:monochromatic} follow from integrating those in Eq.~\eqref{eq:transport} over direction. For the spontaneous scattering this integration is exact, since $\int_{4\pi}\sigma_s(\nu'\mathbin{\to}\nu,\hat{\Omega}'\cdot\hat{\Omega})\,d\Omega=\sigma_s(\nu'\mathbin{\to}\nu)$ for every $\hat{\Omega}'$. The induced scattering, however, depends on the angular correlation of the radiation field at the two photon energies. Consistent with the diffusion approximation, we evaluate the induced scattering terms for a nearly isotropic radiation field, replacing $I_\nu(\hat{\Omega})$ by $cE_\nu/(4\pi)$ and $p_\nu I_\nu(\hat{\Omega})$ by $n_\nu$ in them. The direction integrals of the Compton integrand of Eq.~\eqref{eq:compton_integrand} then become
\begin{equation}
\int_{4\pi}d\Omega\int_{4\pi}d\Omega'\,\mathcal{C}[\sigma;I,\tilde{I}]
\simeq
c\,\mathcal{C}_E[\sigma;E,n]
\label{eq:compton_integrand_isotropic}
\end{equation}
where
\begin{multline}
\mathcal{C}_E[\sigma;E,n]
\equiv
\left(\frac{\nu}{\nu'}\right)\sigma(\nu'\mathbin{\to}\nu)E_{\nu'}\left(1+n_\nu\right)
\\
-\sigma(\nu\mathbin{\to}\nu')E_\nu\left(1+n_{\nu'}\right)
\label{eq:compton_integrand_E}
\end{multline}
is a function of $\nu$ and $\nu'$, $\sigma(\nu'\mathbin{\to}\nu)$ is the angle-integrated kernel of Eq.~\eqref{eq:angle_integrated_kernel}, $E$ is the spectral energy density of $I$, and $n$ is the angular-mean occupation number of $\tilde{I}$. This gives the Compton terms of Eq.~\eqref{eq:monochromatic}, which are $c\int_0^\infty d\nu'\,\mathcal{C}_E[\sigma_s;E,n]$.

Including Compton scattering, the evolution equation for the total gas energy density $E_m$, accounting for both radiative energy and momentum exchange, is
\begin{multline}
\frac{\partial E_m}{\partial t}
+\nabla\cdot\left[(E_m+P)\vec{v}\right]
=
-\vec{v}\cdot\int_0^\infty
\lambda_\nu\nabla E_\nu\,d\nu
\\
+c\int_0^\infty
\kappa_\nu\left(E_\nu-b_\nu U_r\right)d\nu
\\
-c\int_0^\infty d\nu\int_0^\infty d\nu'\,
\mathcal{C}_E[\sigma_s;E,n]
\label{eq:material_total_energy_spectral}
\end{multline}
where $P$ is the gas pressure.
The momentum density obeys
\begin{equation}
\frac{\partial(\rho\vec{v})}{\partial t}
+\nabla\cdot(\rho\vec{v}\vec{v})
=
-\nabla P
-\int_0^\infty\lambda_\nu\nabla E_\nu\,d\nu
\label{eq:momentum_spectral}
\end{equation}
The \textsc{rich} code advances the hydrodynamic and radiation terms in separate steps (Section~\ref{sec:implementation}). We therefore omit the hydrodynamic terms below and focus on the quantities evolved during the radiation step.

When the total gas energy equation is separated into internal-energy and kinetic-energy contributions, the gas internal-energy density $u_m$ satisfies the nearly isotropic form of Eq.~\eqref{eq:material_energy_transport},
\begin{multline}
\frac{\partial u_m}{\partial t}
=
c\int_0^\infty\kappa_\nu E_\nu\,d\nu
-c\kappa_PU_r
\\
-c\int_0^\infty d\nu\int_0^\infty d\nu'\,
\mathcal{C}_E[\sigma_s;E,n]
\label{eq:material_energy_spectral}
\end{multline}
and the gas kinetic-energy density $K$ evolves due to radiation--matter momentum exchange as
\begin{equation}
\frac{\partial K}{\partial t}
=
-\vec{v}\cdot
\int_0^\infty\lambda_\nu\nabla E_\nu\,d\nu
\label{eq:kinetic_energy_spectral}
\end{equation}

During the radiation step we apply the semi-implicit treatment of Section~\ref{sec:semi_implicit}. With the same nearly isotropic evaluation of the induced scattering terms and of $I_\nu^n$, the generalized Fleck factor has the same form as Eq.~\eqref{eq:fleck_continuous}, with the Compton coupling coefficient
\begin{equation}
\kappa_C=
\int_0^\infty d\nu\int_0^\infty d\nu'\,
\mathcal{C}_E\!\left[\frac{\partial\sigma_s^n}{\partial U_r};E^n,n^n\right]
\label{eq:fleck_diffusion}
\end{equation}
and the integral over direction of Eq.~\eqref{eq:reemission_spectrum} becomes
\begin{equation}
\phi_\nu=
\frac{
\kappa_\nu b_\nu
+\int_0^\infty d\nu'\,
\mathcal{C}_E\!\left[\partial\sigma_s^n/\partial U_r;E^n,n^n\right]
}{\kappa_P+\kappa_C}
\label{eq:reemission_spectrum_diffusion}
\end{equation}
Integrating Eq.~\eqref{eq:transport_semi_implicit} over direction, using the FLD closure for the radiation flux, and retaining the radiation pressure work and Doppler terms of Eq.~\eqref{eq:monochromatic}, gives
\begin{multline}
\frac{\partial E_\nu}{\partial t}
=
-\frac{1-R_{2,\nu}}{2}E_\nu\nabla\cdot\vec{v}
+\nabla\cdot\left(D_\nu\nabla E_\nu\right)
\\
+\left(\nabla\cdot\vec{v}\right)
\frac{\partial}{\partial\nu}
\left[\frac{\nu E_\nu(1-R_{2,\nu})}{2}\right]
+c\kappa_\nu\left(b_\nu U_r^n-E_\nu\right)
\\
-(1-f)\,c\kappa_PU_r^n\phi_\nu
+c\int_0^\infty d\nu'\,
\mathcal{C}_E[\sigma_s^n;E,n^n]
\\
+(1-f)\,c\phi_\nu
\Bigg[
\int_0^\infty\kappa_{\nu'}E_{\nu'}\,d\nu'
\\
-\int_0^\infty d\nu'\int_0^\infty d\nu''\,
\mathcal{C}_E[\sigma_s^n;E,n^n](\nu',\nu'')
\Bigg]
\label{eq:monochromatic_semi_implicit}
\end{multline}
while the gas internal energy evolves according to the nearly isotropic form of Eq.~\eqref{eq:material_semi_implicit}. The advection term is treated in the hydrodynamic step.

\section{The Multigroup Diffusion Equation}
\label{sec:multigroup_equations}
To obtain the multigroup diffusion equation, we discretize the monochromatic equations of Section~\ref{sec:monochromatic_equations} in photon energy. We divide the photon-energy spectrum into $G$ groups, with group $g$ spanning the interval $[\nu_{g-1/2},\nu_{g+1/2}]$. The group radiation energy density $E_g$ and dimensionless group Planck weight $b_g$ are defined by integrating their respective spectral quantities (Eqs.~\eqref{eq:spectral_energy_density} and \eqref{eq:b_nu}) over group $g$
\begin{equation}
E_g=\int_{\nu_{g-1/2}}^{\nu_{g+1/2}}E_\nu\,d\nu,
\qquad
b_g=\int_{\nu_{g-1/2}}^{\nu_{g+1/2}}b_\nu\,d\nu
\label{eq:group_variables}
\end{equation}
The group Planck weights $b_g$ are evaluated as described in Appendix~\ref{sec:computing_group_planck_integrals}. The group width $\Delta_g$ and group center $\nu_g$ are
\begin{equation}
\Delta_g=\nu_{g+1/2}-\nu_{g-1/2},
\qquad
\nu_g=\sqrt{\nu_{g+1/2}\,\nu_{g-1/2}}
\label{eq:group_grid}
\end{equation}
We assume that the photon-energy grid spans a sufficiently broad range that, for all relevant gas temperatures, the group Planck weights approximately satisfy the normalization in Eq.~\eqref{eq:b_nu},
\begin{equation}
\sum_{g=1}^{G}b_g(T)\simeq1
\label{eq:group_planck_normalization}
\end{equation}
Integrating Eq.~\eqref{eq:monochromatic_semi_implicit} over group $g$ gives
\begin{multline}
\frac{\partial E_g}{\partial t}
=
-\frac{1-R_{2,g}}{2}E_g\nabla\cdot\vec{v}
\\
+\nabla\cdot\left(D_g\nabla E_g\right)
+c\kappa_{P,g}\left(b_gU_r^n-E_g\right)
\\
\shoveright{+\left(\nabla\cdot\vec{v}\right)
\int_{\nu_{g-1/2}}^{\nu_{g+1/2}}
\frac{\partial}{\partial\nu}
\left[
\frac{\nu E_\nu(1-R_{2,\nu})}{2}
\right]d\nu}
\\
-(1-f)\,c\kappa_PU_r^n\phi_g
+c\sum_{g'}S_{g'g}E_{g'}
\\
+(1-f)\,c\phi_g
\Bigg[
\sum_{g'}\kappa_{P,g'}E_{g'}
-\sum_{g'g''}S_{g'g''}E_{g'}
\Bigg]
\label{eq:multigroup}
\end{multline}
where $f$ and $\phi_g$ are the group counterparts of Eqs.~\eqref{eq:fleck_continuous} and \eqref{eq:reemission_spectrum_diffusion}, defined below, $\kappa_P$ is the total Planck mean opacity of Eq.~\eqref{eq:total_planck_opacity}, and the matrix element $S_{g'g}$ describes the redistribution of radiation from group $g'$ to group $g$ due to Compton scattering, and is defined by
\begin{multline}
S_{g'g}=
\frac{\nu_g}{\nu_{g'}}\,\sigma_s(g'\mathbin{\to}g;\rho,T)(1+n_g)
\\
-\delta_{g'g}\sum_{g''}
\sigma_s(g'\mathbin{\to}g'';\rho,T)(1+n_{g''})
\label{eq:scattering_matrix}
\end{multline}
Here $\delta_{g'g}$ is the Kronecker delta and $\sigma_s(g'\mathbin{\to}g)$ is the group-averaged Compton scattering cross section from group $g'$ to group $g$, described in Section~\ref{sec:compton_coefficients}. 
The group occupation number is approximated as
\begin{equation}
n_g=\frac{h^3c^3}{8\pi\nu_g^3\Delta_g}E_g
\label{eq:occupation_group}
\end{equation}
The matrix $S_{g'g}$ is the group counterpart of the integrand $\mathcal{C}_E$ of Eq.~\eqref{eq:compton_integrand_E}, such that $\sum_{g'}S_{g'g}E_{g'}$ approximates the group integral $\int_{\nu_{g-1/2}}^{\nu_{g+1/2}}d\nu\int_0^\infty d\nu'\,\mathcal{C}_E[\sigma_s;E,n]$ of the monochromatic Compton source, with the induced-scattering factor taken as constant within each group.

The derivative $\partial S_{g'g}/\partial U_r$ (Section~\ref{sec:temperature_derivatives}) includes the temperature dependence of the group Planck weights and of the detailed-balance correction (Section~\ref{sec:compton_coefficients}), so it is not exactly the group integral of $\partial\sigma_s/\partial U_r$. We therefore linearize the group Compton source directly,
\begin{multline}
S_{g'g}(U_r^{n+1})E_{g'}^{n+1} \\
\simeq
S_{g'g}(U_r^n)E_{g'}^{n+1}
+
\left(U_r^{n+1}-U_r^n\right)
\left.\frac{\partial S_{g'g}}{\partial U_r}\right|^n E_{g'}^n
\label{eq:matrix_linearization}
\end{multline}
where unsuperscripted matrix elements and occupation numbers are evaluated at $t^n$. The multigroup generalized Fleck factor $f$ then has the form of Eq.~\eqref{eq:fleck_continuous}, with the Compton coupling coefficient
\begin{equation}
\kappa_C=
\sum_{g'g''}
\left.\frac{\partial S_{g'g''}}{\partial U_r}\right|^n
E_{g'}^n
\label{eq:generalized_fleck}
\end{equation}
and
\begin{equation}
\phi_g=
\frac{
\kappa_{P,g}b_g
+\sum_{g'}\left.\partial S_{g'g}/\partial U_r\right|^nE_{g'}^n
}{\kappa_P+\kappa_C}
\label{eq:reemission_spectrum_group}
\end{equation}
satisfies $\sum_g\phi_g=1$.
The group Eddington factor $R_{2,g}$, dimensionless radiation gradient
$R_g$, and Levermore--Pomraning flux limiter $\lambda_g$ are
\begin{equation}
R_{2,g}=\lambda_g+\lambda_g^2R_g^2
\label{eq:group_fld_closure}
\end{equation}
\begin{equation}
    \lambda_g=\frac{1}{R_g}
    \left(\coth R_g-\frac{1}{R_g}\right)
    \quad
    R_g=\frac{|\nabla E_g|}{\kappa_{R,g}E_g}
\end{equation}
where $\kappa_{R,g}$ is the group Rosseland mean opacity. The group
diffusion coefficient is
\begin{equation}
D_g=\frac{c\lambda_g}{\kappa_{R,g}}
\label{eq:group_diffusion_coefficient}
\end{equation}
The group Planck and Rosseland mean opacities are defined as
\begin{equation}
\kappa_{P,g}=
\frac{
\int_{\nu_{g-1/2}}^{\nu_{g+1/2}}
\kappa_\nu B_\nu\,d\nu
}{
\int_{\nu_{g-1/2}}^{\nu_{g+1/2}}
B_\nu\,d\nu
}
\label{eq:planck_group_opacity}
\end{equation}
and
\begin{equation}
\kappa_{R,g}^{-1}=
\frac{
\int_{\nu_{g-1/2}}^{\nu_{g+1/2}}
\kappa_{t,\nu}^{-1}
\left(\partial B_\nu/\partial T\right)d\nu
}{
\int_{\nu_{g-1/2}}^{\nu_{g+1/2}}
\left(\partial B_\nu/\partial T\right)d\nu
}
\label{eq:rosseland_group_opacity}
\end{equation}
The total Planck mean opacity is
\begin{equation}
\kappa_P=\sum_{g'} \kappa_{P,g'}b_{g'}
\label{eq:total_planck_opacity}
\end{equation}

Summing the absorption--emission and Compton terms of Eq.~\eqref{eq:multigroup} over all groups gives the corresponding gas internal-energy equation with the opposite sign,
\begin{multline}
\frac{\partial u_m}{\partial t}
=
cf
\Bigg[
\sum_{g'} \kappa_{P,g'}E_{g'}
-\kappa_PU_r^n
-\sum_{g'g''}S_{g'g''}E_{g'}
\Bigg]
\label{eq:material_energy}
\end{multline}
The last term, multiplied by $cf$, is the net rate of change of the gas energy due to Compton scattering.

\section{Compton Scattering Coefficients}
\label{sec:compton_coefficients}

The group center $\nu_g$ of Eq.~\eqref{eq:group_grid} is an average photon energy of group $g$. It sets the group occupation number $n_g$ of Eq.~\eqref{eq:occupation_group} and the photon number density $E_g/\nu_g$ of group $g$, so that the total photon number density is
\begin{equation}
N_\gamma=\sum_g\frac{E_g}{\nu_g}
\label{eq:photon_number}
\end{equation}
The scattering matrix of Eq.~\eqref{eq:scattering_matrix} moves photons between the group centers. A transfer from group $g$ to group $g'\neq g$ removes energy from group $g$ at the rate $c\,\sigma_s(g\mathbin{\to}g')(1+n_{g'})E_g$ and deposits energy in group $g'$ at $(\nu_{g'}/\nu_g)$ times this rate, so that the same number of photons leaves group $g$ and enters group $g'$. Thus $c\,\sigma_s(g\mathbin{\to}g')(1+n_{g'})$ is a rate per photon of group $g$. A scattering from $\nu$ in group $g$ to $\nu'$ in group $g'$ changes the photon energy by $\nu'-\nu$, whereas a transfer between the group centers carries $\nu_{g'}-\nu_g$. We therefore count each such scattering with the weight $(\nu'-\nu)/(\nu_{g'}-\nu_g)$, and average the kernel over the photon distribution $B_\nu/\nu$ within the source group, normalized to $B_g/\nu_g$, the photon number density of group $g$ for a Planck spectrum with $B_g=b_gU_r$. For $g'\neq g$, the group-averaged Compton scattering cross section is
\begin{multline}
\sigma_s(g\mathbin{\to}g';\rho,T)=
\frac{\nu_g}{B_{g}}
\int_{g}d\nu\,\frac{B_\nu}{\nu}\int_{g'} d\nu'\,
\\
\times\sigma_s(\nu\mathbin{\to}\nu';\rho,T)\,
\frac{\nu'-\nu}{\nu_{g'}-\nu_g}
\label{eq:sigma_s_group}
\end{multline}
where $\int_g$ denotes integration over group $g$. The weight $(\nu'-\nu)/(\nu_{g'}-\nu_g)$ is positive for all $\nu$ in group $g$ and $\nu'$ in group $g'\neq g$, so $\sigma_s(g\mathbin{\to}g')\geq0$. Scatterings within a group, which change the photon energy without moving photons between groups, are neglected. The energy exchange of the transfer, per unit $cE_g(1+n_{g'})$, is
\begin{multline}
\sigma_s(g\mathbin{\to}g')\left(\frac{\nu_{g'}}{\nu_g}-1\right)
\\
=\frac{1}{B_{g}}
\int_{g}d\nu\int_{g'} d\nu'\,
B_\nu\,\sigma_s(\nu\mathbin{\to}\nu')\,\frac{\nu'-\nu}{\nu}
\label{eq:sigma_s_energy_exchange}
\end{multline}
which equals the energy exchange of the monochromatic kernel averaged over a Planck spectrum within group $g$, and does not depend on the value of $\nu_g$ within the group. By construction, the Compton source conserves $N_\gamma$ for any occupation numbers,
\begin{equation}
\sum_g\frac{1}{\nu_g}\sum_{g'}S_{g'g}E_{g'}=0
\label{eq:photon_number_conservation}
\end{equation}

The microscopic coefficients
$\sigma_s^{\mathrm{mic}}(g\mathbin{\to}g',T)$ and their temperature derivatives are computed with \textsc{mace} \citep{compton_kernel_integration_paper}, which integrates the thermal Compton kernel over pairs of energy groups with arbitrary weight functions of the initial and final photon energies, here $B_\nu/\nu$ and $(\nu'-\nu)/(\nu_{g'}-\nu_g)$ of Eq.~\eqref{eq:sigma_s_group}. The resulting coefficients are tabulated on a grid spanning the full range of gas temperatures relevant to the simulations and interpolated linearly in temperature at runtime.
The macroscopic scattering coefficients used in the radiation and gas equations are obtained by multiplying the microscopic cross sections by the local electron number density. For simplicity we assume that the gas consists entirely of ionized hydrogen, so that
\begin{equation}
N_e=\rho N_A
\label{eq:electron_density}
\end{equation}
where $N_A$ is the Avogadro constant. For a fully ionized gas of another composition, $N_e$ in Eq.~\eqref{eq:electron_density} is divided by the mean molecular weight per electron and the method is otherwise unchanged. The macroscopic coefficients are then
\begin{equation}
\sigma_s(g\mathbin{\to}g';\rho,T)
=
N_e\,\sigma_s^{\mathrm{mic}}(g\mathbin{\to}g',T)
\label{eq:macroscopic_compton_coefficients}
\end{equation}
Since the electron number density is independent of temperature, the
temperature derivatives of the macroscopic coefficients are obtained
directly from the tabulated microscopic derivatives,
\begin{equation}
\frac{\partial \sigma_s}{\partial T}
=
N_e\frac{\partial \sigma_s^{\mathrm{mic}}}{\partial T}
\label{eq:macroscopic_compton_derivatives}
\end{equation}

\subsection{Enforcing Detailed Balance}
\label{sec:detailed_balance}
For an isotropic Planck field, $I_\nu=cB_\nu/(4\pi)$, the occupation number of every state is the equilibrium occupation number
\begin{equation}
n_\nu^B=
\frac{h^3c^3}{8\pi\nu^3}B_\nu
=
\frac{1}{\exp(\nu/k_BT)-1}
\label{eq:equilibrium_occupation_spectral}
\end{equation}
Detailed balance requires that, in such a field, the scattering contributions between every pair of photon energies $\nu'$ and $\nu$ cancel. The kernel satisfies this at each scattering angle \citep{TILL2020108990,POMRANING19721047}, and integrating over $\mu_0$ with Eq.~\eqref{eq:angle_integrated_kernel} gives
\begin{multline}
\left(\frac{\nu}{\nu'}\right)
\sigma_s(\nu'\mathbin{\to}\nu;\rho,T)
B_{\nu'}
\left(1+n_\nu^B\right)
\\
=
\sigma_s(\nu\mathbin{\to}\nu';\rho,T)
B_\nu
\left(1+n_{\nu'}^B\right)
\label{eq:detailed_balance_spectral}
\end{multline}

For the multigroup operator, the equilibrium group occupation number is
\begin{equation}
n_g^B=
\frac{h^3c^3}{8\pi\nu_g^3\Delta_g}B_g
\label{eq:equilibrium_occupation_group}
\end{equation}
In equilibrium, detailed balance requires that the numbers of photons exchanged between every pair of groups balance,
\begin{multline}
\sigma_s(g'\mathbin{\to}g;\rho,T)
\frac{B_{g'}}{\nu_{g'}}
\left(1+n_g^B\right)
\\
=
\sigma_s(g\mathbin{\to}g';\rho,T)
\frac{B_g}{\nu_g}
\left(1+n_{g'}^B\right)
\label{eq:detailed_balance_group}
\end{multline}
Since each transferred photon carries $\nu_g$ out of group $g$ and $\nu_{g'}$ into group $g'$, Eq.~\eqref{eq:detailed_balance_group} also balances the energy exchange of the pair in Eq.~\eqref{eq:scattering_matrix}.

Equation~\eqref{eq:detailed_balance_spectral} holds exactly for the monochromatic Compton kernel. In constructing the multigroup operator, however, the induced-scattering factor is approximated as constant within each group, such that $n_\nu$ is replaced by the group occupation number $n_g$. Consequently, the group-averaged cross sections of a pair of groups, $\sigma_s(g\mathbin{\to}g')$ and $\sigma_s(g'\mathbin{\to}g)$, do not in general satisfy Eq.~\eqref{eq:detailed_balance_group} exactly.

We therefore enforce Eq.~\eqref{eq:detailed_balance_group} pairwise at the local gas temperature, after the tabulated coefficients have been interpolated to it. For each pair of groups, it can
be used either to reconstruct the reverse coefficient from the
forward coefficient or vice versa. Defining
\begin{equation}
\mathcal{R}_{g'g}(T)
=
\frac{
\nu_{g'}B_g\left(1+n_{g'}^B\right)
}{
\nu_gB_{g'}\left(1+n_g^B\right)
}
\label{eq:db_factor}
\end{equation}
the two equivalent relations are
\begin{equation}
\sigma_s^{\mathrm{mic}}
(g'\mathbin{\to}g,T)
=
\mathcal{R}_{g'g}(T)\,
\sigma_s^{\mathrm{mic}}
(g\mathbin{\to}g',T)
\label{eq:reverse_from_forward}
\end{equation}
and
\begin{equation}
\sigma_s^{\mathrm{mic}}
(g\mathbin{\to}g',T)
=
\mathcal{R}_{g'g}^{-1}(T)\,
\sigma_s^{\mathrm{mic}}
(g'\mathbin{\to}g,T)
\label{eq:forward_from_reverse}
\end{equation}
For each pair, the correction direction is chosen from the ratio of the two sides of Eq.~\eqref{eq:detailed_balance_group} evaluated with the raw coefficients,
\begin{equation}
F_{g'g}(T)
=
\frac{
\sigma_s^{\mathrm{mic}}(g'\mathbin{\to}g,T)
}{
\mathcal{R}_{g'g}(T)\,
\sigma_s^{\mathrm{mic}}(g\mathbin{\to}g',T)
}
\label{eq:db_residual}
\end{equation}
If $F_{g'g}>1$ we apply Eq.~\eqref{eq:reverse_from_forward}, and if $F_{g'g}<1$ we apply Eq.~\eqref{eq:forward_from_reverse}. The coefficient that exceeds its detailed-balance value is thus multiplied by $F_{g'g}^{-1}$ or $F_{g'g}$, respectively, which is smaller than one, while the other coefficient is left unchanged. This choice avoids evaluating a ratio with an extremely small denominator, which can lead to floating-point errors.

\subsection{Scattering-Coefficient Derivatives}
\label{sec:temperature_derivatives}

When the reverse coefficient is reconstructed from the forward
coefficient according to Eq.~\eqref{eq:reverse_from_forward},
its temperature derivative is
\begin{multline}
\frac{\partial
\sigma_s^{\mathrm{mic}}
(g'\mathbin{\to}g,T)}
{\partial T}
=
\mathcal{R}_{g'g}
\Bigg[
\frac{\partial
\sigma_s^{\mathrm{mic}}
(g\mathbin{\to}g',T)}
{\partial T}
\\
+
\sigma_s^{\mathrm{mic}}
(g\mathbin{\to}g',T)
\frac{\partial\ln\mathcal{R}_{g'g}}{\partial T}
\Bigg]
\label{eq:reverse_derivative}
\end{multline}
The logarithmic derivative of the detailed-balance ratio is
\begin{multline}
\frac{\partial\ln\mathcal{R}_{g'g}}{\partial T}
=
\frac{\partial}{\partial T}
\ln\left[
B_g\left(1+n_{g'}^B\right)
\right]
\\
-
\frac{\partial}{\partial T}
\ln\left[
B_{g'}\left(1+n_g^B\right)
\right]
\label{eq:db_log_derivative}
\end{multline}
When the forward coefficient is instead reconstructed
from the reverse coefficient according to
Eq.~\eqref{eq:forward_from_reverse}, its temperature derivative is
\begin{multline}
\frac{\partial
\sigma_s^{\mathrm{mic}}
(g\mathbin{\to}g',T)}
{\partial T}
=
\mathcal{R}_{g'g}^{-1}
\Bigg[
\frac{\partial
\sigma_s^{\mathrm{mic}}
(g'\mathbin{\to}g,T)}
{\partial T}
\\
-
\sigma_s^{\mathrm{mic}}
(g'\mathbin{\to}g,T)
\frac{\partial\ln\mathcal{R}_{g'g}}{\partial T}
\Bigg]
\label{eq:forward_derivative}
\end{multline}
Since $n_g^B\propto B_g$ (Eq.~\eqref{eq:equilibrium_occupation_group}), the derivatives in Eq.~\eqref{eq:db_log_derivative} require only $\partial B_g/\partial T$. Defining $x_{g\pm1/2}=\nu_{g\pm1/2}/(k_BT)$, we evaluate it as
\begin{multline}
    \frac{\partial B_g}{\partial T}
    =
    a_RT^3
    \Bigg[
    4b_g(T)
    +\frac{15}{\pi^4}
    \Bigg(
    \frac{x_{g-1/2}^4}{e^{x_{g-1/2}}-1}
    \\
    -\frac{x_{g+1/2}^4}{e^{x_{g+1/2}}-1}
    \Bigg)
    \Bigg]
    \label{eq:group_planck_temperature_derivative}
\end{multline}

The derivative $\partial S_{g'g}/\partial T$ is given by Eq.~\eqref{eq:scattering_matrix} with $\sigma_s$ replaced by $\partial\sigma_s/\partial T$ and the occupation numbers $n_g$ of the radiation held fixed.

\section{Numerical Scheme}
\label{sec:implementation}

The \textsc{rich} code solves the hydrodynamic and multigroup radiation equations
using an operator-split approach on a moving Voronoi mesh. The numerical implementation of multigroup FLD in \textsc{rich}, including the spatial discretization and radiation--matter coupling, is described in detail by \citet{Giron_2026}, building on the gray RHD implementation of \citet{steinberg_streamdisk_2024}. The moving-mesh hydrodynamics is described elsewhere \citep{Steinberg+15,yalinewich_rich_2015,Steinberg+16}. Here we present only the modifications required to incorporate Compton scattering into the radiation step.

\subsection{Compton Coupling in the Linear System}
\label{sec:cell_matrix}

During the radiation step, \textsc{rich} advances the group radiation energy density $E_{g,i}$ in every cell $i$ and for every photon-energy group $g$. There are $N\times G$ unknowns in total, where $N$ is the number of cells and $G$ the number of groups. We integrate Eq.~\eqref{eq:multigroup} over the time step, with the radiation energy densities evaluated at the end of the time step. The multigroup equations then form the linear system
\begin{equation}
\vec{M}\vec{E}^{n+1}=\vec{b}
\label{eq:linear_system}
\end{equation}
where $\vec{E}^{n+1}$ contains the group radiation energy densities for all cells and
photon-energy groups, and $\vec{b}$ is the vector of right-hand sides.

For destination group $g$ in cell $i$, the multigroup equation is
\begin{multline}
V_i\left(1+c\Delta t\,\kappa_{P,g,i}\right)E_{g,i}^{n+1}
+\mathcal{L}_{g,i}\left[\vec{E}^{n+1}\right]
\\
-V_ic\Delta t\sum_{g'}S_{g'g,i}^nE_{g',i}^{n+1}
\\
-V_i\mathcal{A}_{g,i}
\left[
\sum_{g'}\kappa_{P,g',i}E_{g',i}^{n+1}
-\sum_{g'g''}S_{g'g'',i}^nE_{g',i}^{n+1}
\right]
\\
=V_iE_{g,i}^n
+V_iU_{r,i}^n
\left(c\Delta t\,\kappa_{P,g,i}b_{g,i}-\kappa_{P,i}\mathcal{A}_{g,i}\right)
\label{eq:cell_equation}
\end{multline}
Here a subscript $i$ denotes a quantity in cell $i$, $V_i$ is the volume of cell $i$, and $\mathcal{L}_{g,i}[\vec{E}^{n+1}]$ contains the non-Compton terms, which are the diffusion, compression, and Doppler terms. These terms are implemented as in \citet{Giron_2026} and are given in Appendix~\ref{sec:discrete_multigroup_equation}.\footnote{The implementation in \textsc{rich} is unchanged. Appendix~\ref{sec:discrete_multigroup_equation} follows the code and corrects errors in the discrete equations printed by \citet{Giron_2026}: the face opacity (their Eq.~23), the indexing of $R_2$ in the Doppler operator (their Eqs.~30, 31, and 33), and two signs in the expansion operator (their Eq.~33).} The coupling coefficient is
\begin{multline}
\mathcal{A}_{g,i}
=
\frac{(c\Delta t)^2}{\overline{c}_{v,i}}\,
f_i
\left[
\kappa_{P,g,i}b_{g,i}
+
\sum_{g'}
\left.
\frac{\partial S_{g'g}}{\partial U_r}
\right|_i^n
E_{g',i}^n
\right]
\\
=c\Delta t\left(1-f_i\right)\phi_{g,i}
\label{eq:coupling_coefficient}
\end{multline}
where $\phi_{g,i}$ is $\phi_g$ of Eq.~\eqref{eq:reemission_spectrum_group} in cell $i$. The left-hand side of Eq.~\eqref{eq:cell_equation} defines the matrix
$\vec{M}$. The
right-hand side contains only quantities known at the beginning of the
time step and defines $\vec{b}$.

\subsection{Numerical Solution of the Multigroup Equation}
\label{sec:linear_solve}

The matrix $\vec{M}$ of Eq.~\eqref{eq:linear_system} has a block-sparse structure. Diffusion couples each group only
to the same group in neighboring cells, while within each cell the different
photon-energy groups are coupled through Compton scattering, the Doppler
terms, and the coupling to the gas energy. The resulting local blocks are
dense, and the full matrix is nonsymmetric.

The \textsc{rich} code solves Eq.~\eqref{eq:linear_system} using an
MPI-parallel, Jacobi-preconditioned BiCGSTAB method \citep{BiCGSTAB}.
The addition of Compton scattering modifies the matrix structure and
coefficients, but does not require changes to the convergence criteria,
conservative defect correction, or convergence acceleration through
opacity limiting, which are described in \citet{Giron_2026}.

After solving the system, the material internal energy is
updated with the new radiation field,
\begin{multline}
u_{m,i}^{n+1}
=
u_{m,i}^n
+
c\Delta t\,f_i
\Bigg[
\sum_{g'} \kappa_{P,g',i}E_{g',i}^{n+1}
-\kappa_{P,i}U_{r,i}^n
\\
-\sum_{g'g''}S_{g'g'',i}^nE_{g',i}^{n+1}
\Bigg]
\label{eq:gas_energy_update}
\end{multline}
With the same $f_i$ as in Eq.~\eqref{eq:coupling_coefficient}, the gas gains exactly the energy that the radiation loses, so the radiation step conserves energy.

The gas momentum is also updated after the radiation step using the
converged radiation field. This radiation-force update is unchanged from
the procedure described in \citet{Giron_2026}.

\subsection{Stability Safeguards}
\label{sec:stability}

The general numerical safeguards used in the multigroup diffusion scheme 
are described in \citet{Giron_2026}. Here we focus on the
additional safeguards introduced by the Compton extension.

The temperature dependence of the Compton operator can, in some states,
produce a negative or weak effective coupling in the generalized Fleck factor. With the Compton coupling coefficient $\kappa_C$ of Eq.~\eqref{eq:generalized_fleck}, we apply a safeguard whenever $f<0$, or whenever $\kappa_C<0$ and $|\kappa_C|>0.1\,\kappa_P$. When this condition is satisfied, we suppress induced Compton scattering
for that cell and time step by setting the group occupation number $n_g$ of Eq.~\eqref{eq:occupation_group}
to zero when constructing the Compton matrices, and recompute $f$ once. The equilibrium occupation numbers $n_g^B$ of the detailed-balance correction are unchanged. The step then proceeds with the recomputed $f$. With $n_g=0$ the Compton operator does not depend on the radiation field, so lagging the induced terms introduces no error, which makes this a more robust choice, although it does not guarantee that the recomputed $f$ is positive. For $n_g=0$ the continuous $\kappa_C$ is non-negative, but the group discretization can leave a small negative value.
The same choice is used
in the post-step material-energy update of Eq.~\eqref{eq:gas_energy_update}.

We additionally cap the occupation number entering the induced-scattering
factors at $n_g=100$. This prevents very large occupation numbers from
producing excessively strong induced Compton coupling. The same cap is
applied to the equilibrium occupation numbers used when enforcing detailed
balance. The temperature derivatives of the detailed-balance-corrected scattering
coefficients are computed with the uncapped equilibrium occupation numbers.

Finally, to speed up the TDE simulation of Section~\ref{sec:tde}, and only in that simulation, we omit the Compton terms from the generalized Fleck factor, the linear system, and the material-energy update in cells where the absorption optical depth per time step satisfies
\begin{equation}
c\Delta t\,\kappa_P\geq 10
\label{eq:compton_cutoff}
\end{equation}

\section{Equilibration Verification}
\label{sec:verification}

We test the local radiation step in a spatially uniform,
one-cell configuration. The radiation field is initialized as a Planck
spectrum at $T_{\rm rad}(0)$, while the material is initialized at the temperature
$T_{\rm mat}(0)$, where $T_{\rm mat}\equiv T$ denotes the material temperature. The radiation temperature
\begin{equation}
T_{\rm rad}=\left(\frac{1}{a_R}\sum_gE_g\right)^{1/4}
\label{eq:radiation_temperature}
\end{equation}
is the temperature of a Planck spectrum with the same radiation energy density. The material density is $\rho=1\,\mathrm{g\,cm^{-3}}$. The gas is an ideal gas of fully ionized hydrogen with $u_m=c_vT$ and $c_v=3\rho N_Ak_B$. Hydrodynamic motion and diffusion are disabled,
so the evolution is governed entirely by Compton scattering and absorption--emission coupling. The multigroup diffusion part of the scheme was verified by \citet{Giron_2026} against a comprehensive suite of gray and multigroup benchmarks, including the self-similar, non-equilibrium supersonic Marshak waves of \citet{krief_self-similar_2024} and \citet{derei_non-equilibrium_2024}. Similar self-similar methods have recently produced new solutions for radiative transfer \citep{krief2021analytic,krief2024unified,heizler2024accurate} and shock hydrodynamics \citep{giron2021solutions,giron2023solutions,krief2023piston}.

The multigroup calculation uses 64 logarithmically spaced photon-energy
groups spanning $1\,\mathrm{eV}$ to $1\,\mathrm{MeV}$. The time step is
adaptive and is increased by 5\% after each step until it reaches the
case-dependent maximum value. The initial and maximum time steps for each
configuration are listed in Table~\ref{tab:equilibration_timesteps}.

\begin{table}
    \centering
    \caption{Time steps for the adaptive equilibration calculations. The cases are the five problems of Section~\ref{sec:equilibration_reference}, in order. Case 2 is case 1 without absorption--emission.}
    \label{tab:equilibration_timesteps}
    \begin{tabular}{@{}ccccc@{}}
        \toprule
        Case & $T_{\rm mat}(0)$ & $T_{\rm rad}(0)$ &
        $\Delta t_{\rm init}$ & $\Delta t_{\rm max}$ \\
        & $[\mathrm{keV}]$ & $[\mathrm{keV}]$ &
        $[\mathrm{s}]$ & $[\mathrm{s}]$ \\
        \midrule
        1 & 20 & 1  & $5\times10^{-14}$ & $1\times10^{-10}$ \\
        2 & 20 & 1  & $5\times10^{-14}$ & $1\times10^{-10}$ \\
        3 & 1  & 10 & $5\times10^{-14}$ & $1\times10^{-10}$ \\
        4 & 50 & 1  & $1\times10^{-14}$ & $1\times10^{-10}$ \\
        5 & 1  & 50 & $1\times10^{-15}$ & $1\times10^{-10}$ \\
        \bottomrule
    \end{tabular}
\end{table}

The opacities of the equilibration calculations are interpolated, log-bilinearly in $T$ and $\rho$, from the TOPS tables \citep{Colgan2016}. The TOPS Planck and Rosseland opacities are capped at $10^3\,\mathrm{cm^2\,g^{-1}}$. To provide a more easily reproducible test, we
also repeat the calculation using an analytic free--free (bremsstrahlung)
absorption coefficient \citep{RybickiLightman1979},
\begin{equation}
\kappa_\nu^{\rm ff}
=
3.7\times10^8\,N_e^2\,T^{-1/2}
\left(\frac{\nu}{h}\right)^{-3}
\left(1-e^{-\nu/k_BT}\right)\bar{g}_{\rm ff}
\label{eq:ff_opacity}
\end{equation}
where $\kappa_\nu^{\rm ff}$ is in $\mathrm{cm^{-1}}$ for $N_e$ in $\mathrm{cm^{-3}}$, $T$ in K, and $\nu/h$ in Hz, $N_e$ is the electron number density of Eq.~\eqref{eq:electron_density}, and $\bar{g}_{\rm ff}$ is the
velocity-averaged Gaunt factor, for which we use
\begin{equation}
\bar{g}_{\rm ff}
=
\max\left(1,\,
\ln\left[\frac{387.61\,T^{3/2}}{(\nu/h)}\right]\right)
\label{eq:gaunt_factor}
\end{equation}

For comparison, we independently integrate the same local radiation--matter
equations using an ordinary differential equation (ODE) solver. The reference
calculation uses the same group-integrated Compton coefficients, opacity
model, nonlinear source terms, and detailed-balance correction as the
\textsc{rich} calculation. This comparison therefore tests the semi-implicit time integration of the radiation step implemented in \textsc{rich}. Because both calculations use the same group coefficients, it does not test the underlying group approximation, which is examined separately in Section~\ref{sec:photon_number_be}.

\subsection{Reference Equilibration Problems}
\label{sec:equilibration_reference}

\begin{figure}
    \centering
    \includegraphics[width=1.0\linewidth]{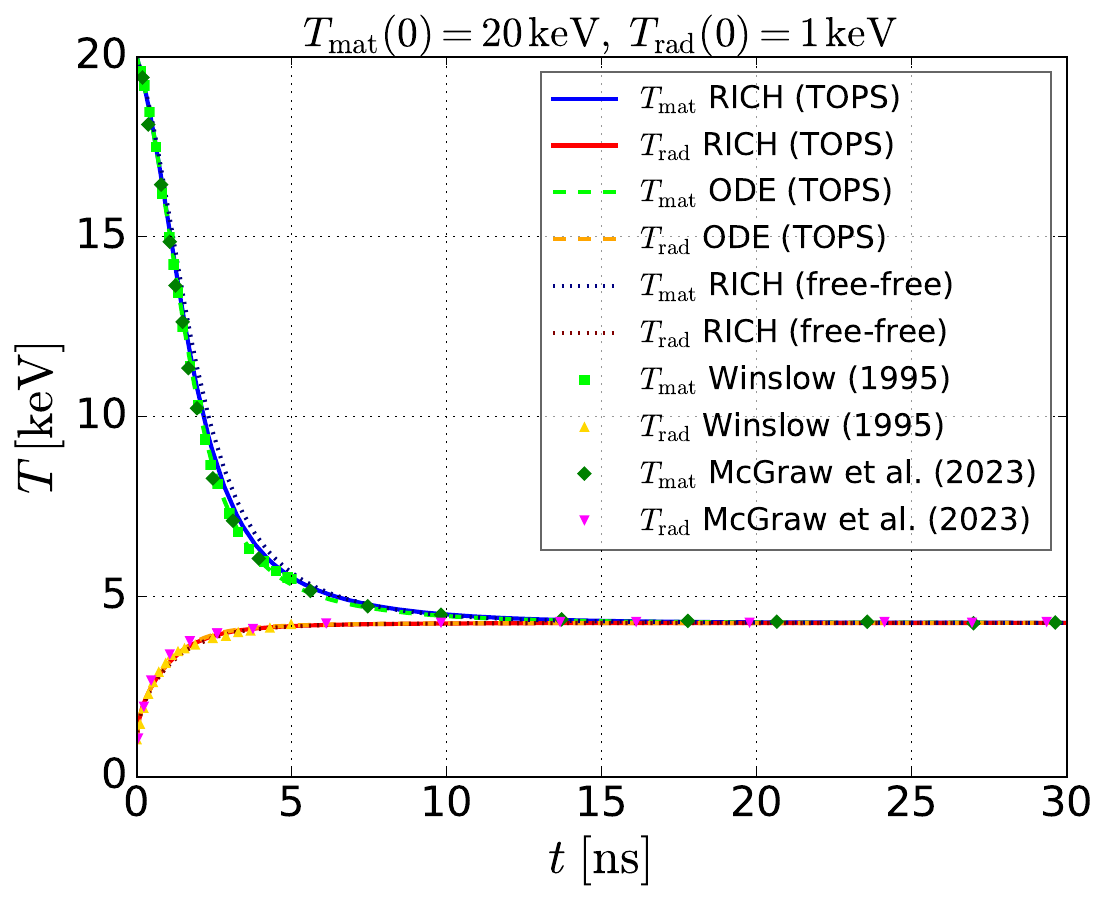}
    \caption{Equilibration test for the Winslow configuration with $T_{\rm mat}(0)=20\,\mathrm{keV}$ and $T_{\rm rad}(0)=1\,\mathrm{keV}$, including Compton scattering and absorption--emission. Solid curves are the 64-group \textsc{rich} solutions with TOPS opacities, dashed curves are the independently integrated reference solutions, and dotted curves are the \textsc{rich} solutions with the analytic free--free opacity. Symbols are digitized data from \citet{WINSLOW1995262} and \citet{MCGRAW2023111980}.}
    \label{fig:equilibration_winslow}
\end{figure}

The first and second problems follow the setup of
\citet{WINSLOW1995262}, with $T_{\rm mat}(0)=20\,\mathrm{keV}$ and
$T_{\rm rad}(0)=1\,\mathrm{keV}$. \citet{WINSLOW1995262} treats Compton scattering in the Fokker--Planck approximation, using the Kompaneets expression \citep{Kompaneets1957} with a relativistic correction. In the first problem both absorption--emission and Compton scattering are included. Figure~\ref{fig:equilibration_winslow} compares the material and radiation temperatures obtained with \textsc{rich} to the independent ODE solution and digitized data from \citet{WINSLOW1995262} and \citet{MCGRAW2023111980}. The \textsc{rich} and ODE solutions agree with each other and with the digitized data on both the equilibrium temperature and the time-dependent evolution.

\begin{figure}
    \centering
    \includegraphics[width=1.0\linewidth]{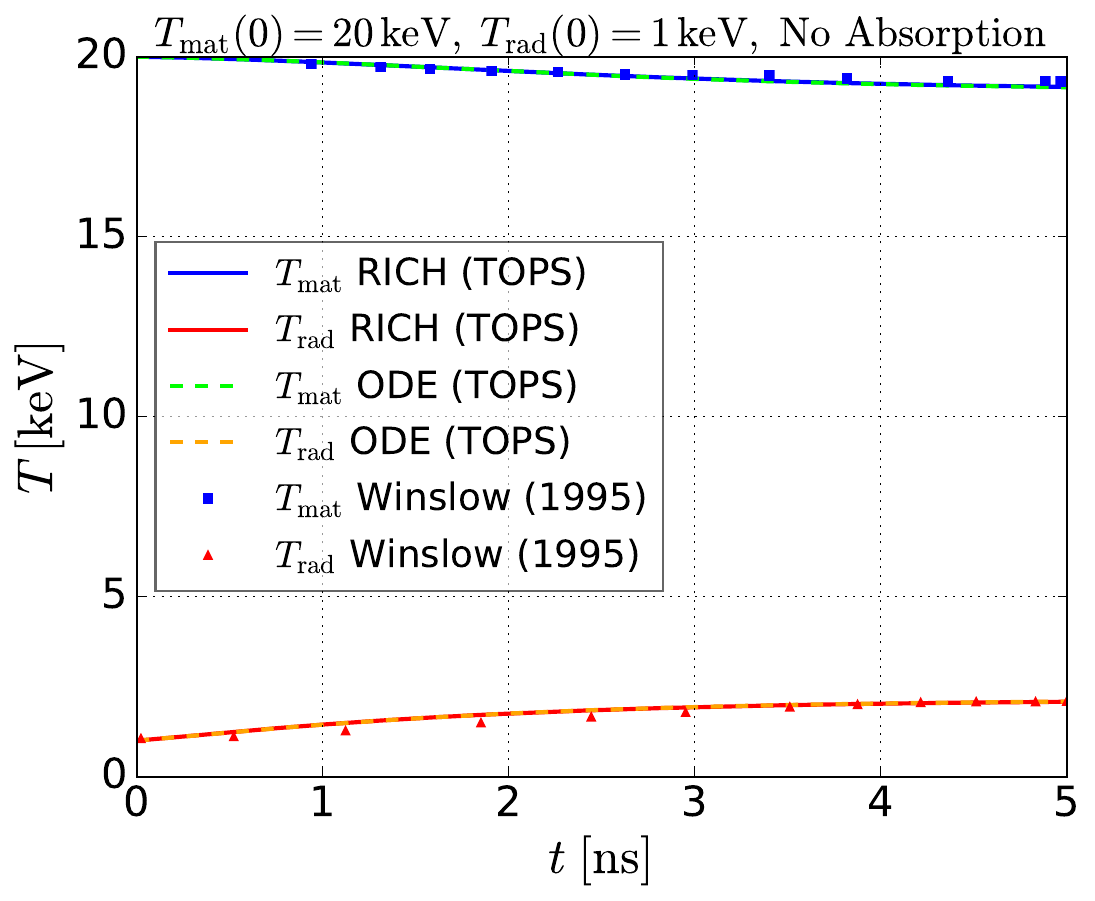}
    \caption{Equilibration test for the Winslow configuration with $T_{\rm mat}(0)=20\,\mathrm{keV}$ and $T_{\rm rad}(0)=1\,\mathrm{keV}$, with Compton scattering and without absorption--emission. Solid curves are the 64-group \textsc{rich} solutions and dashed curves are the independently integrated reference solutions. Symbols are digitized data from \citet{WINSLOW1995262}.}
    \label{fig:equilibration_winslow_only_compton}
\end{figure}

The second problem uses the same configuration as the first, but absorption--emission is turned off. The material and radiation therefore exchange energy only through Compton scattering. Figure~\ref{fig:equilibration_winslow_only_compton} compares the material and radiation temperatures obtained with \textsc{rich} to the independent ODE solution and digitized data from \citet{WINSLOW1995262}. The \textsc{rich} and ODE solutions agree, and both are close to the digitized data.

\begin{figure}
    \centering
    \includegraphics[width=1.0\linewidth]{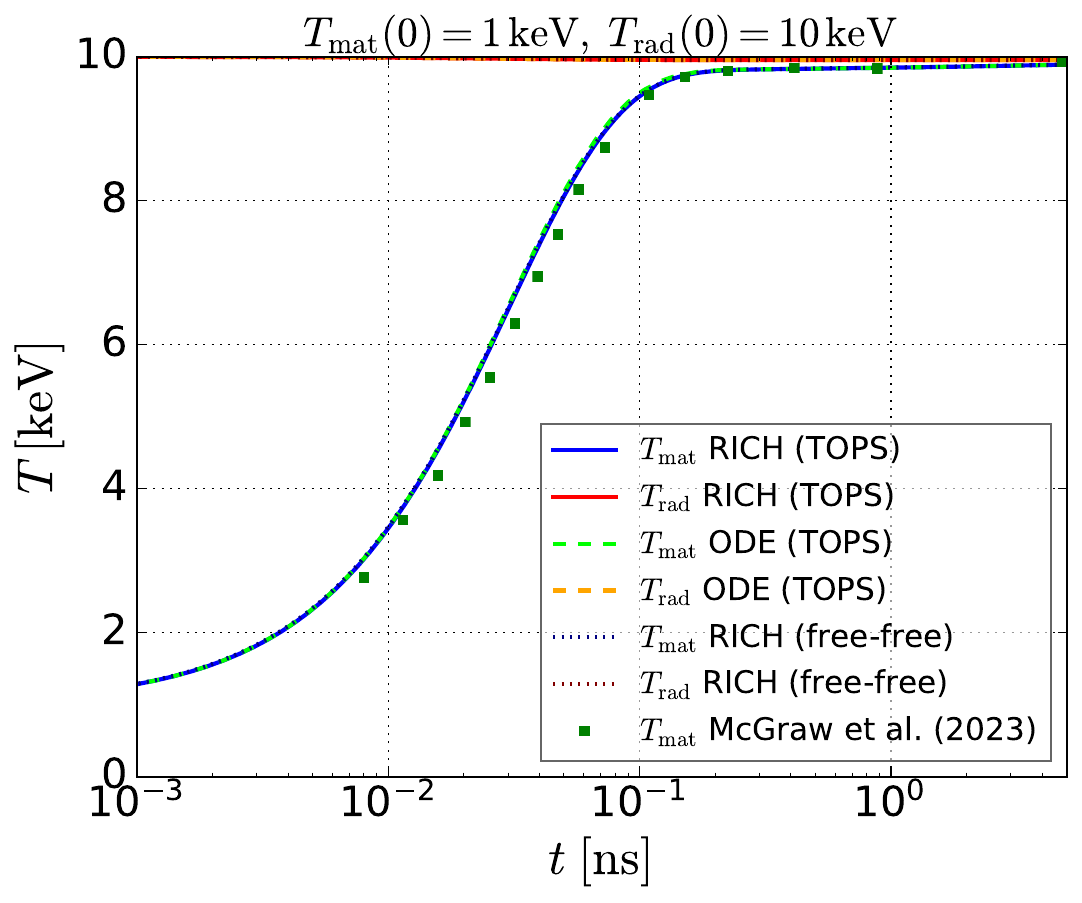}
    \caption{Equilibration test for the hotter-radiation configuration with $T_{\rm mat}(0)=1\,\mathrm{keV}$ and $T_{\rm rad}(0)=10\,\mathrm{keV}$. Solid curves are the 64-group \textsc{rich} solutions with TOPS opacities, dashed curves are the independently integrated reference solutions, and dotted curves are the \textsc{rich} solutions with the analytic free--free opacity. Symbols are digitized data from \citet{MCGRAW2023111980}.}
    \label{fig:equilibration_mcgraw}
\end{figure}

The third problem follows the hotter-radiation case used by
\citet{MCGRAW2023111980}, with $T_{\rm mat}(0)=1\,\mathrm{keV}$ and
$T_{\rm rad}(0)=10\,\mathrm{keV}$. In this case, the radiation field
initially heats the material, providing a complementary test in which the
direction of the net radiation--matter energy exchange is reversed. 
Figure~\ref{fig:equilibration_mcgraw} shows that the \textsc{rich} solution closely follows the reference ODE
evolution throughout the heating phase and is consistent with the digitized data from \citet{MCGRAW2023111980}.

\begin{figure*}
    \centering
    \subfloat[$T_{\rm mat}(0)=50\,\mathrm{keV}$, $T_{\rm rad}(0)=1\,\mathrm{keV}$\label{fig:equilibration_t50_r1}]{%
        \includegraphics[width=0.49\textwidth]{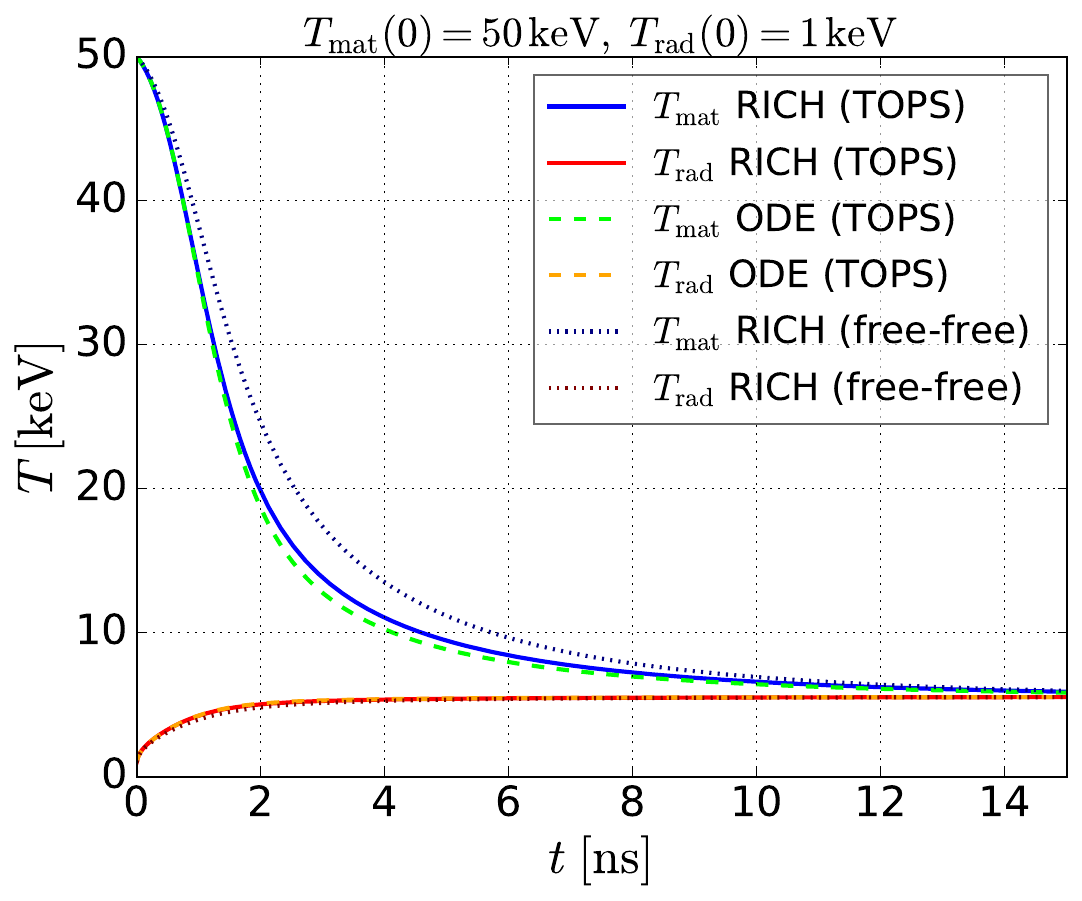}}
    \hfill
    \subfloat[$T_{\rm mat}(0)=1\,\mathrm{keV}$, $T_{\rm rad}(0)=50\,\mathrm{keV}$\label{fig:equilibration_t1_r50}]{%
        \includegraphics[width=0.49\textwidth]{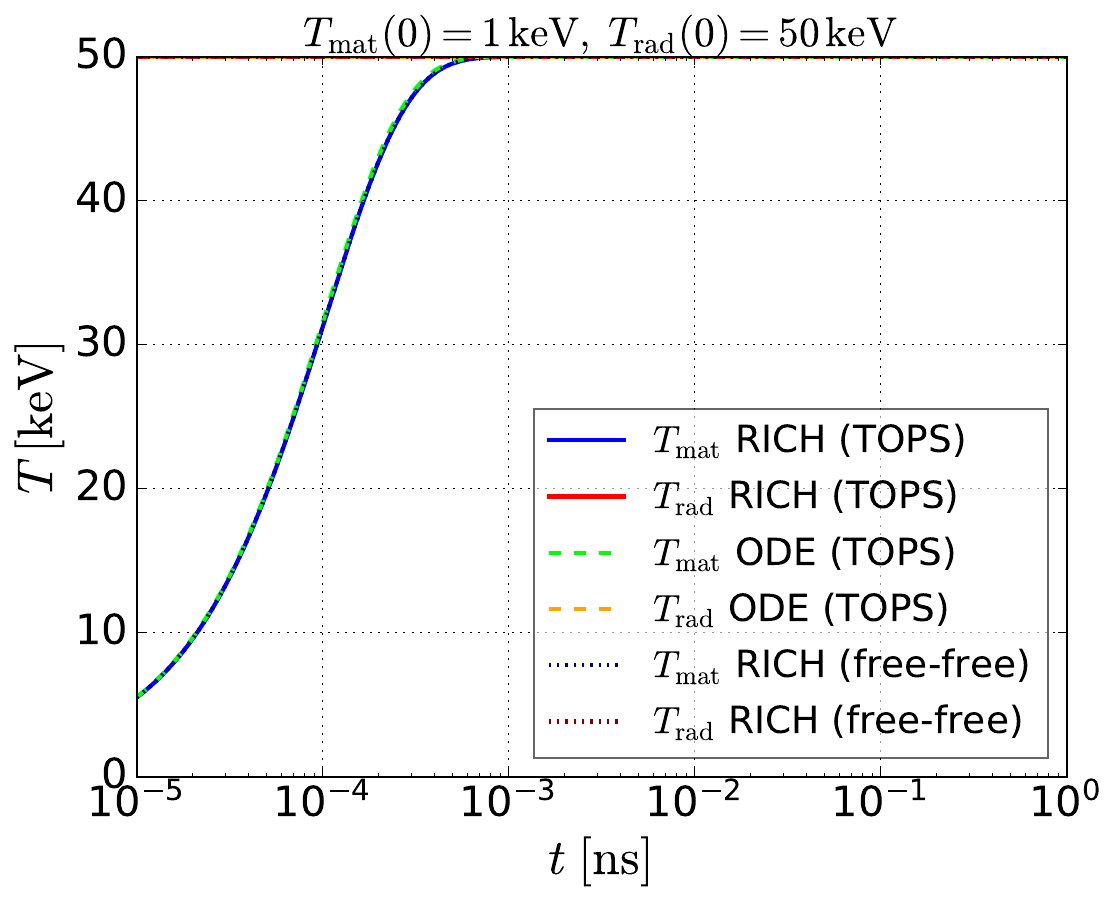}}
    \caption{Equilibration tests with extreme initial temperature ratios. Solid curves are the 64-group \textsc{rich} material and radiation temperatures with TOPS opacities, dashed curves are the independently integrated reference solutions, and dotted curves are the \textsc{rich} solutions with the analytic free--free opacity.}
    \label{fig:equilibration_extreme}
\end{figure*}

We extend the equilibration test to substantially larger initial temperature contrasts. We consider two complementary configurations: $T_{\rm mat}(0)=50\,\mathrm{keV}$ and $T_{\rm rad}(0)=1\,\mathrm{keV}$, which produces strong Compton cooling of the material, and $T_{\rm mat}(0)=1\,\mathrm{keV}$ and $T_{\rm rad}(0)=50\,\mathrm{keV}$, which produces strong Compton heating. Figures~\ref{fig:equilibration_t50_r1} and \ref{fig:equilibration_t1_r50} show the results. In both configurations the \textsc{rich} solutions follow the reference ODE solutions.

\subsection{Photon-Number Conservation and Bose--Einstein Relaxation}
\label{sec:photon_number_be}

Compton scattering conserves the number of photons. Without absorption and emission, as in the second problem, the radiation should therefore relax toward a Bose--Einstein spectrum 
\begin{equation}
E_\nu^{\rm BE}=\frac{8\pi\nu^3}{h^3c^3}
\frac{1}{e^{(\nu-\mu)/k_BT}-1}
\label{eq:bose_einstein}
\end{equation}
at the gas temperature $T$, with a chemical potential $\mu\leq0$~\citep{RybickiLightman1979}. The group operator of Eq.~\eqref{eq:scattering_matrix} conserves the total photon number density $N_\gamma$ of Eq.~\eqref{eq:photon_number}, as shown by Eq.~\eqref{eq:photon_number_conservation}. Let $E_g^{\rm BE}(T,\mu)$ be the group integrals of Eq.~\eqref{eq:bose_einstein}. The Bose--Einstein state has the same total energy as the initial state
\begin{multline}
c_vT+\sum_gE_g^{\rm BE}(T,\mu)
=c_vT_{\rm mat}(0)+\sum_gB_g\left(T_{\rm rad}(0)\right)
\label{eq:be_energy}
\end{multline}
and the same $N_\gamma$
\begin{equation}
\sum_g\frac{E_g^{\rm BE}(T,\mu)}{\nu_g}
=\sum_g\frac{B_g\left(T_{\rm rad}(0)\right)}{\nu_g}
\label{eq:be_photon_number}
\end{equation}
For the second problem, the numerical solution of Eqs.~\eqref{eq:be_energy} and \eqref{eq:be_photon_number} is $T=19.045\,\mathrm{keV}$ and $\mu/k_BT=-8.66$. A Planck spectrum with the same total energy would instead have $T=4.27\,\mathrm{keV}$.

\begin{figure*}
    \centering
    \subfloat[Radiation spectra\label{fig:equilibration_be_spectrum}]{%
        \includegraphics[width=0.49\textwidth]{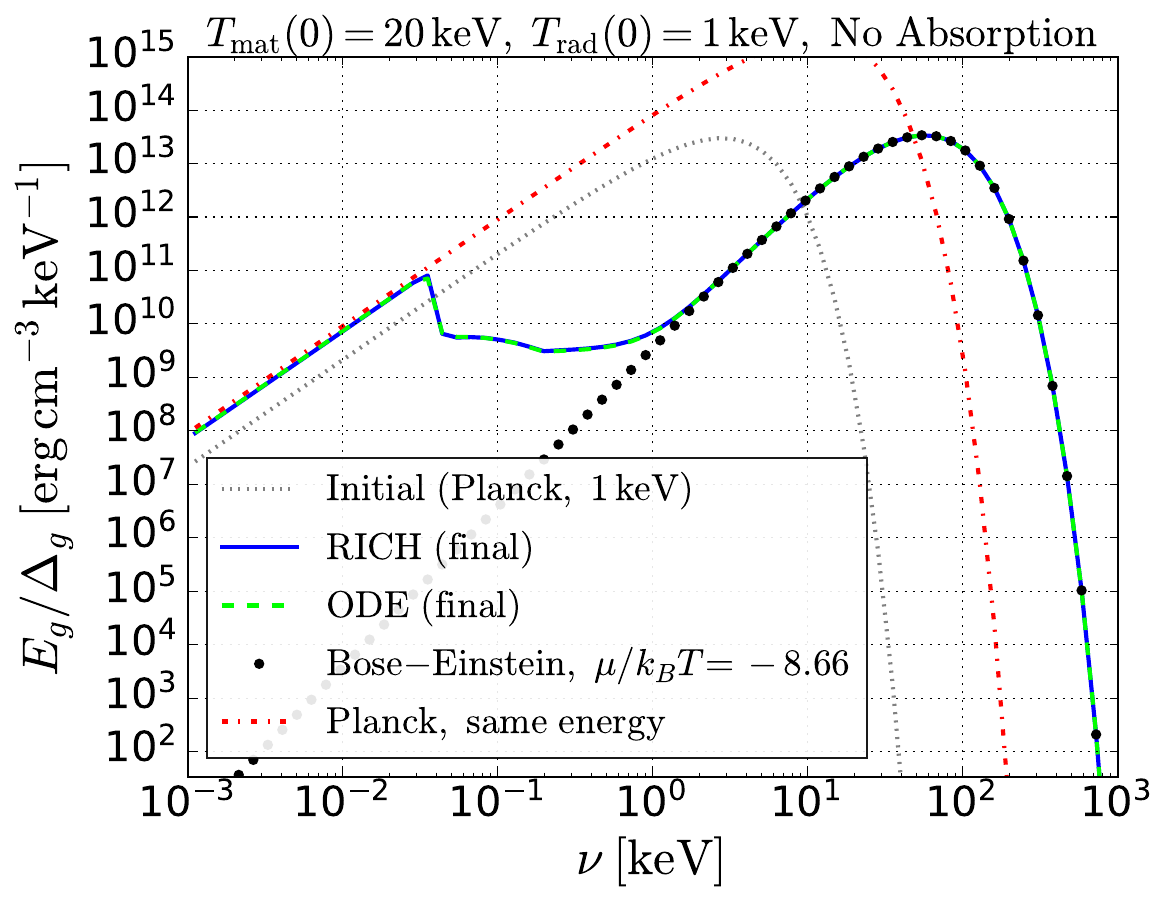}}
    \hfill
    \subfloat[Photon-number and total-energy drifts\label{fig:equilibration_photon_number}]{%
        \includegraphics[width=0.49\textwidth]{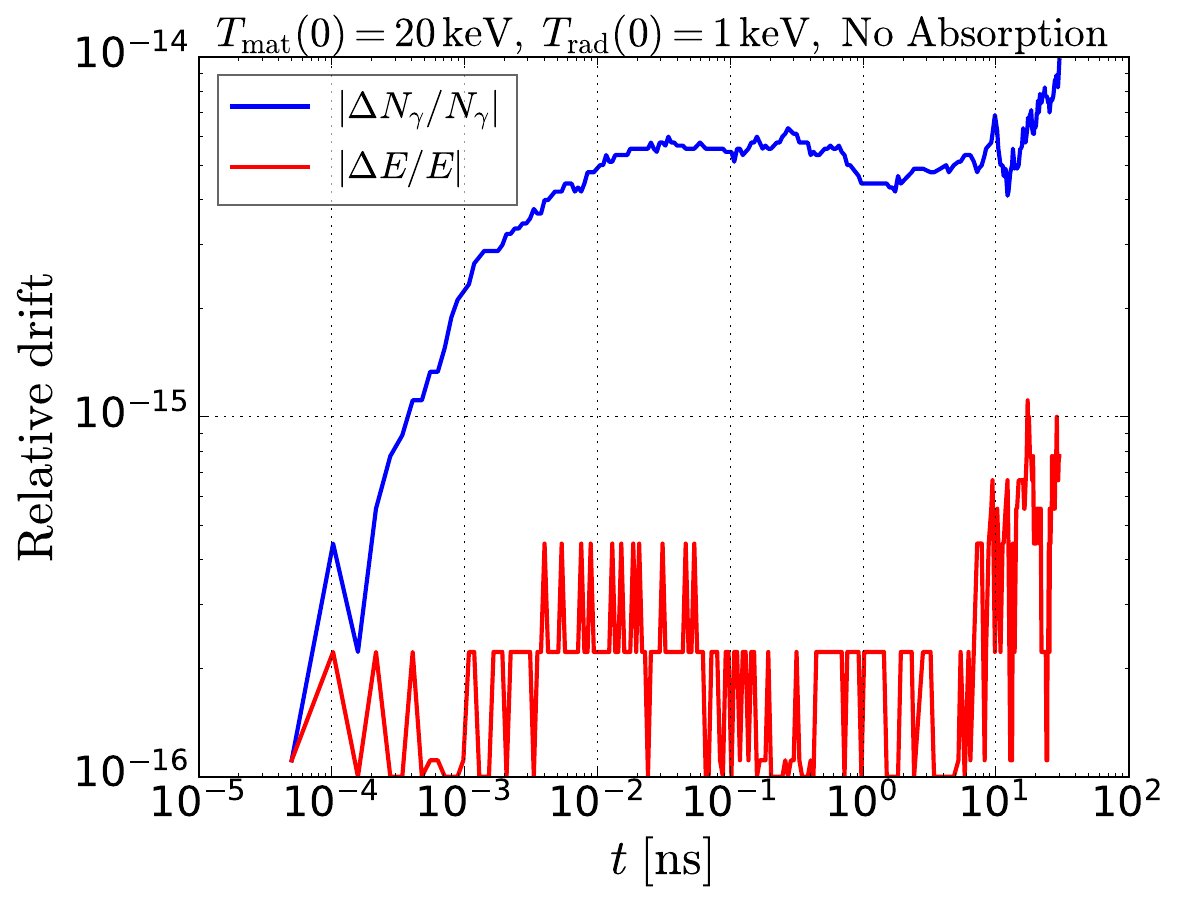}}
    \caption{Compton-only configuration with $T_{\rm mat}(0)=20\,\mathrm{keV}$ and $T_{\rm rad}(0)=1\,\mathrm{keV}$. (a) The initial Planck spectrum, the final 64-group \textsc{rich} and ODE spectra at $t=3\times10^{-8}\,\mathrm{s}$, the Bose--Einstein spectrum that conserves the initial energy and photon number, and the Planck spectrum at $T=4.27\,\mathrm{keV}$ that conserves the initial total energy. (b) Relative drifts of the total photon number density $N_\gamma$~\eqref{eq:photon_number} and of the total energy. The curves are the 64-group \textsc{rich} solution.}
    \label{fig:equilibration_be}
\end{figure*}

Figure~\ref{fig:equilibration_be_spectrum} shows the final spectra at $t=3\times10^{-8}\,\mathrm{s}$, and Figure~\ref{fig:equilibration_photon_number} shows the relative drifts of $N_\gamma$ and of the total energy. In \textsc{rich}, $N_\gamma$ is conserved to $10^{-14}$ and the total energy to $10^{-15}$. The final gas temperature is $19.045\,\mathrm{keV}$, in agreement with the Bose--Einstein state. At low photon energies the spectrum still exceeds the Bose--Einstein spectrum, because the cap $n_g,\,n_g^B\leq100$ on the occupation numbers (Section~\ref{sec:stability}) slows the relaxation of the low-energy groups.

\subsection{Time-Step Stability}
\label{sec:timestep_stability}

\begin{figure*}
    \centering
    \subfloat[$T_{\rm mat}(0)=20\,\mathrm{keV}$, $T_{\rm rad}(0)=1\,\mathrm{keV}$\label{fig:equilibration_timestep_winslow}]{%
        \includegraphics[width=0.49\textwidth]{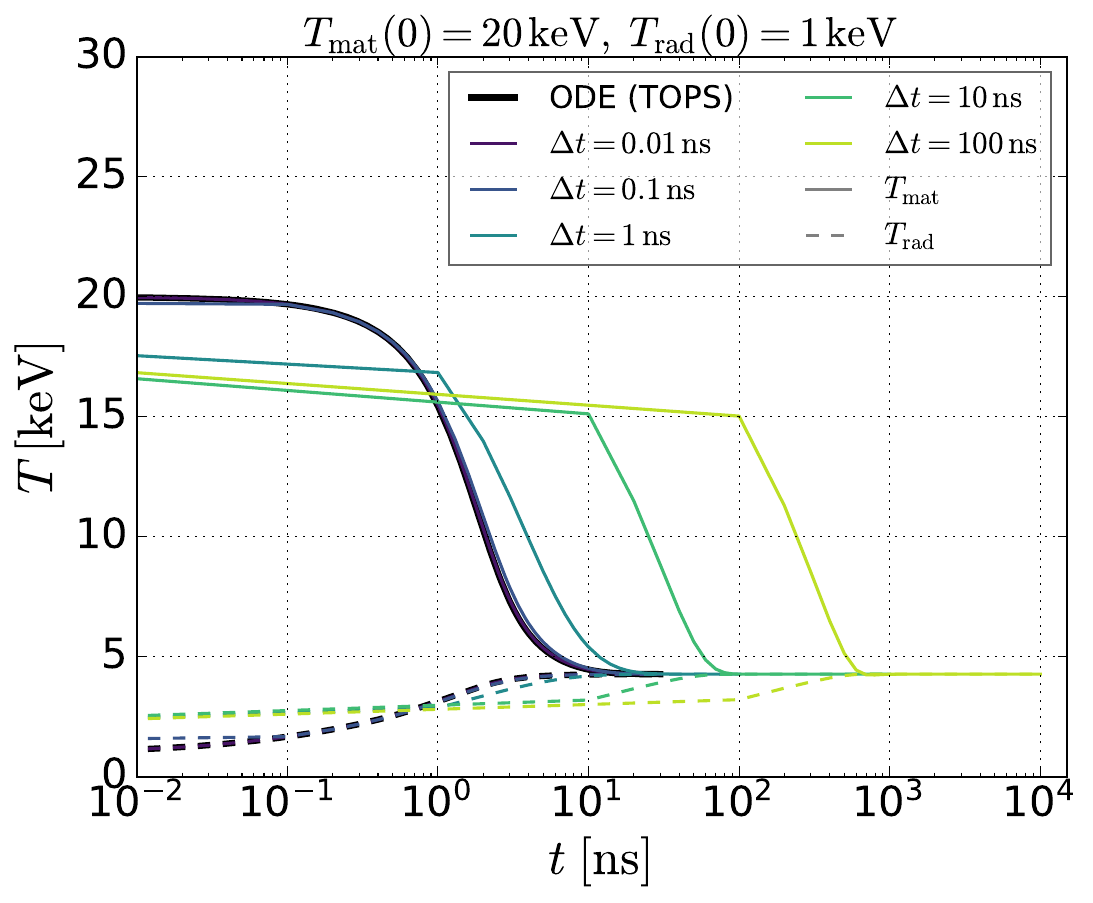}}
    \hfill
    \subfloat[$T_{\rm mat}(0)=1\,\mathrm{keV}$, $T_{\rm rad}(0)=10\,\mathrm{keV}$\label{fig:equilibration_timestep_mcgraw}]{%
        \includegraphics[width=0.49\textwidth]{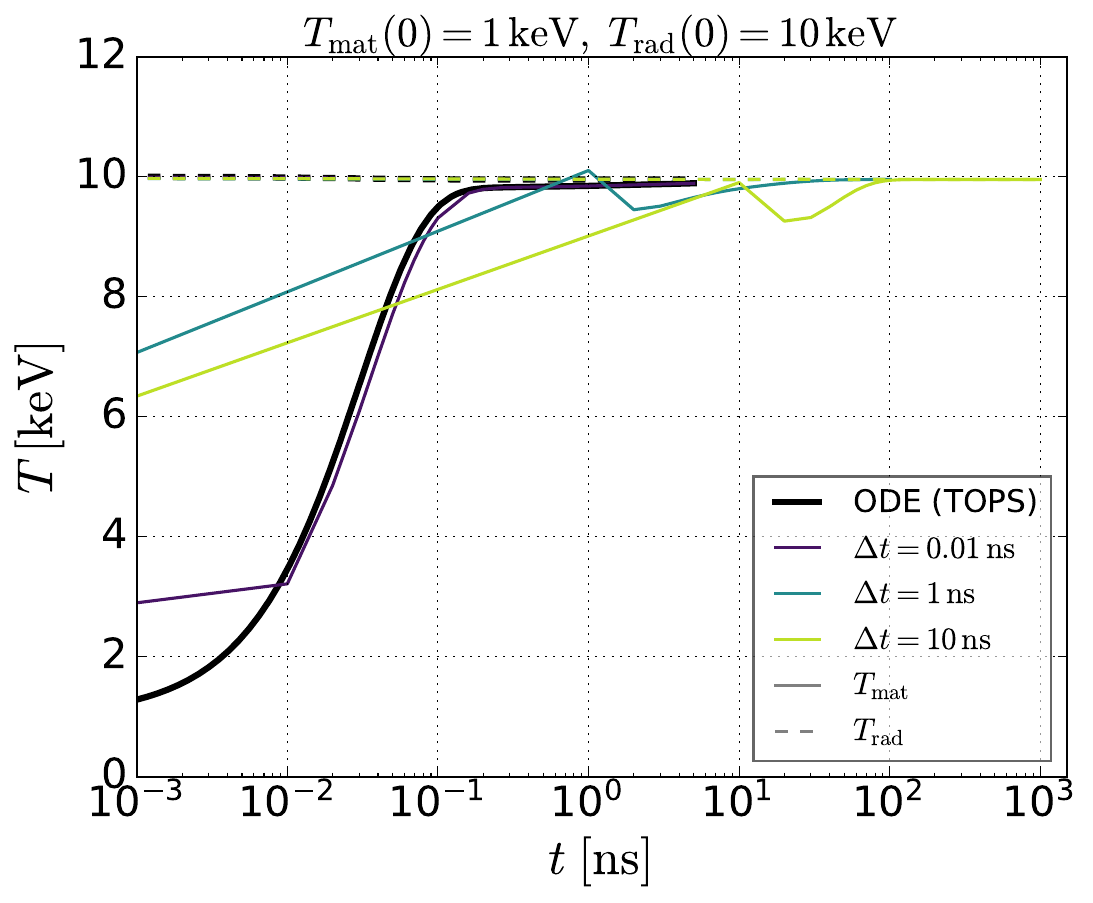}}
    \caption{Fixed-time-step equilibration tests. Black curves show the reference solution. Colored solid and dashed curves show the material and radiation temperatures, respectively, for the fixed time steps listed in the legends.}
    \label{fig:equilibration_timestep}
\end{figure*}

We test the stability of the semi-implicit scheme using prescribed fixed time steps, including time steps that do not resolve the Compton equilibration timescale. Figure~\ref{fig:equilibration_timestep} compares the resulting evolution with the reference ODE solutions for the first and third problems. In these problems, the material and radiation temperatures equilibrate in about $10^{-8}$ and $10^{-10}\,\mathrm{s}$, respectively (Figures~\ref{fig:equilibration_winslow} and \ref{fig:equilibration_mcgraw}), so the largest fixed steps exceed the Compton equilibration timescale by factors of 10 and 100. As the time step is reduced, the \textsc{rich} solutions converge toward the reference curve. For the fixed steps plotted here, the solutions remain finite and approach the reference equilibrium state. These tests indicate that resolving the Compton equilibration timescale is required for accuracy, but not for stability, in the configurations considered. They do not establish unconditional stability or positivity for other thermodynamic states.

\section{Tidal Disruption Event Simulation}
\label{sec:tde}

We apply the new multigroup Compton module to a three-dimensional simulation of a tidal disruption event (TDE). Here we post-process a single simulation with Compton scattering to identify where and when Compton scattering is important in the flow. We first describe the light curve and the flow, and then define each Compton diagnostic and show its results in turn: the spectral change rate, the coupling fraction, the direction of the redistribution, and the X-ray production and emission. We close with the limitations of the analysis.

\subsection{Simulation Setup}
\label{sec:tde_setup}

The simulation follows the disruption of a star of mass $M_\star=0.5\,M_\odot$ by a black hole of mass $M_{\rm BH}=10^6\,M_\odot$ with penetration factor $\beta=r_t/r_p=1$, where $r_t=R_\star(M_{\rm BH}/M_\star)^{1/3}\approx59.2\,R_\odot$ is the tidal radius, $R_\star=0.47\,R_\odot$ the stellar radius, and $r_p$ the pericenter distance. The simulation parameters are listed in Table~\ref{tab:tde_parameters}. The star is initialized as an $n=1.5$ polytrope, and the gas follows an equation of state that includes hydrogen and helium recombination \citep{Tomida_2013,Tomida+15,Pejcha2016,steinberg_streamdisk_2024}. The black-hole potential is the \citet{PaczynskiWiita1980} pseudo-Newtonian potential. Times are given in units of the fallback time of the most bound debris and are measured from the pericenter passage of the star 
\begin{equation}
t_{\rm fb}=\frac{\pi}{\sqrt{2}}\left(\frac{R_\star^3}{GM_\star}\right)^{1/2}\left(\frac{M_{\rm BH}}{M_\star}\right)^{1/2}\approx26.6\,\mathrm{days}
\label{eq:fallback_time}
\end{equation}
where $G$ is the gravitational constant. 

The star is initially placed on a parabolic orbit at $3r_t$ from the black hole, at $t=-1.1\times10^{-3}\,t_{\rm fb}$, and is resolved by $1.75\times10^7$ cells. We run the simulation until $t\approx1.16\,t_{\rm fb}$ ($\approx30.7\,\mathrm{days}$). For a more detailed description of the simulation setup, initial conditions, and adaptive mesh refinement criteria see \citet{Martire+25}.

\begin{table}
    \centering
    \caption{Parameters of the TDE simulation.}
    \label{tab:tde_parameters}
    \begin{tabular}{lc}
        \toprule
        Parameter & Value \\
        \midrule
        Stellar mass $M_\star$ & $0.5\,M_\odot$ \\
        Stellar radius $R_\star$ & $0.47\,R_\odot$ \\
        Polytropic index $n$ & $1.5$ \\
        Black-hole mass $M_{\rm BH}$ & $10^6\,M_\odot$ \\
        Penetration factor $\beta$ & $1$ \\
        Tidal radius $r_t=r_p$ & $59.2\,R_\odot$ \\
        Fallback time $t_{\rm fb}$ & $26.6\,\mathrm{days}$ \\
        Smoothing radius $R_{\rm smooth}$ & $29.6\,R_\odot$ \\
        Initial distance & $3\,r_t$ \\
        Initial time & $-1.1\times10^{-3}\,t_{\rm fb}$ \\
        Final time & $\approx1.16\,t_{\rm fb}$ \\
        \bottomrule
    \end{tabular}
\end{table}

We use tabulated gas opacities calculated with the STAR code \citep{Krief2016} and interpolate them according to the local gas density and temperature. STAR is based on the super-transition-array method \citep{krief2015effect,krief2015variance,krief2018new,krief2018effect}. The Compton coefficients use the fully ionized electron density of Eq.~\eqref{eq:electron_density}, whereas the equation of state includes recombination. 

We use ten photon-energy groups spanning $0.01\,\mathrm{eV}$ to $20\,\mathrm{keV}$, listed in Table~\ref{tab:tde_groups}. The group boundaries are chosen to separate approximately the optical, UV, EUV, supersoft, and X-ray bands relevant to quasi-thermal TDE emission. Nonthermal radio and gamma-ray emission are not included.

\begin{table}
    \centering
    \caption{Photon-energy groups used in the TDE simulation.}
    \label{tab:tde_groups}
    \begin{tabular}{lcc}
        \toprule
        Group & $E_{\rm low}\;[\mathrm{eV}]$ & $E_{\rm high}\;[\mathrm{eV}]$ \\
        \midrule
        i & 0.01 & 1.68 \\
        r & 1.68 & 2.48 \\
        b & 2.48 & 3.35 \\
        U & 3.35 & 4.13 \\
        UVW2 & 4.13 & 7.75 \\
        H & 7.75 & 15 \\
        He & 15 & 100 \\
        SS & 100 & 300 \\
        Soft X-ray & 300 & 1000 \\
        Hard X-ray & 1000 & 20000 \\
        \bottomrule
    \end{tabular}
\end{table}

For computational efficiency, we smooth the gravitational potential inside the smoothing radius $R_{\rm smooth}=r_t/2\approx29.6\,R_\odot$, as in \citet{steinberg_streamdisk_2024} and \citet{Martire+25}. We refer to the sphere $r<R_{\rm smooth}$, where $r$ is the distance from the black hole, as the center region and exclude it from all diagnostics.

\begin{figure}
    \centering
    \includegraphics[width=1.0\linewidth]{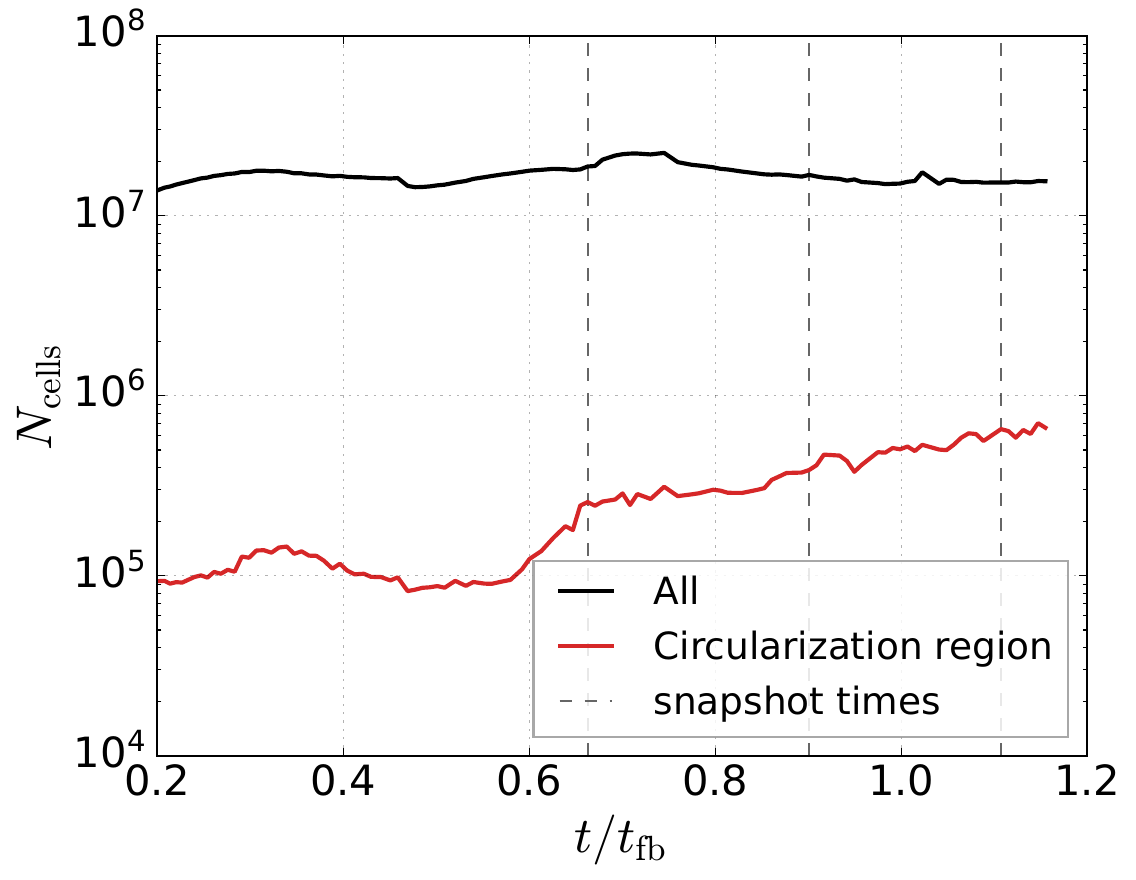}
    \caption{Number of cells in the entire computational domain (black) and in the circularization region (red), from $0.2\,t_{\rm fb}$. The dashed vertical lines mark the times of the orbital-plane maps and of the Mollweide maps (Figure~\ref{fig:tde_mollweide}).}
    \label{fig:tde_cells}
\end{figure}

Figure~\ref{fig:tde_cells} shows the number of cells in the computational domain and near the black hole. We define the circularization region as the gas at cylindrical radii $R_{\rm smooth}<R<2r_p$ about the black hole and heights $|z|<0.3R$ above the orbital plane, where $2r_p$ is the circularization radius of the debris and $0.3$ is a representative aspect ratio of the accretion flow that forms after a TDE \citep{Meza+25}. The domain contains a maximum of $\approx2\times10^7$ cells, and the circularization region a maximum of $\approx7\times10^5$ cells, as the refinement follows the gas that accumulates near the black hole.

\subsection{Light Curve}
\label{sec:tde_lightcurve}

We compute the light curve from the radiation at the photosphere of each group, which we locate along radial rays from the black hole in the directions $\hat{\Omega}$ of a HEALPix grid with $N_{\mathrm{side}}=8$ \citep{HEALPix}. Along each ray, $\ell$ is the distance from the black hole, and the Rosseland optical depth of group $g$ is integrated inward
\begin{equation}
\tau_{R,g}(\ell)
=
\int_{\ell}^{\infty}
\kappa_{R,g}\,d\ell'
\label{eq:tau_rosseland}
\end{equation}
where $\kappa_{R,g}$ includes electron scattering. The group photosphere $r_{{\rm ph},g}(\hat{\Omega})$ is the distance of the innermost point with $\tau_{R,g}<2/3$ \citep{Martire+26} (if the optical depth is still below $2/3$ at $\ell=r_p$, we set $r_{{\rm ph},g}=r_p$). At the photosphere the radiation field is still nearly isotropic, and its outward flux is $cE_g/4$, so the isotropic-equivalent luminosity of group $g$ is
\begin{equation}
L_g^{\rm ph}(\hat{\Omega},t)
=
\pi r_{{\rm ph},g}^2\,c\,E_g(r_{{\rm ph},g})
\label{eq:luminosity_photosphere}
\end{equation}
The sky-averaged luminosity $\overline{L}_g^{\rm ph}(t)$ is the mean over all rays.
\begin{figure}
    \centering
    \includegraphics[width=1.0\linewidth]{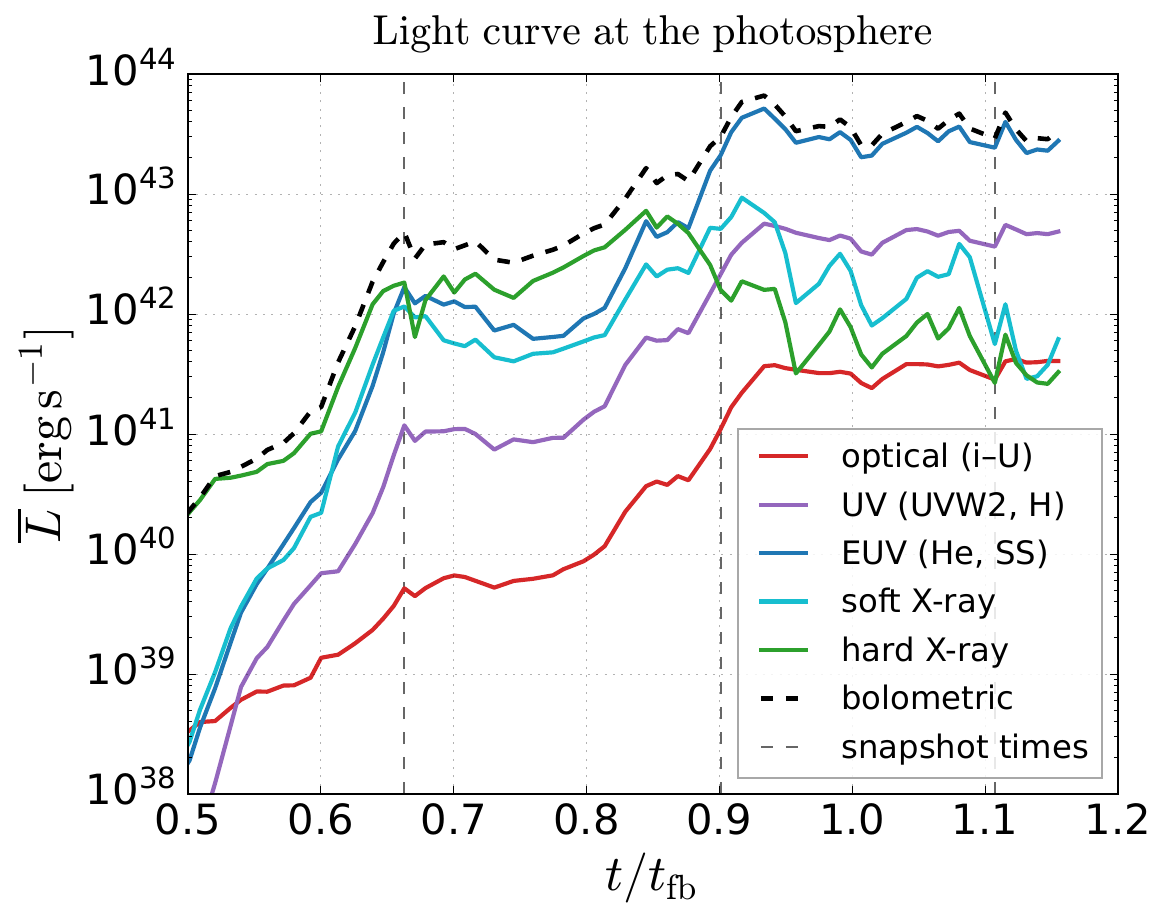}
    \caption{Sky-averaged luminosities $\overline{L}^{\rm ph}$ (Eq.~\eqref{eq:luminosity_photosphere}) for the optical (i to U), UV (UVW2 and H), EUV (He and SS), and soft and hard X-ray bands of Table~\ref{tab:tde_groups}, and the bolometric luminosity. The dashed vertical lines mark $0.66$, $0.90$, and $1.11\,t_{\rm fb}$, the times of the orbital-plane maps and of the Mollweide maps (Figure~\ref{fig:tde_mollweide}).}
    \label{fig:tde_lightcurve}
\end{figure}
Figure~\ref{fig:tde_lightcurve} shows the light curves from $0.5\,t_{\rm fb}$, when the returning debris reaches the pericenter. The X-ray luminosity (soft and hard), $\overline{L}_X^{\rm ph}$, rises rapidly and reaches $\approx 10^{42}\,\mathrm{erg\,s^{-1}}$ at $0.66\,t_{\rm fb}$, marking the beginning of an X-ray flash.  Early X-ray flashes were also found in previous \textsc{rich} simulations of TDEs, where they are attributed to shocks near pericenter \citep{steinberg_streamdisk_2024,Giron_2026}. During the flash the X-rays carry most of the bolometric luminosity, while after it most of the bolometric luminosity emerges in the EUV \citep{Dai+18,LuKumar18,Thomsen2022,Giron_2026}, which continues to rise.

\begin{figure*}
    \centering
    \subfloat[$\rho$, $0.66\,t_{\rm fb}$]{\includegraphics[width=0.32\textwidth]{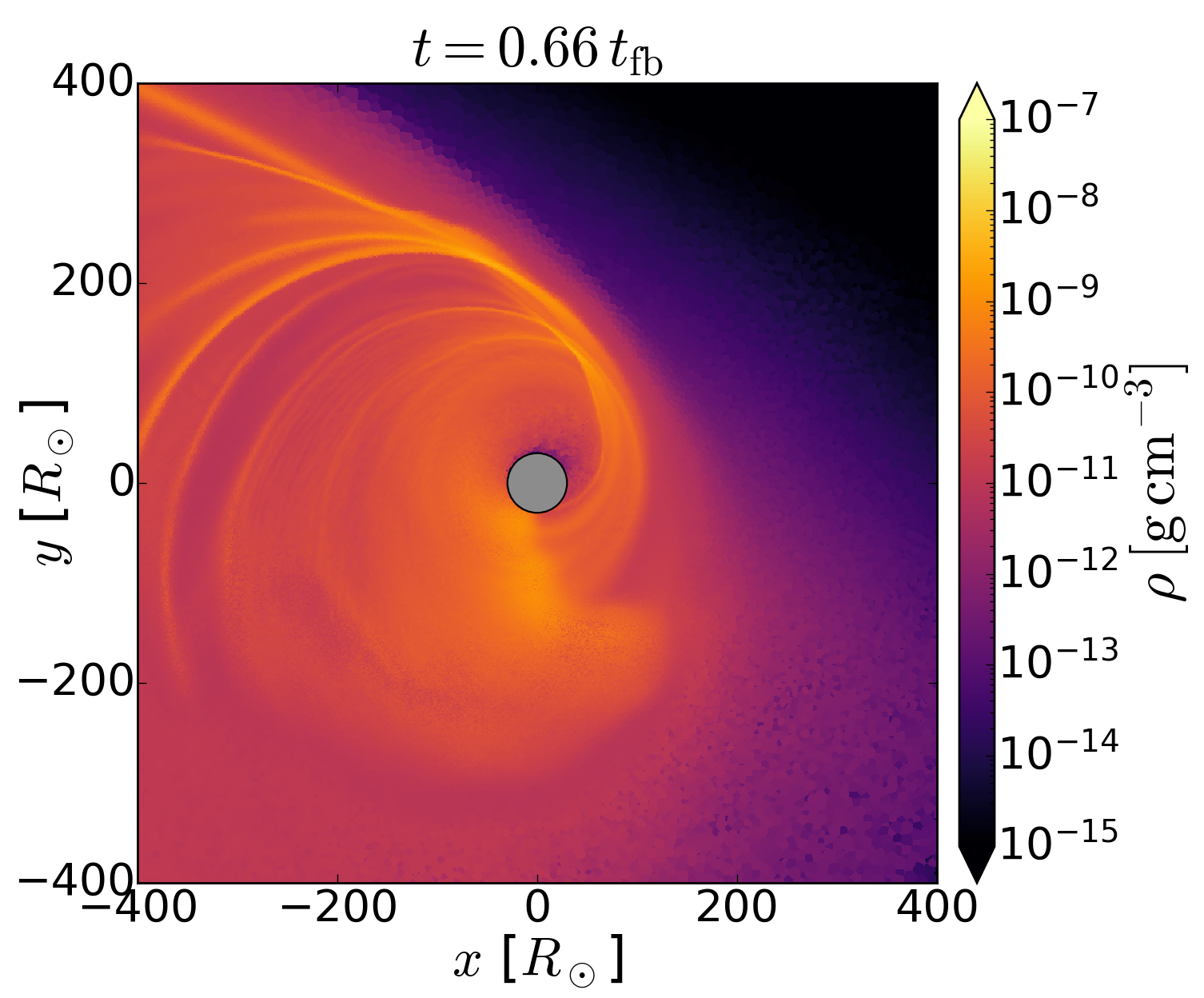}}
    \hfill
    \subfloat[$\rho$, $0.90\,t_{\rm fb}$]{\includegraphics[width=0.32\textwidth]{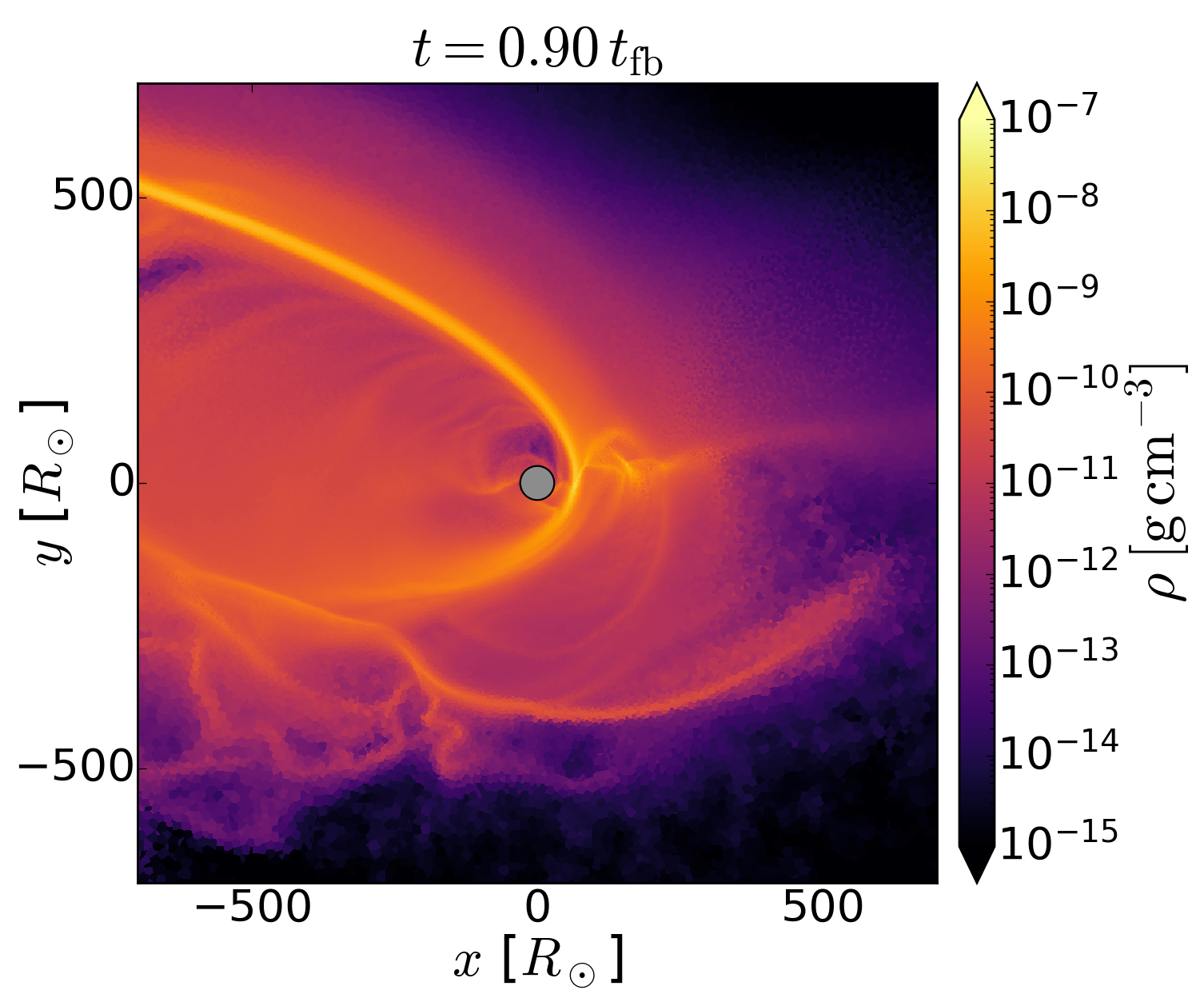}}
    \hfill
    \subfloat[$\rho$, $1.11\,t_{\rm fb}$]{\includegraphics[width=0.32\textwidth]{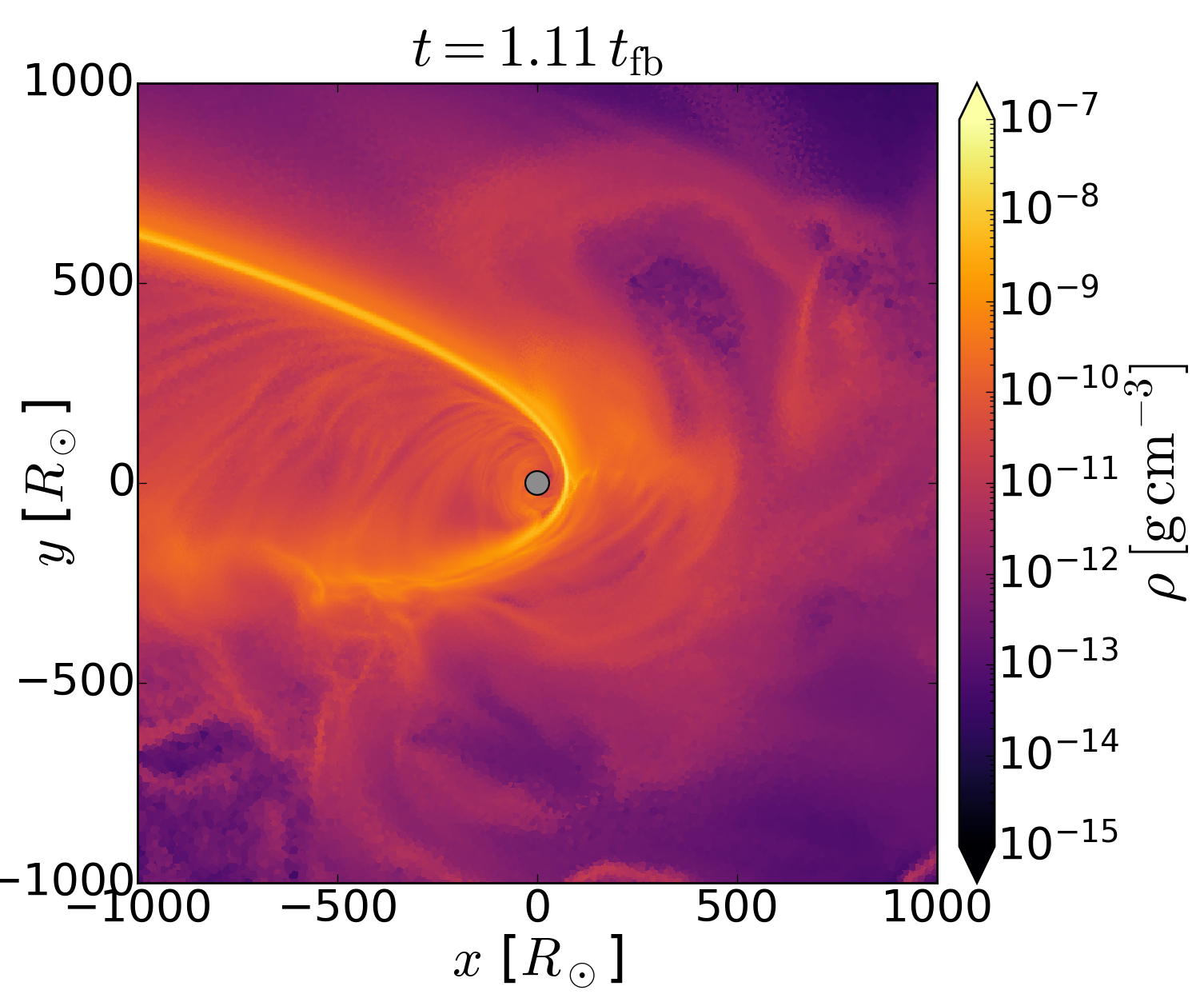}}
    \caption{Gas density in the orbital plane at $0.66$, $0.90$, and $1.11\,t_{\rm fb}$. The gray disk is the center region $r<R_{\rm smooth}$.}
    \label{fig:tde_density}
\end{figure*}

Figure~\ref{fig:tde_density} shows the gas density in the orbital plane at the beginning of the X-ray flash ($0.66\,t_{\rm fb}$), at its end ($0.90\,t_{\rm fb}$), and in the EUV-dominated phase that follows ($1.11\,t_{\rm fb}$). The flow has the structure found in previous simulations of TDEs \citep{Shiokawa+15,BonnerotStone21,Andalman+22,steinberg_streamdisk_2024}. The returning stream circles the black hole clockwise and converges at the pericenter, where it is vertically compressed in the nozzle shock. Outflowing gas fills the region beyond the stream and the disk, as in the optically thick winds of previous work \citep{Martire+25,Martire+26}.

\subsection{Spectral Change Rate}
\label{sec:tde_spectral}

The density, temperature, and radiation field vary strongly across the flow, so the importance of Compton scattering cannot be described by a single global measure. We therefore use local diagnostics, evaluated with the Compton operator of Eq.~\eqref{eq:scattering_matrix}, from the gas state and radiation field. These are explicit, instantaneous rates at the snapshot state. They are not the energy exchanged over a time step, which also includes the generalized Fleck factor and is evaluated with the end-of-step radiation field (Eq.~\eqref{eq:gas_energy_update}). We define the total radiation energy density and the local rate of energy change in group $g$ due to Compton scattering as
\begin{equation}
E_{\rm rad}=\sum_{g'}E_{g'},
\qquad
C_g=c\sum_{g'}S_{g'g}E_{g'}
\label{eq:tde_compton_source}
\end{equation}
The time over which the net Compton sources would redistribute the local radiation energy density is
\begin{equation}
t_{\rm spec}=\frac{2E_{\rm rad}}{\sum_{g'}|C_{g'}|}
\label{eq:spectral_timescale}
\end{equation}
A transfer of energy from one group to another contributes twice to $\sum_{g'}|C_{g'}|$, once through the loss from the source group and once through the gain in the destination group, and the factor of two compensates for this double counting. We compare $t_{\rm spec}$ with the dynamical time, over which the gas moves a distance comparable to its distance from the black hole,
\begin{equation}
t_{\rm dyn}=\frac{r}{|\vec{v}|}
\label{eq:dynamical_time}
\end{equation}
Where $t_{\rm dyn}/t_{\rm spec}\geq1$, Compton scattering changes the local spectrum faster than the gas moves.

\begin{figure*}
    \centering
    \subfloat[$t_{\rm dyn}/t_{\rm spec}$, $0.66\,t_{\rm fb}$]{\includegraphics[width=0.32\textwidth]{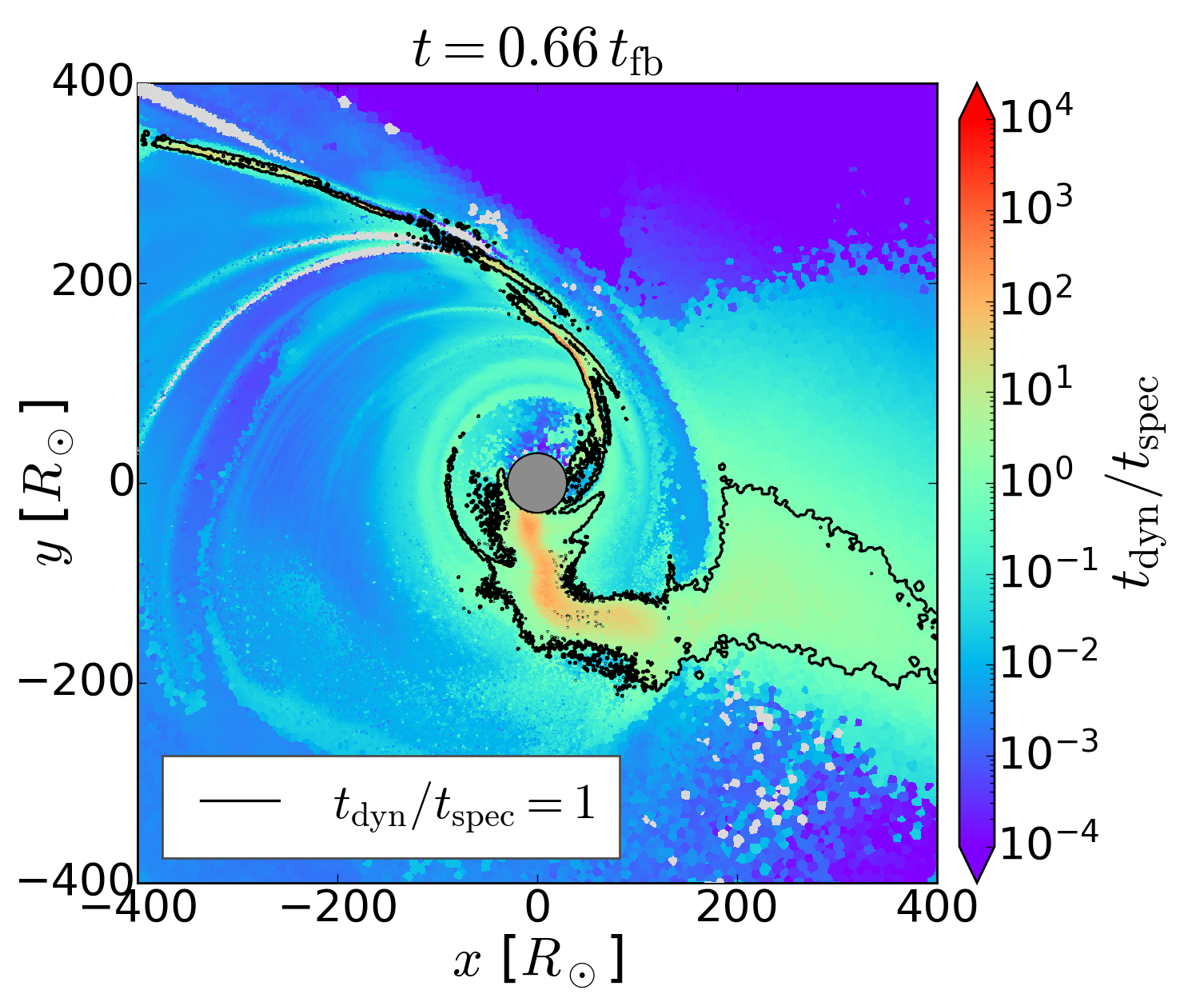}}
    \hfill
    \subfloat[$t_{\rm dyn}/t_{\rm spec}$, $0.90\,t_{\rm fb}$]{\includegraphics[width=0.32\textwidth]{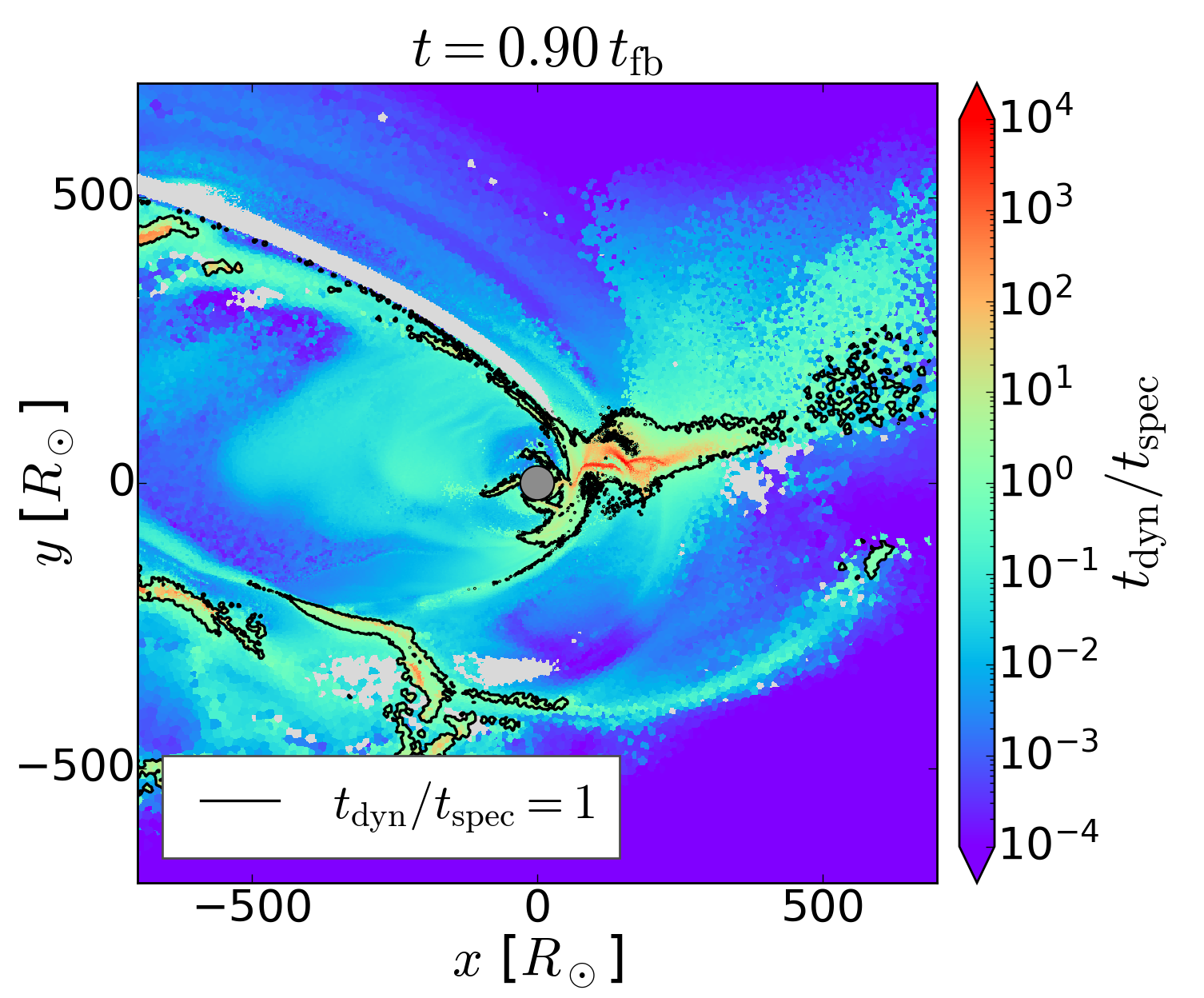}}
    \hfill
    \subfloat[$t_{\rm dyn}/t_{\rm spec}$, $1.11\,t_{\rm fb}$]{\includegraphics[width=0.32\textwidth]{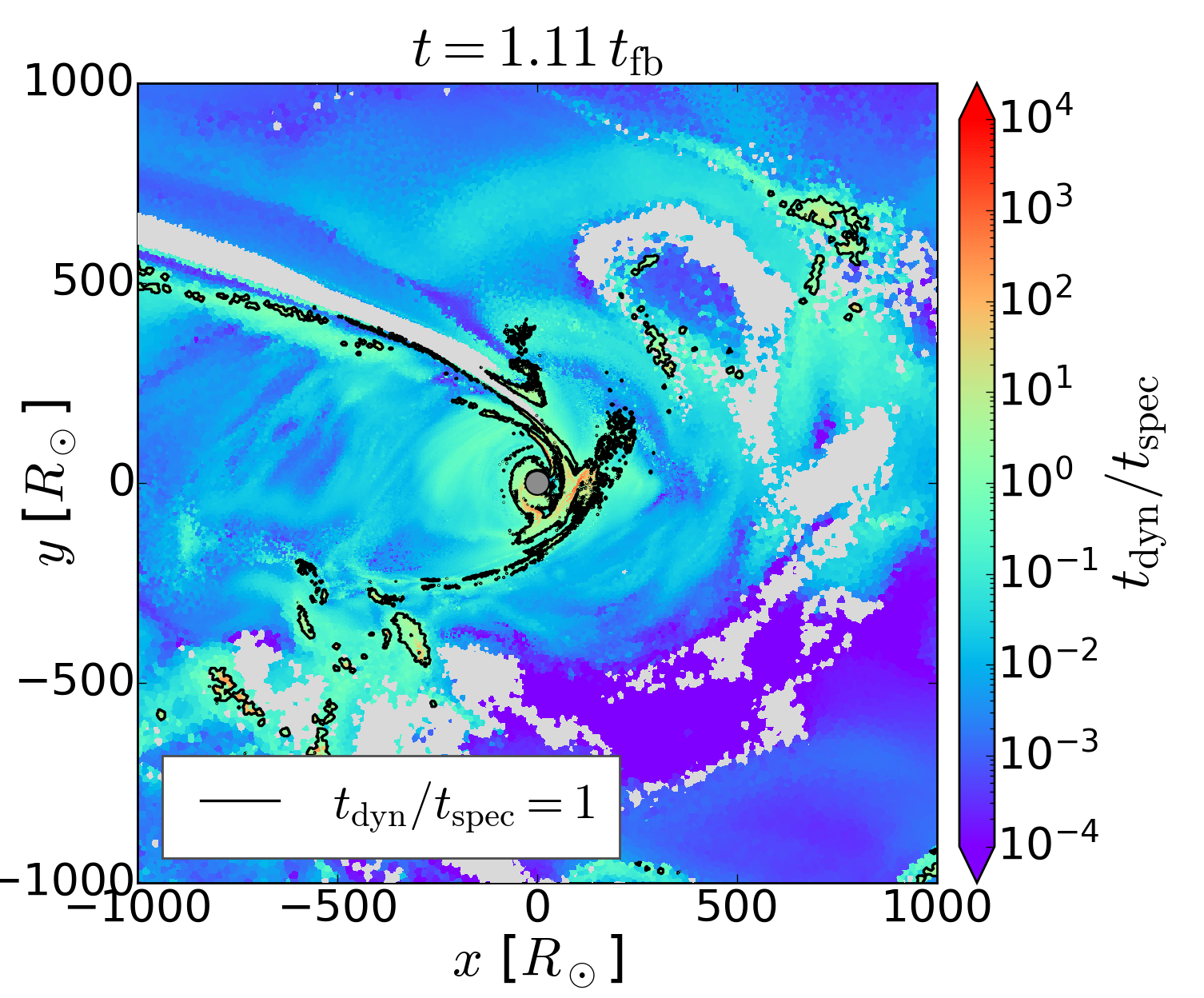}}
    \caption{The ratio $t_{\rm dyn}/t_{\rm spec}$ (Eqs.~\eqref{eq:spectral_timescale},~\eqref{eq:dynamical_time}) in the orbital plane at $0.66$, $0.90$, and $1.11\,t_{\rm fb}$, with a black contour at $t_{\rm dyn}/t_{\rm spec}=1$. The gray disk is the center region, and light gray cells are gas in which Compton scattering is omitted by the cutoff of Eq.~\eqref{eq:compton_cutoff}.}
    \label{fig:tde_active}
\end{figure*}

Figure~\ref{fig:tde_active} shows where $t_{\rm dyn}/t_{\rm spec}\geq1$. At $0.66\,t_{\rm fb}$, this holds in the nozzle shock region, in the outflowing gas that leaves it toward the positive $x$ direction below the black hole, and along the edges of the returning stream. At $0.90\,t_{\rm fb}$ the ratio reaches up to $10^4$ in the nozzle shock region and in the outflowing gas that leaves it along the positive $x$ axis. At $1.11\,t_{\rm fb}$ it exceeds unity mainly in the disk near $R_{\rm smooth}$ and along the returning stream close to the black hole. 

\subsection{Coupling Fraction}
\label{sec:tde_coupling}

Rapid spectral redistribution does not necessarily mean that Compton scattering dominates the energy exchanged between the radiation and the gas. The gas heating rates by Compton scattering and by absorption and emission are
\begin{equation}
Q_C=-\sum_{g'}C_{g'},
\qquad
Q_{\rm abs}=c\sum_{g'}\kappa_{P,g'}
\left(E_{g'}-b_{g'}U_r\right)
\label{eq:coupling_rates}
\end{equation}
which are the explicit forms of the absorption--emission and Compton terms of Eq.~\eqref{eq:material_energy}. We define the Compton coupling fraction as
\begin{equation}
f_C=\frac{|Q_C|}{|Q_C|+|Q_{\rm abs}|}
\label{eq:coupling_fraction}
\end{equation}
We say that Compton scattering dominates the coupling where $f_C>0.5$, that is, where the net gas heating or cooling by Compton scattering exceeds that by absorption and emission.

\begin{figure*}
    \centering
    \subfloat[$f_C$, $0.66\,t_{\rm fb}$]{\includegraphics[width=0.32\textwidth]{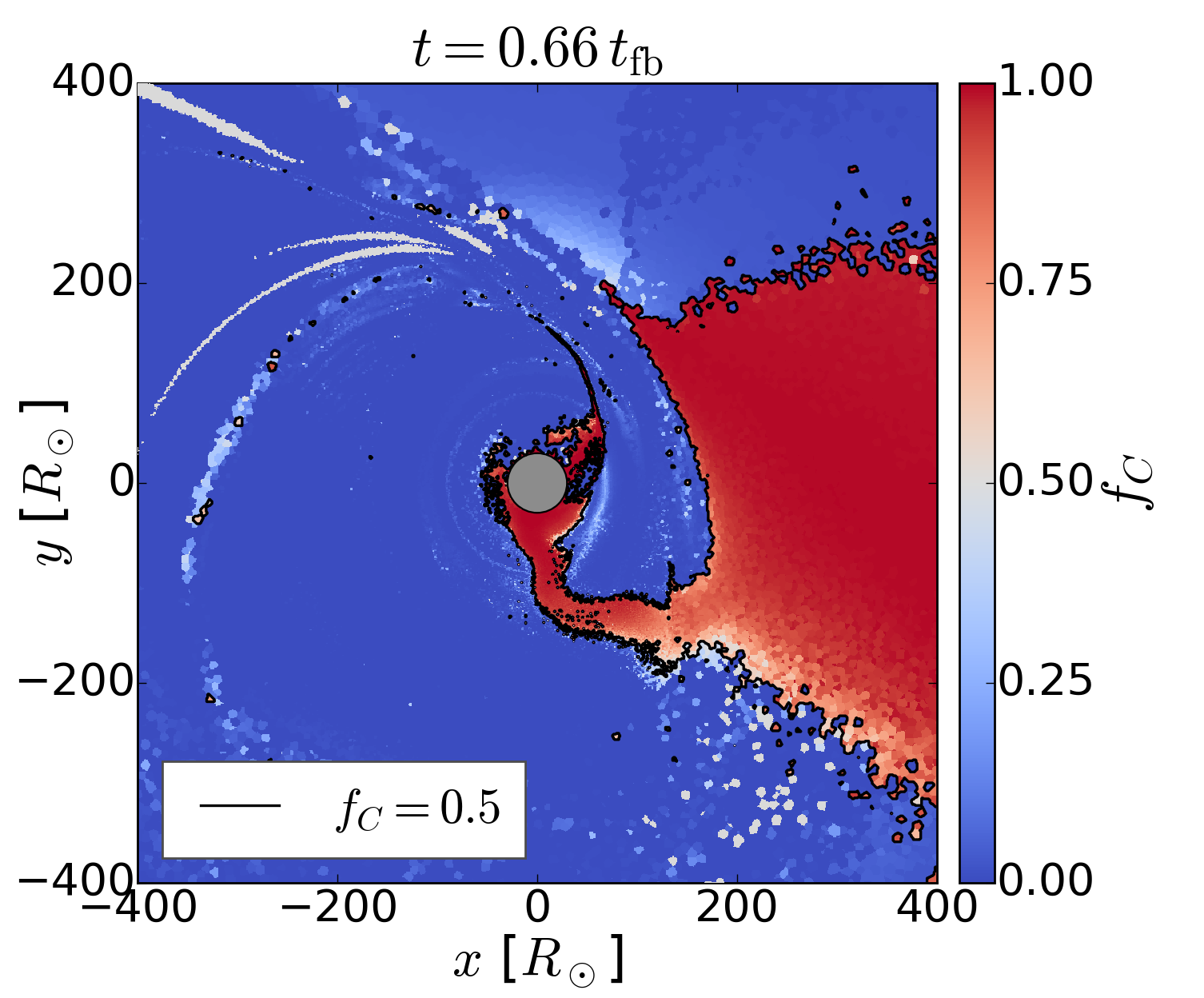}}
    \hfill
    \subfloat[$f_C$, $0.90\,t_{\rm fb}$]{\includegraphics[width=0.32\textwidth]{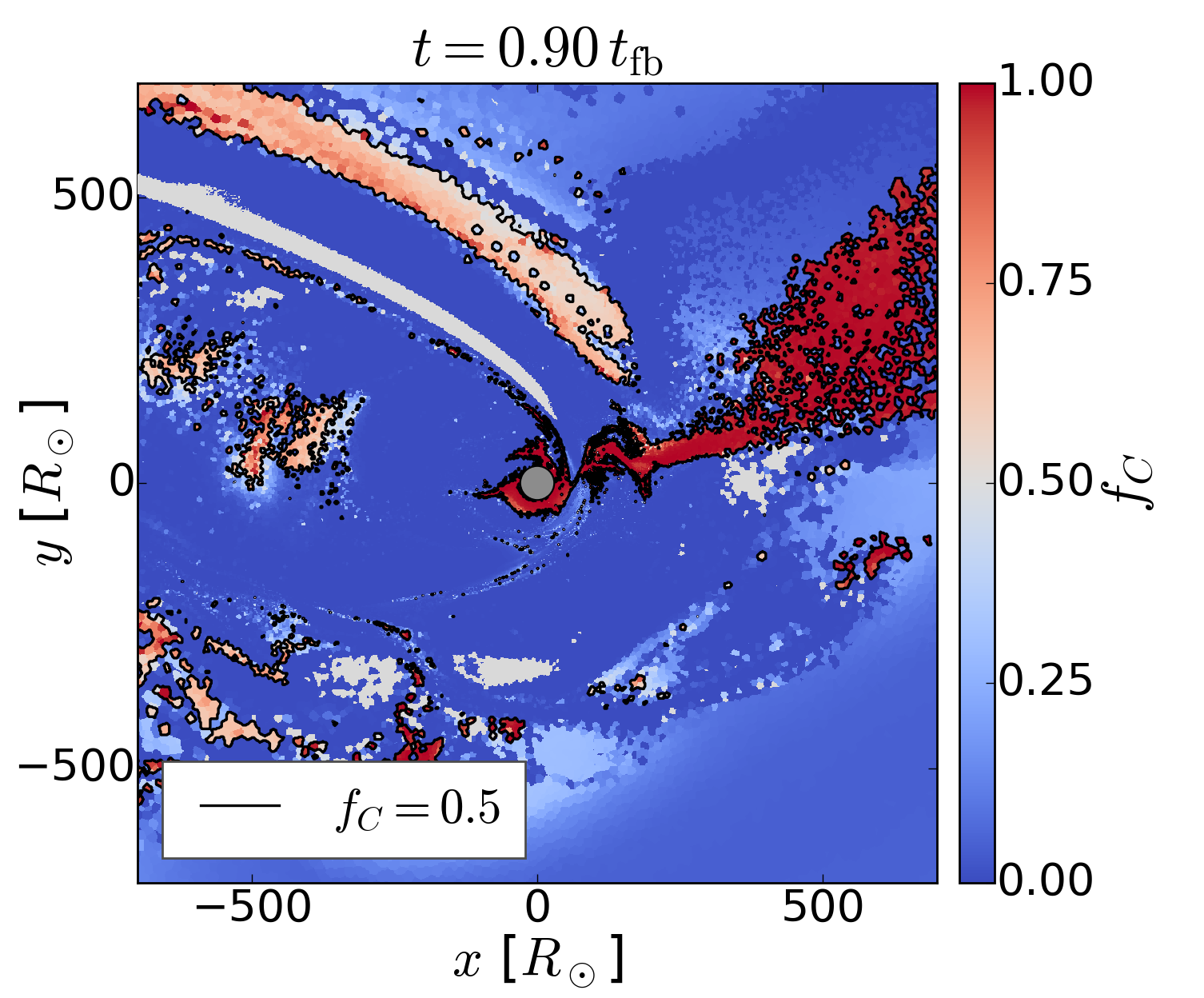}}
    \hfill
    \subfloat[$f_C$, $1.11\,t_{\rm fb}$]{\includegraphics[width=0.32\textwidth]{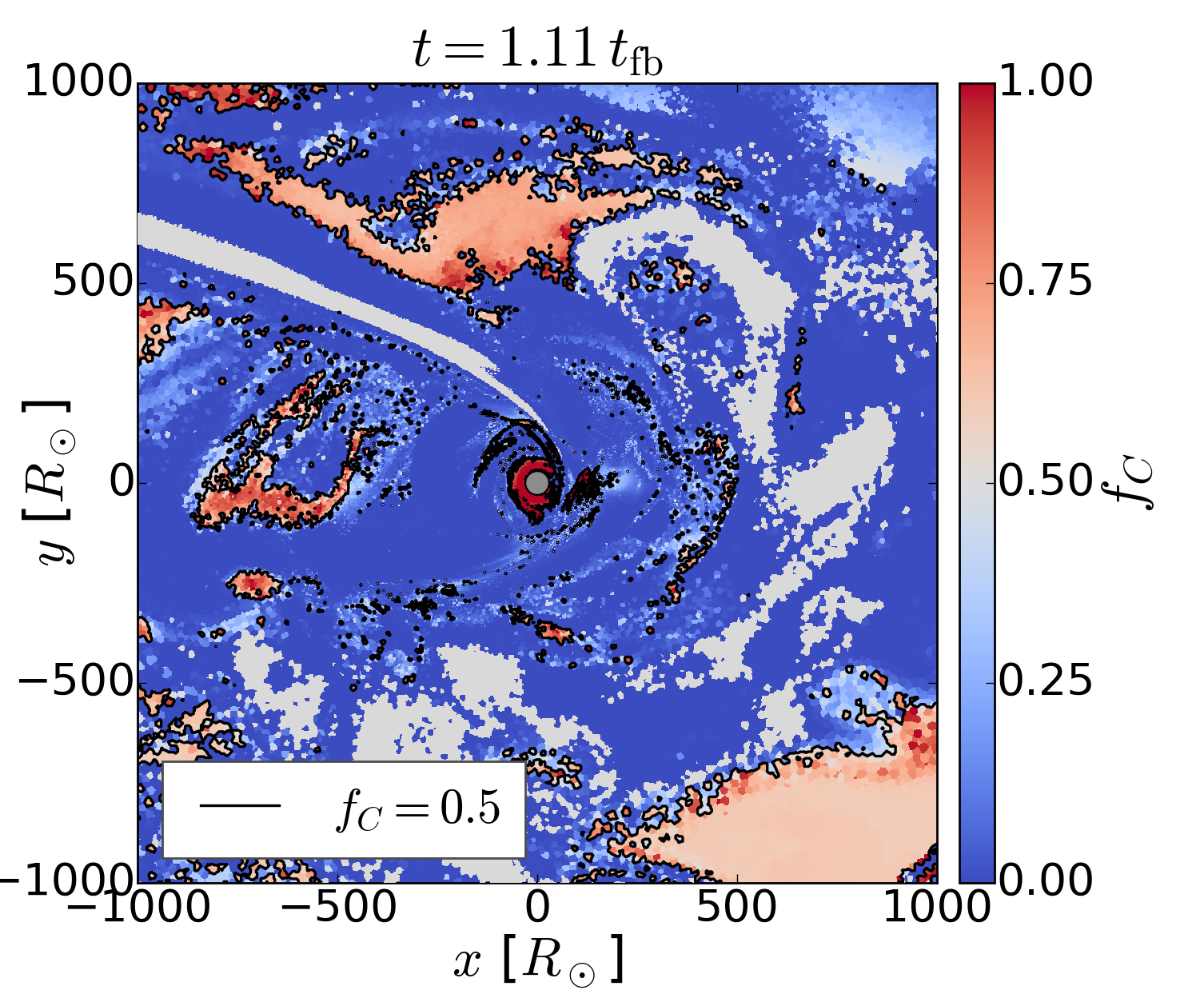}}
    \caption{$f_C$ of Eq.~\eqref{eq:coupling_fraction} in the orbital plane at $0.66$, $0.90$, and $1.11\,t_{\rm fb}$, with a black contour at $f_C=0.5$. The gray disk is the center region, and light gray cells are gas in which Compton scattering is omitted by the cutoff of Eq.~\eqref{eq:compton_cutoff}.}
    \label{fig:tde_coupling}
\end{figure*}

Figure~\ref{fig:tde_coupling} shows that during the X-ray flash, $f_C\approx1$ in the nozzle shock region and in the outflowing gas that leaves it. This gas also has $t_{\rm dyn}/t_{\rm spec}\geq1$ (Figure~\ref{fig:tde_active}), so Compton scattering both dominates the energy exchange and rapidly changes the spectrum, and may therefore affect the dynamics near the nozzle shock. As denser gas accumulates near the black hole, absorption and emission take over, and at $1.11\,t_{\rm fb}$ Compton scattering dominates near the black hole only in the disk near $R_{\rm smooth}$. Elsewhere it dominates in tenuous gas far from the black hole, where both the Compton and the absorption--emission rates are small.

\begin{figure*}
    \centering
    \subfloat[Outside the center region\label{fig:tde_mass_fractions}]{\includegraphics[width=0.49\textwidth]{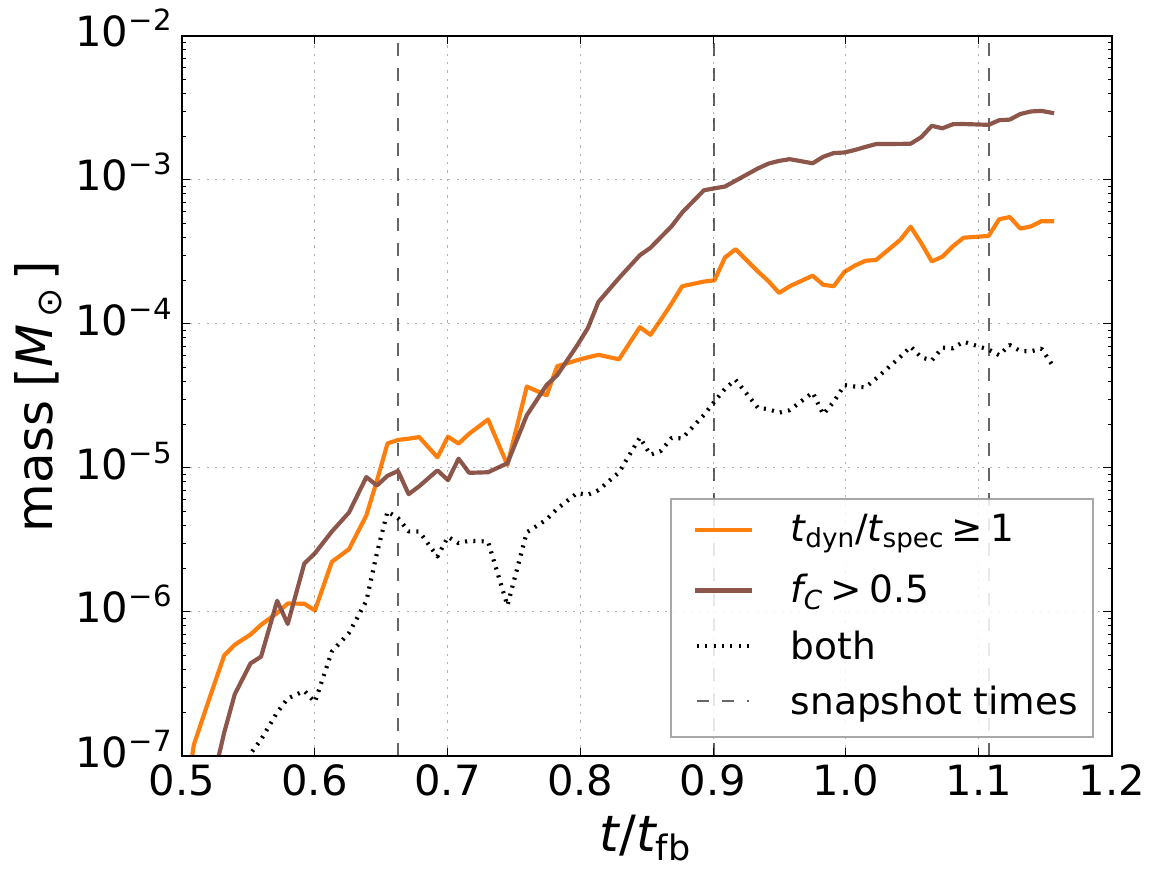}}
    \hfill
    \subfloat[Disk\label{fig:tde_disk}]{\includegraphics[width=0.49\textwidth]{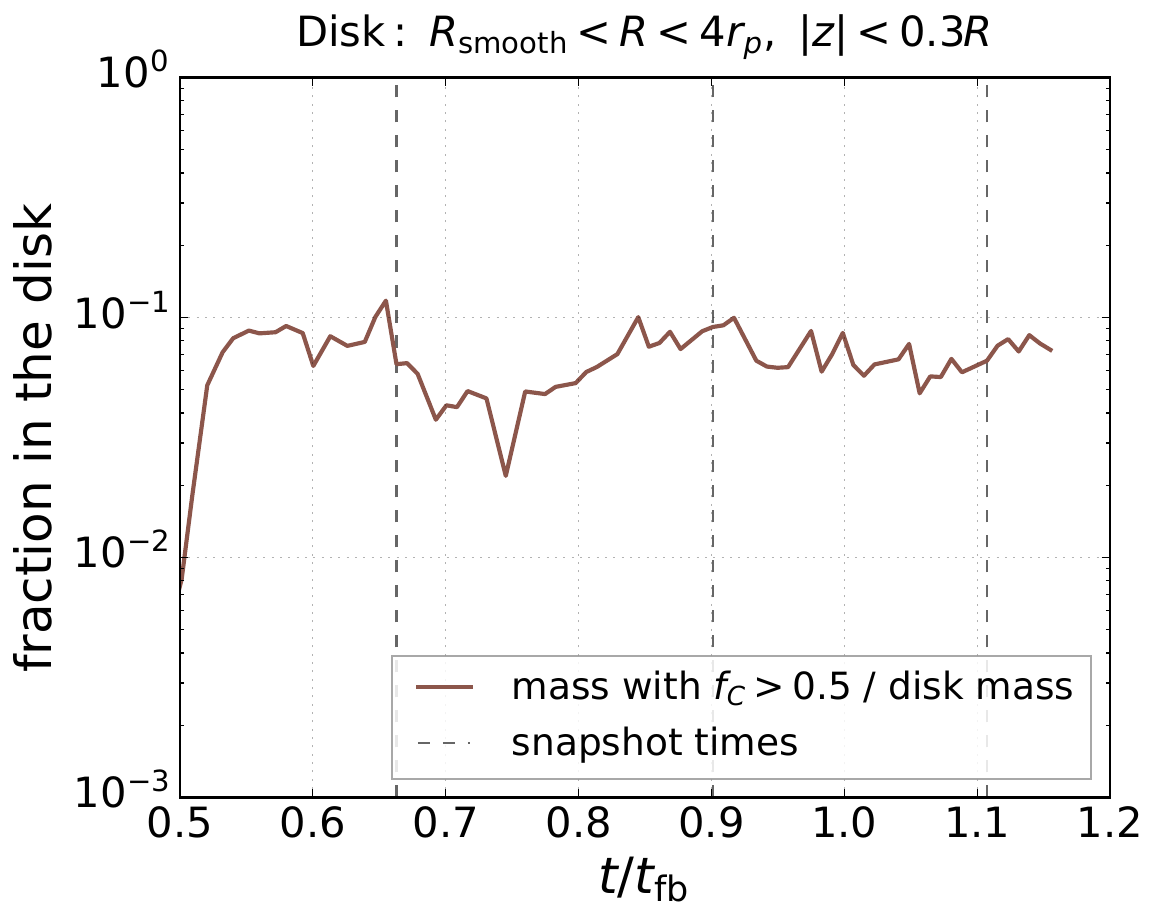}}
    \caption{Time evolution of (a) the mass of the gas outside the center region with $t_{\rm dyn}/t_{\rm spec}\geq1$, with $f_C>0.5$, and with both, and (b) the disk mass fraction with $f_C>0.5$. The dashed vertical lines are as in Figure~\ref{fig:tde_lightcurve}.}
    \label{fig:tde_coupling_mass}
\end{figure*}

Figure~\ref{fig:tde_mass_fractions} shows that the masses of the gas outside the center region with $t_{\rm dyn}/t_{\rm spec}\geq1$ and with $f_C>0.5$ grow with time but remain below $4\times10^{-3}\,M_\odot$, less than $1\%$ of the gas mass. Most of the gas with $f_C>0.5$ is far from the black hole, so we also measure the coupling in the disk.

We define the disk as the gas at $R_{\rm smooth}<R<4r_p$ and $|z|<0.3R$, out to twice the circularization radius. In the disk, we measure the fraction of the mass in which Compton scattering dominates the coupling.
Figure~\ref{fig:tde_disk} shows that the picture is different in the disk, where Compton scattering dominates the coupling in a few percent of the mass, compared with less than $1\%$ of the gas outside the center region.

\subsection{Direction of the Redistribution}
\label{sec:tde_direction}

The sign of $Q_C$ distinguishes Compton heating of the gas, $Q_C>0$, from inverse-Compton cooling, $Q_C<0$. Since the radiation gains energy where the gas is cooled, the sign of $Q_C$ also gives the net direction of the redistribution, but not how one-sided it is. A transfer from group $g$ to group $g'$ changes the radiation energy by $\nu_{g'}/\nu_g-1$ times the energy it removes from group $g$ (Section~\ref{sec:compton_coefficients}). The rates at which the radiation gains energy in upscattering transitions and loses energy in downscattering transitions between distinct groups are therefore
\begin{equation}
\begin{aligned}
U_C&=c\sum_{g}\sum_{g'>g}\left(\frac{\nu_{g'}}{\nu_g}-1\right)E_g(1+n_{g'})\,\sigma_s(g\mathbin{\to}g'),\\
D_C&=c\sum_{g}\sum_{g'<g}\left(1-\frac{\nu_{g'}}{\nu_g}\right)E_g(1+n_{g'})\,\sigma_s(g\mathbin{\to}g')
\end{aligned}
\label{eq:direction_rates}
\end{equation}
Since $\sigma_s(g\mathbin{\to}g')\geq0$, $U_C$ and $D_C$ are nonnegative. The inverse-Compton preference
\begin{equation}
A_{\rm IC}=\frac{U_C-D_C}{U_C+D_C}
\label{eq:inverse_compton_preference}
\end{equation}
is positive where upscattering dominates the resolved redistribution and negative where downscattering dominates. Transfers within a group exchange no energy, so Eqs.~\eqref{eq:tde_compton_source}, \eqref{eq:coupling_rates}, and~\eqref{eq:direction_rates} give $Q_C=D_C-U_C$ in every cell.

The Compton temperature $T_C$ of the local radiation field is the gas temperature at which Compton scattering neither heats nor cools the gas. In the nonrelativistic limit without induced scattering, $k_BT_C$ is one quarter of the mean photon energy $\int\nu E_\nu\,d\nu/\int E_\nu\,d\nu$ \citep{Kompaneets1957,RybickiLightman1979,WINSLOW1995262}. We define $T_C$ of a cell as the gas temperature at which $Q_C$ of Eq.~\eqref{eq:coupling_rates}, evaluated with the group energies $\{E_{g,i}\}$ of the cell, changes sign from heating to cooling as the gas temperature increases. Writing $T_{\rm gas}\equiv T$ for the gas temperature, the gas is cooled by Compton scattering where $T_{\rm gas}/T_C>1$. 

\begin{figure*}
    \centering
    \subfloat[$A_{\rm IC}$, $0.66\,t_{\rm fb}$]{\includegraphics[width=0.32\textwidth]{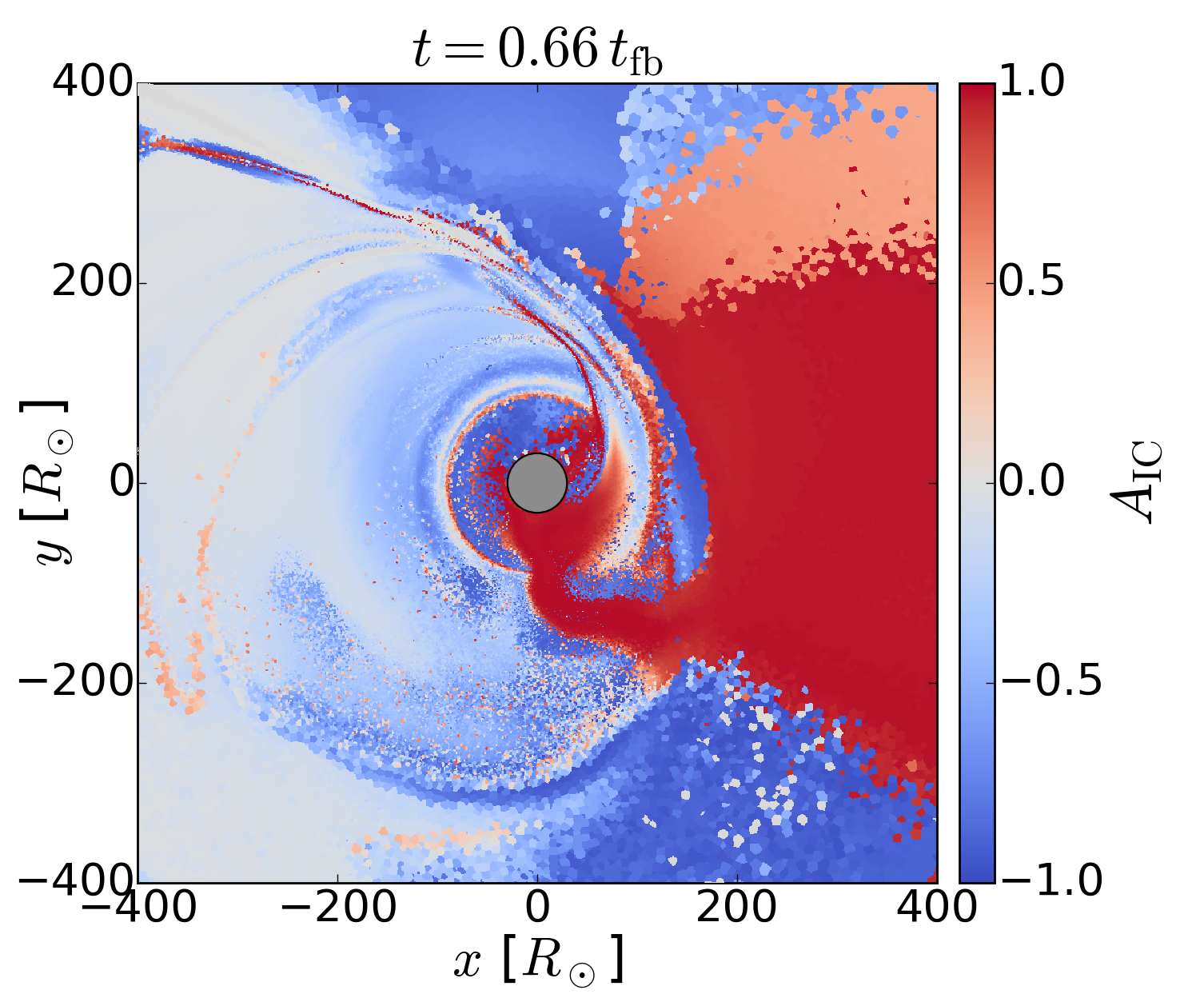}}
    \hfill
    \subfloat[$A_{\rm IC}$, $0.90\,t_{\rm fb}$]{\includegraphics[width=0.32\textwidth]{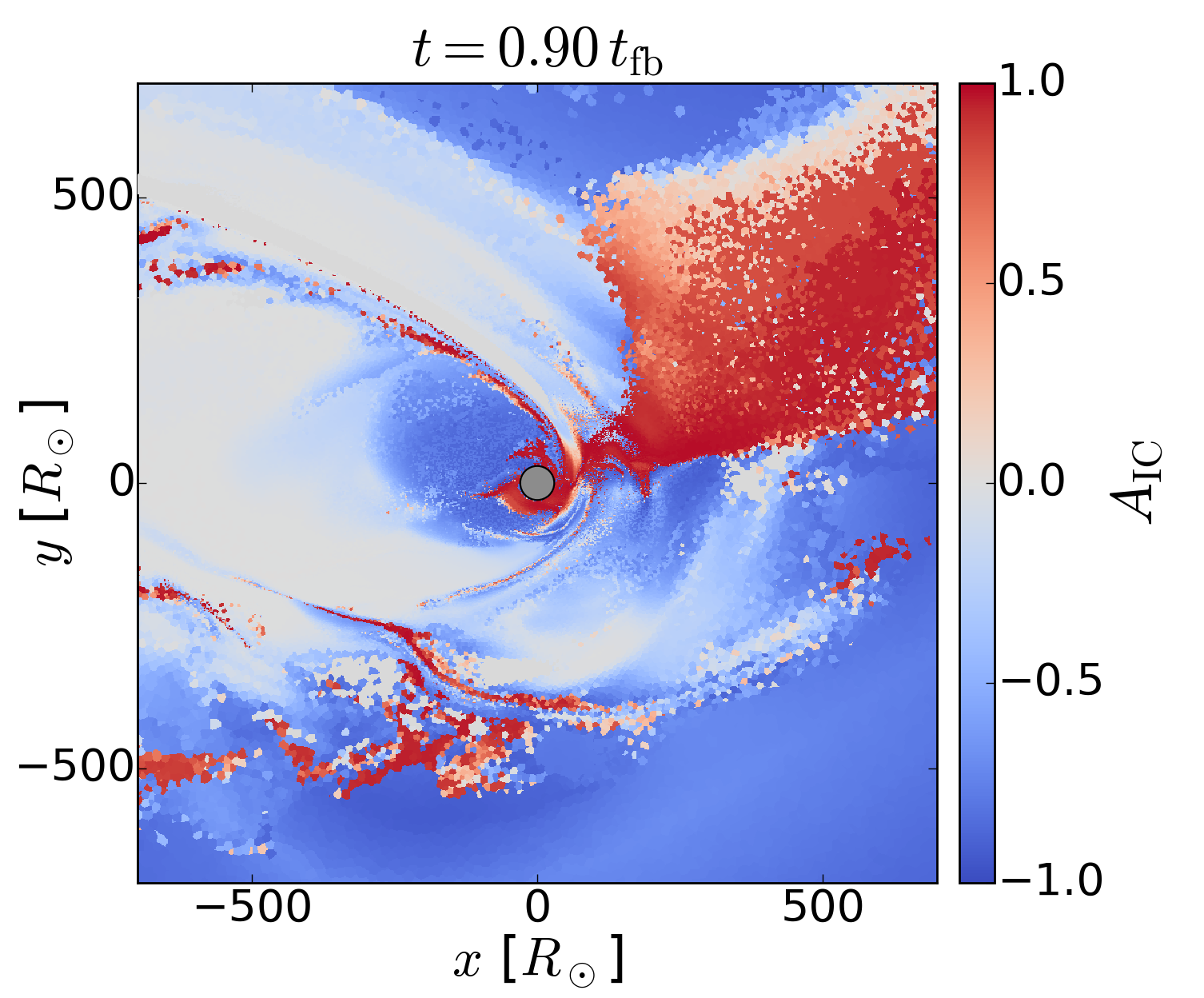}}
    \hfill
    \subfloat[$A_{\rm IC}$, $1.11\,t_{\rm fb}$]{\includegraphics[width=0.32\textwidth]{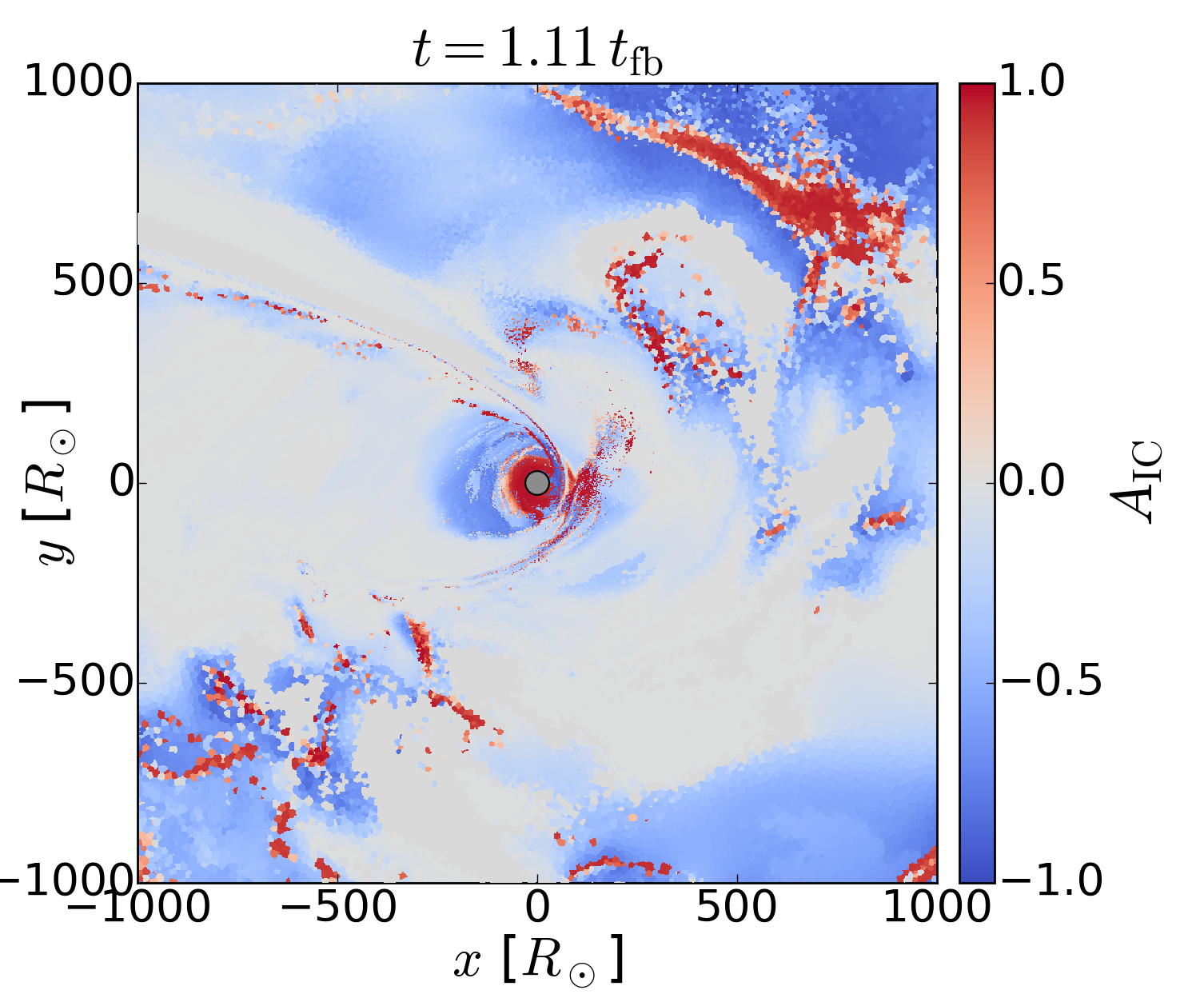}}\\
    \subfloat[$T_{\rm gas}/T_C$, $0.66\,t_{\rm fb}$]{\includegraphics[width=0.32\textwidth]{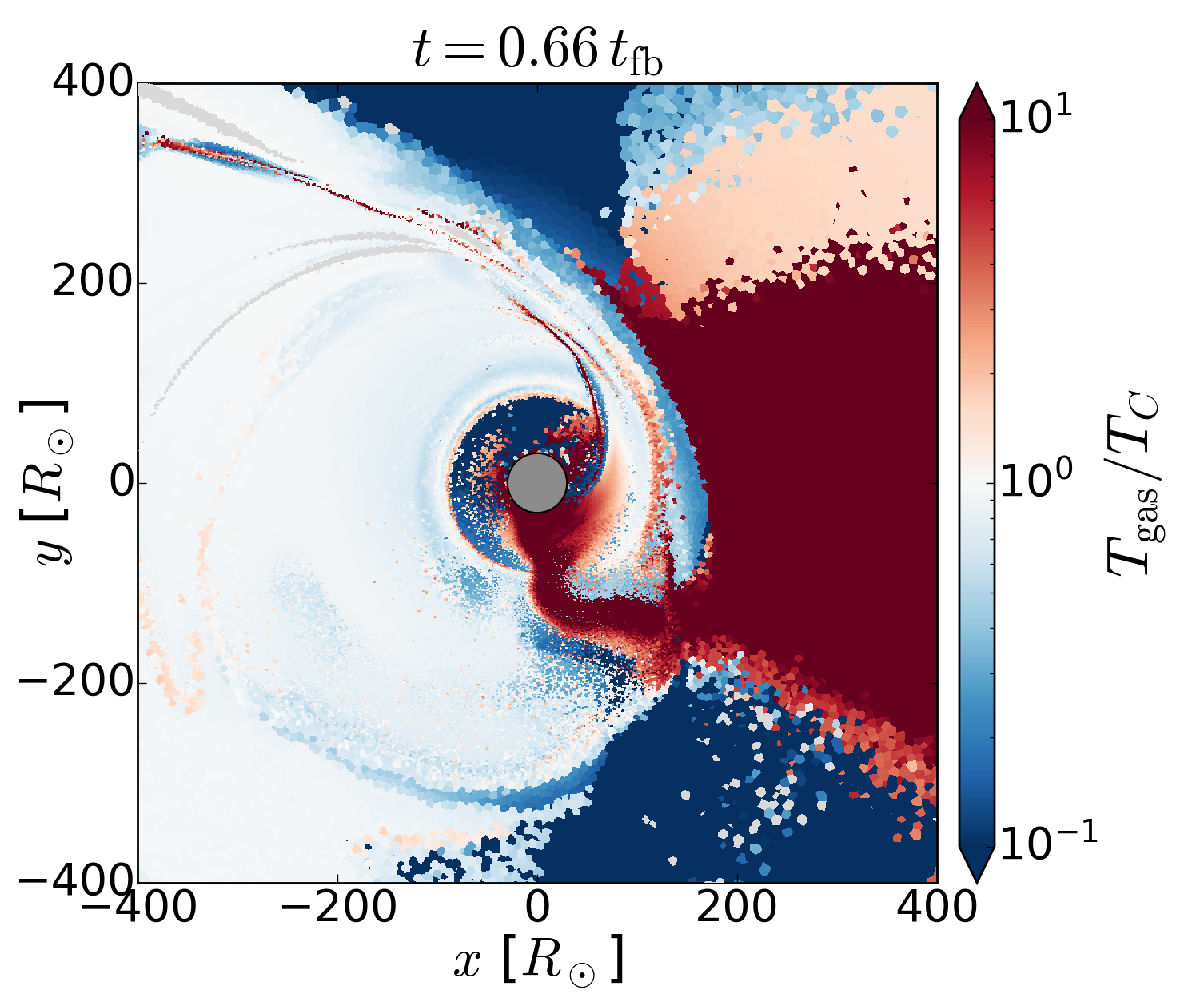}}
    \hfill
    \subfloat[$T_{\rm gas}/T_C$, $0.90\,t_{\rm fb}$]{\includegraphics[width=0.32\textwidth]{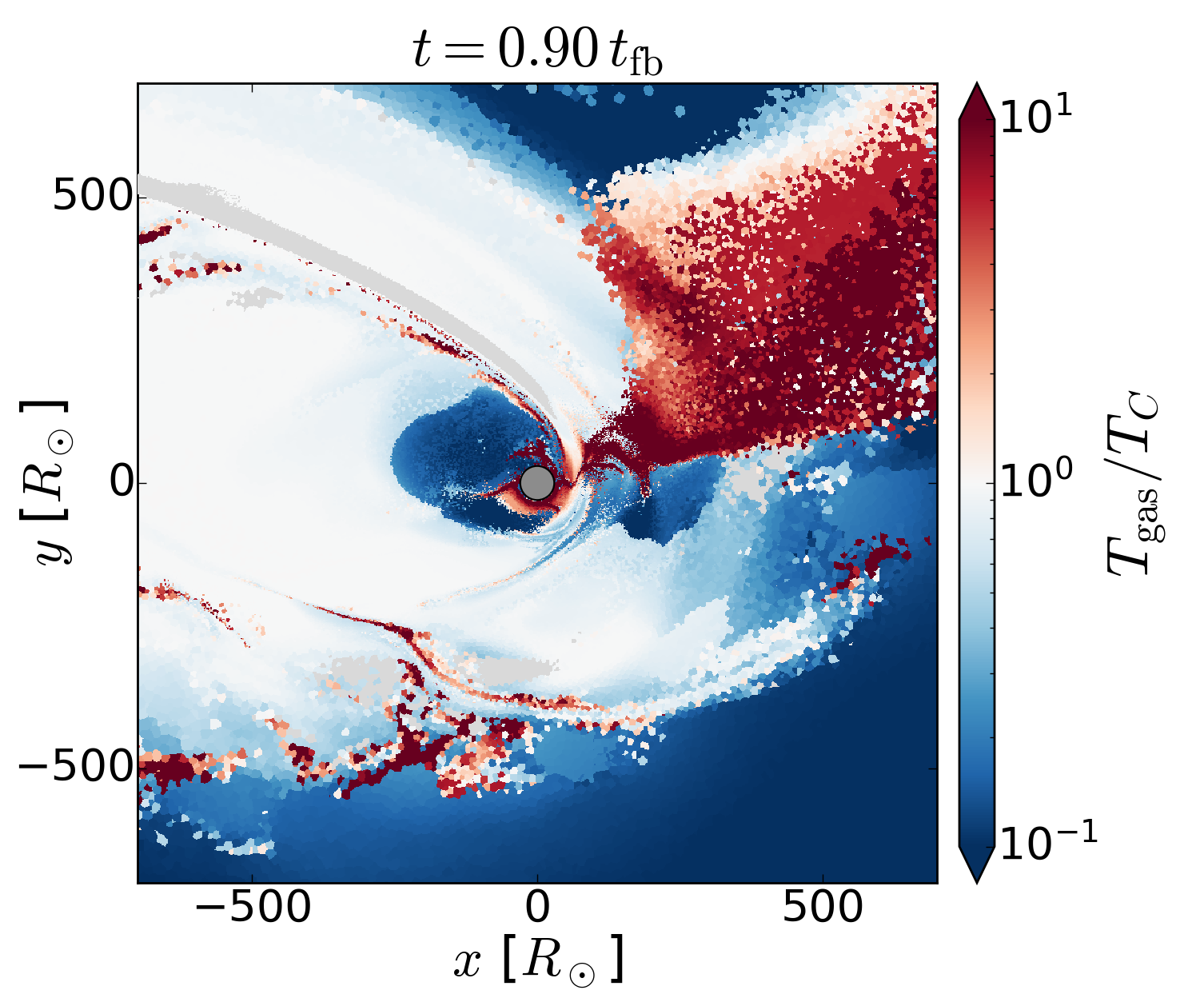}}
    \hfill
    \subfloat[$T_{\rm gas}/T_C$, $1.11\,t_{\rm fb}$]{\includegraphics[width=0.32\textwidth]{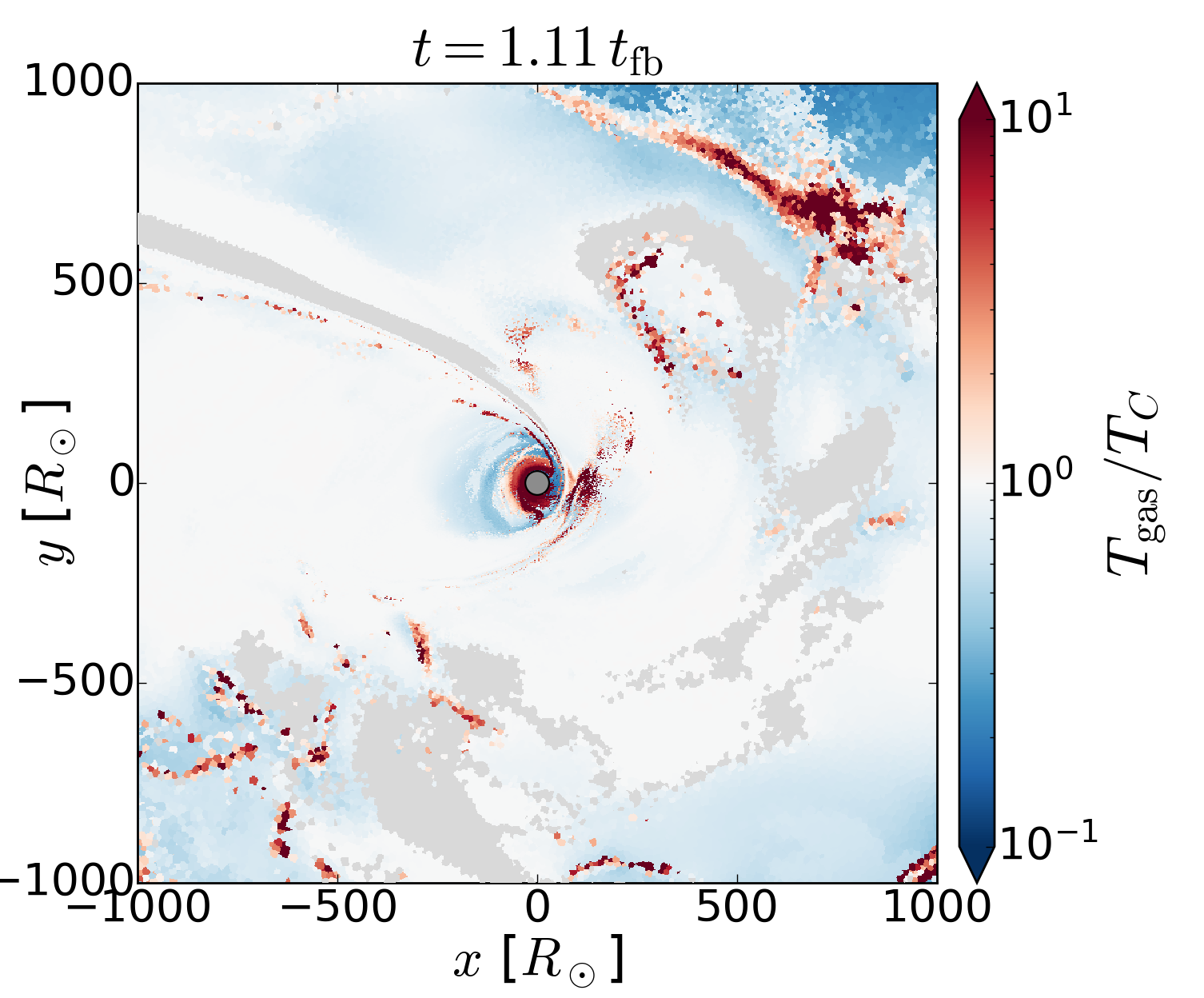}}
    \caption{Top: $A_{\rm IC}$ of Eq.~\eqref{eq:inverse_compton_preference}. Bottom: $T_{\rm gas}/T_C$, with $T_C$ defined in Section~\ref{sec:tde_direction}. Both rows show the orbital plane at $0.66$, $0.90$, and $1.11\,t_{\rm fb}$. Red marks upscattering and gas that is hotter than the Compton temperature. The gray disk is the center region, and light gray cells are gas in which Compton scattering is omitted by the cutoff of Eq.~\eqref{eq:compton_cutoff}.}
    \label{fig:tde_direction}
\end{figure*}

Figure~\ref{fig:tde_direction} shows the direction of the redistribution. The top row shows how one-sided the redistribution is, and the bottom row how far the gas is from the Compton temperature. Where Compton scattering dominates near the black hole (the nozzle shock region and the outflowing gas at $0.66$ and $0.90\,t_{\rm fb}$, and the disk near $R_{\rm smooth}$ at $1.11\,t_{\rm fb}$), the gas is much hotter than the Compton temperature, so Compton scattering cools it and the redistribution is almost entirely toward higher photon energies, $A_{\rm IC}\approx1$. Elsewhere near the black hole, in parts of the gas just outside the center region at $0.66\,t_{\rm fb}$ and in the disk away from the nozzle shock at $0.90\,t_{\rm fb}$, the gas is colder than the Compton temperature. There Compton scattering heats the gas and downscatters the radiation, consistent with heating by X-rays produced in the hotter gas nearby. The distant gas in which Compton scattering dominates at $0.90$ and $1.11\,t_{\rm fb}$ is also colder than the Compton temperature and is thus heated. At $1.11\,t_{\rm fb}$ most of the flow is close to the Compton temperature.

\subsection{X-Ray Production and Emission}
\label{sec:tde_xray}

The net rates at which Compton scattering and absorption--emission change the X-ray energy are
\begin{equation}
\dot E_{X,C}=\sum_{g\in X}C_{g},
\qquad
\dot E_{X,\rm th}=\sum_{g\in X}c\kappa_{P,g}\left(b_{g}U_r-E_{g}\right)
\label{eq:xray_sources}
\end{equation}
where $X$ denotes the soft and hard X-ray groups of Table~\ref{tab:tde_groups}. To estimate which part of this energy can escape, we also locate the thermalization radius of each group along the rays of Section~\ref{sec:tde_lightcurve}. The effective optical depth of group $g$, integrated inward like $\tau_{R,g}$, is \citep{RybickiLightman1979,steinberg_streamdisk_2024}
\begin{equation}
\tau_{{\rm eff},g}(\ell)
=
\int_{\ell}^{\infty}
\sqrt{3\kappa_{R,g}\kappa_{P,g}}\,d\ell'
\label{eq:tau_eff}
\end{equation}
and the group thermalization radius $r_{{\rm th},g}(\hat{\Omega})$ is the distance of the innermost point of the ray with $\tau_{{\rm eff},g}<1$, outside which photons of group $g$ escape without being absorbed. Where scattering dominates absorption, the thermalization radius lies inside the photosphere. We integrate the net Compton X-ray power over the gas outside the photosphere, and over the scattering layer between the thermalization radius and the photosphere, of each X-ray group,
\begin{equation}
\begin{aligned}
P_{X,C}^{\rm out}&=\sum_{g\in X}\int_{\ell>r_{{\rm ph},g}(\hat{\Omega})}C_g\,dV,\\
P_{X,C}^{\rm sc}&=\sum_{g\in X}\int_{r_{{\rm th},g}(\hat{\Omega})\leq\ell<r_{{\rm ph},g}(\hat{\Omega})}C_g\,dV
\end{aligned}
\label{eq:xray_power}
\end{equation}
X-ray energy produced outside the photosphere can escape directly, while in the scattering layer X-ray photons are scattered before they reach the photosphere, but most of them reach it before being absorbed. X-ray energy produced in the scattering layer can therefore also escape, after it diffuses out, while X-ray energy produced inside the thermalization radius is likely to be absorbed. 

\begin{figure*}
    \centering
    \subfloat[$0.66\,t_{\rm fb}$]{\includegraphics[width=0.32\textwidth]{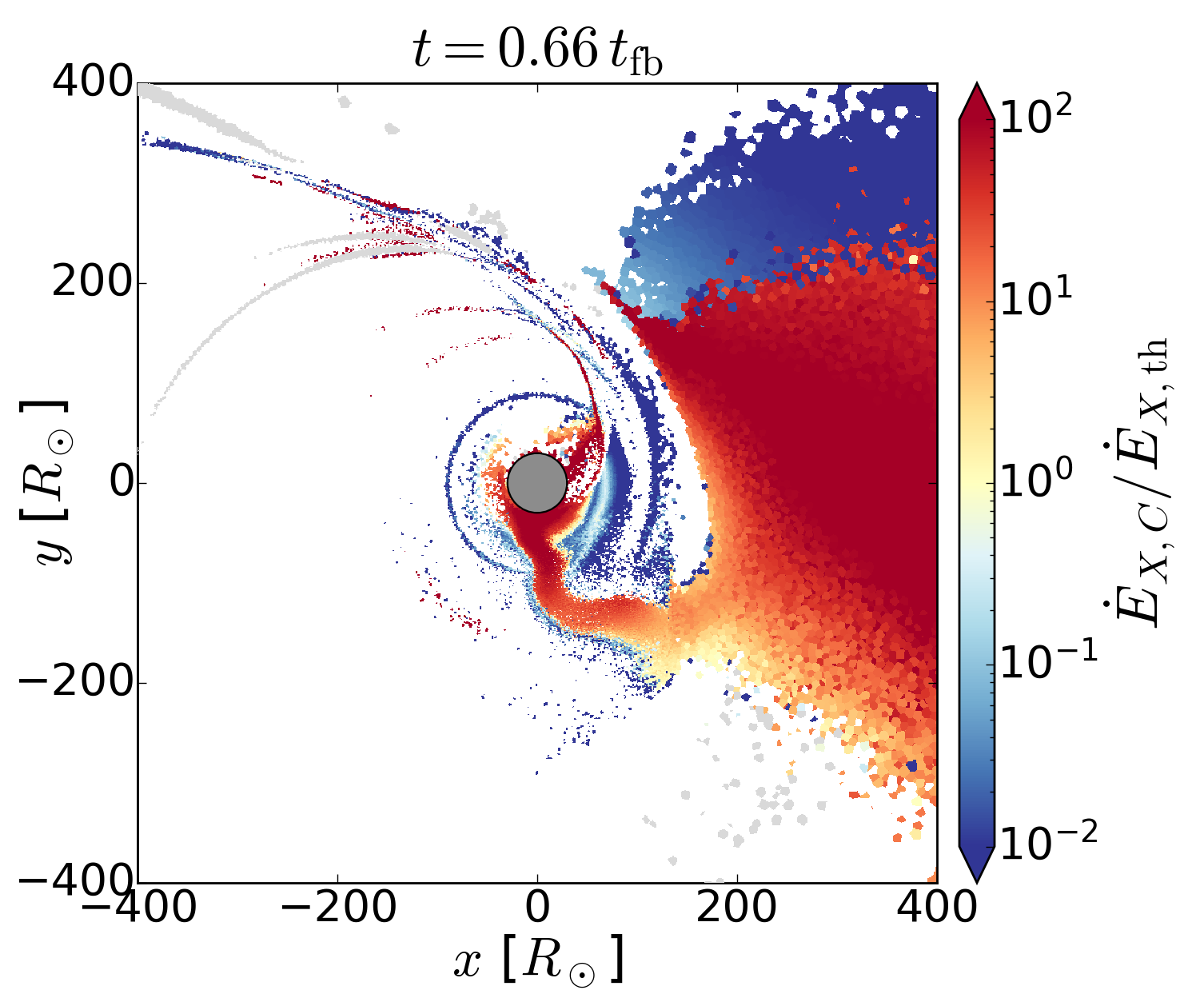}}
    \hfill
    \subfloat[$0.90\,t_{\rm fb}$]{\includegraphics[width=0.32\textwidth]{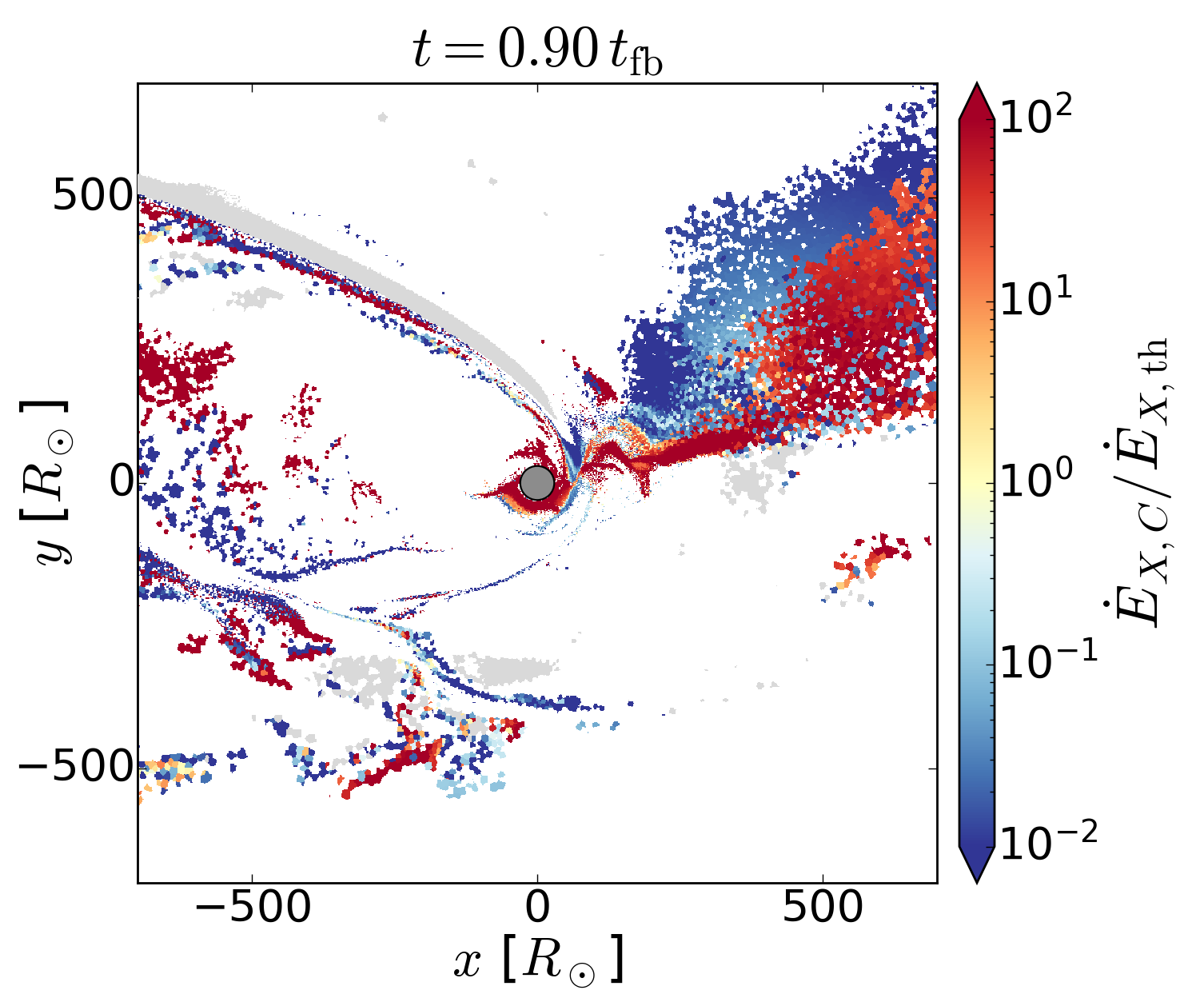}}
    \hfill
    \subfloat[$1.11\,t_{\rm fb}$]{\includegraphics[width=0.32\textwidth]{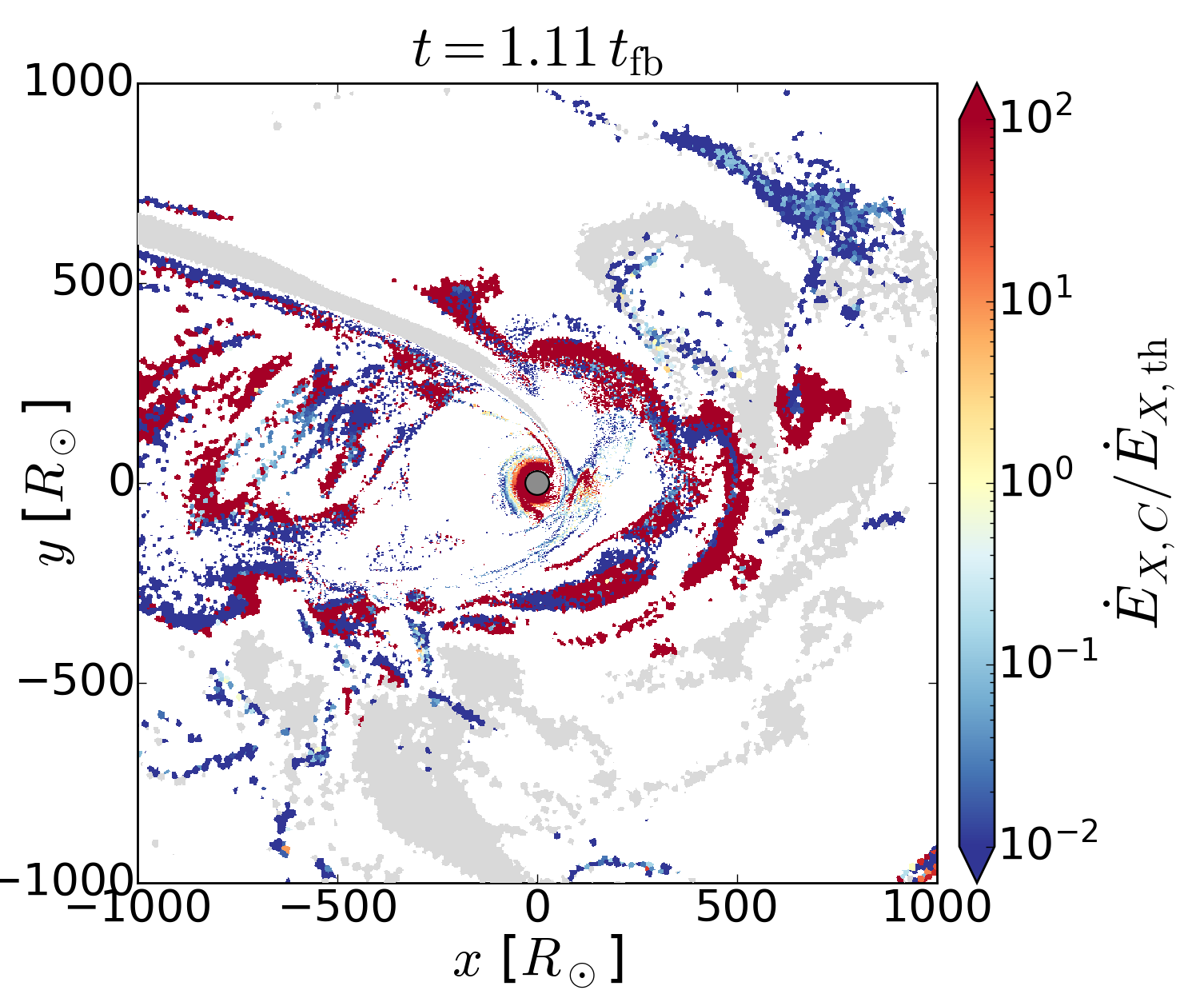}}
    \caption{Ratio $\dot E_{X,C}/\dot E_{X,\rm th}$ of the Compton and thermal X-ray sources of Eq.~\eqref{eq:xray_sources} in the orbital plane at $0.66$, $0.90$, and $1.11\,t_{\rm fb}$. The colors saturate below $10^{-2}$ and above $10^{2}$. Gas in which Compton scattering adds X-ray energy while absorption removes it, $\dot E_{X,\rm th}\leq0$, is shown at the upper limit. White cells are gas in which Compton scattering removes X-ray energy, $\dot E_{X,C}\leq0$. The gray disk is the center region, and light gray cells are gas in which Compton scattering is omitted by the cutoff of Eq.~\eqref{eq:compton_cutoff}.}
    \label{fig:tde_xray}
\end{figure*}

\begin{figure}
    \centering
    \includegraphics[width=1.0\linewidth]{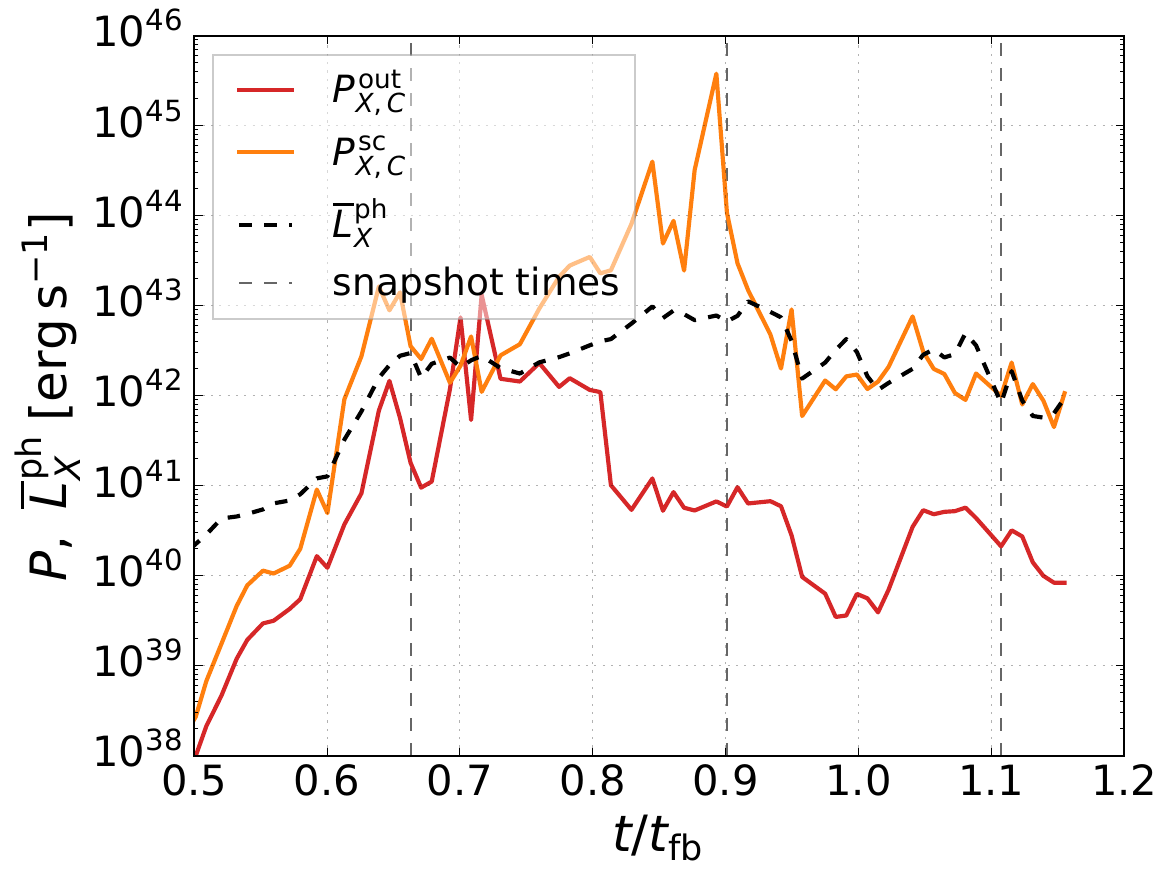}
    \caption{Compton X-ray power of Eq.~\eqref{eq:xray_power} produced outside the X-ray photospheres ($P_{X,C}^{\rm out}$) and in the scattering layer between the X-ray thermalization radii and photospheres ($P_{X,C}^{\rm sc}$), and the sky-averaged X-ray luminosity $\overline{L}_X^{\rm ph}$ of Eq.~\eqref{eq:luminosity_photosphere}. The dashed vertical lines are as in Figure~\ref{fig:tde_lightcurve}.}
    \label{fig:tde_xray_power}
\end{figure}

\begin{figure*}
    \centering
    \subfloat[Optical and UV, $0.66\,t_{\rm fb}$]{\includegraphics[width=0.46\textwidth]{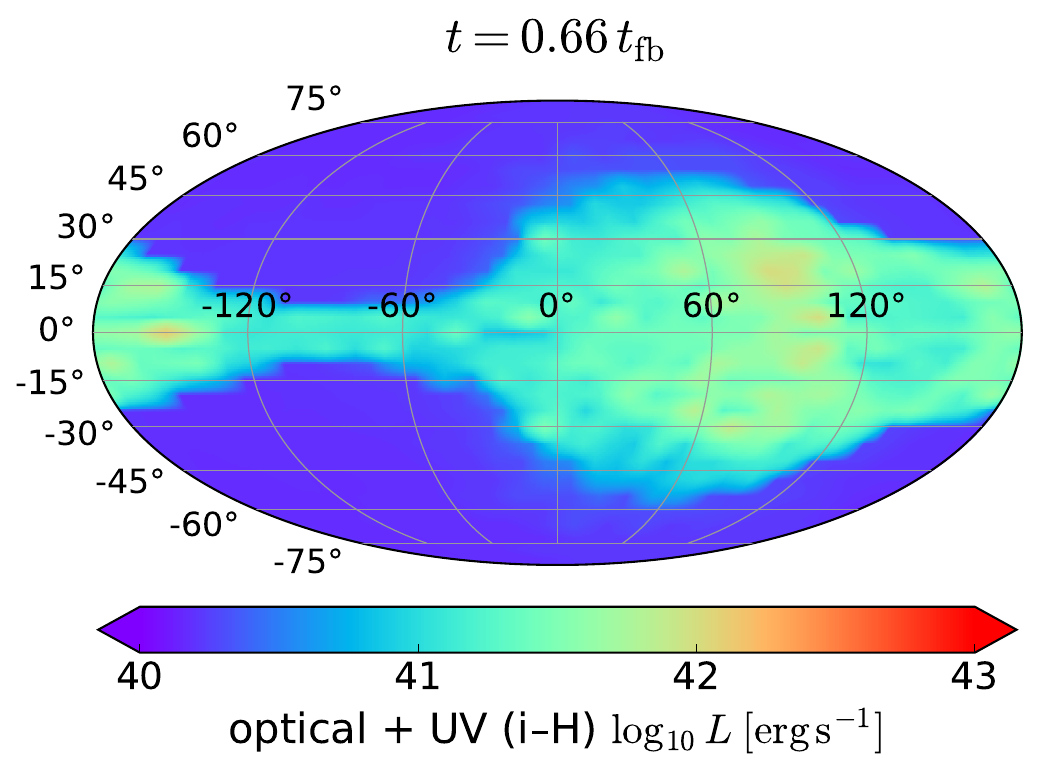}}
    \hfill
    \subfloat[X-ray, $0.66\,t_{\rm fb}$]{\includegraphics[width=0.46\textwidth]{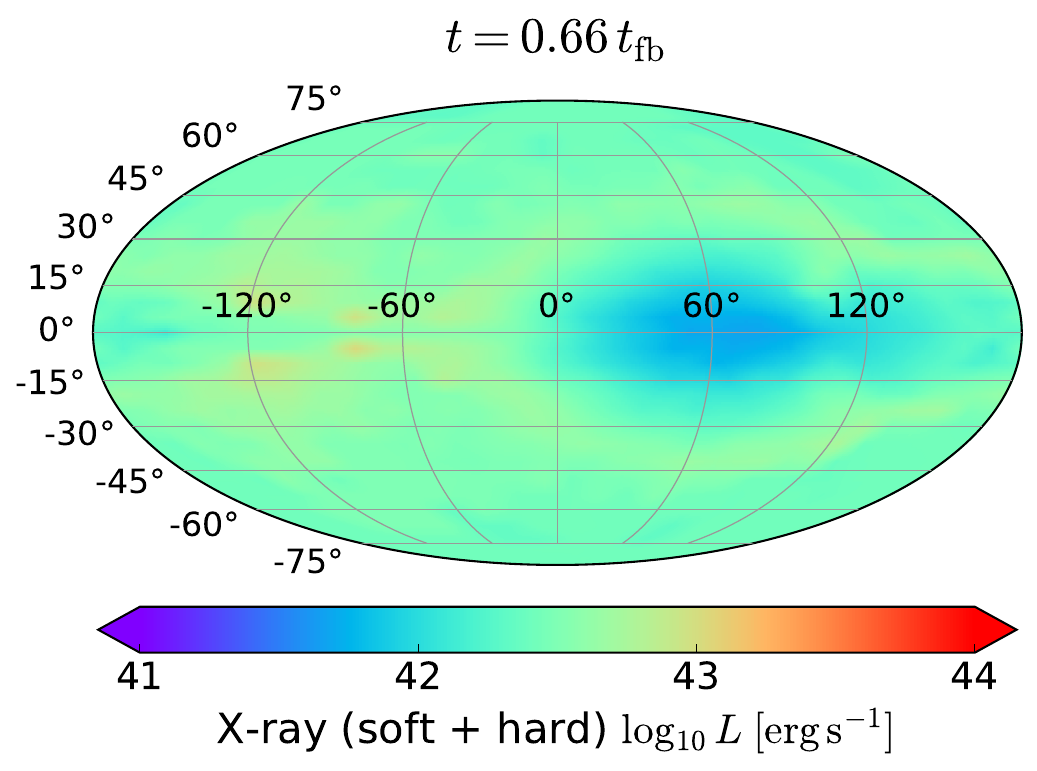}}\\
    \subfloat[Optical and UV, $0.90\,t_{\rm fb}$]{\includegraphics[width=0.46\textwidth]{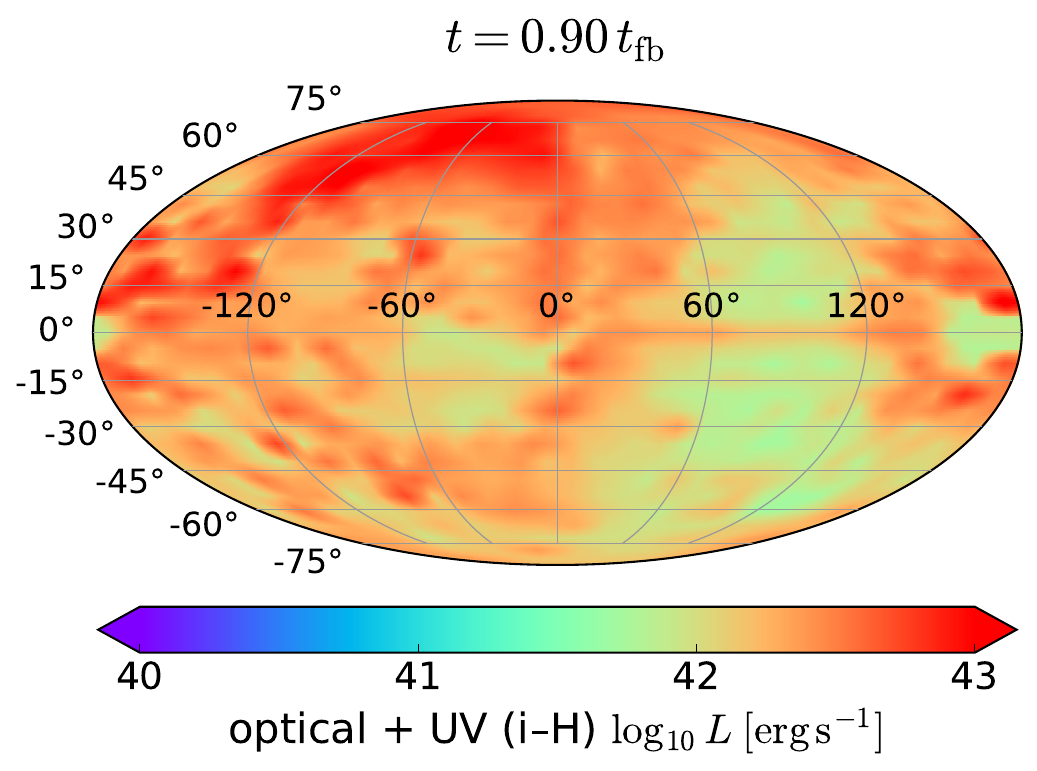}}
    \hfill
    \subfloat[X-ray, $0.90\,t_{\rm fb}$]{\includegraphics[width=0.46\textwidth]{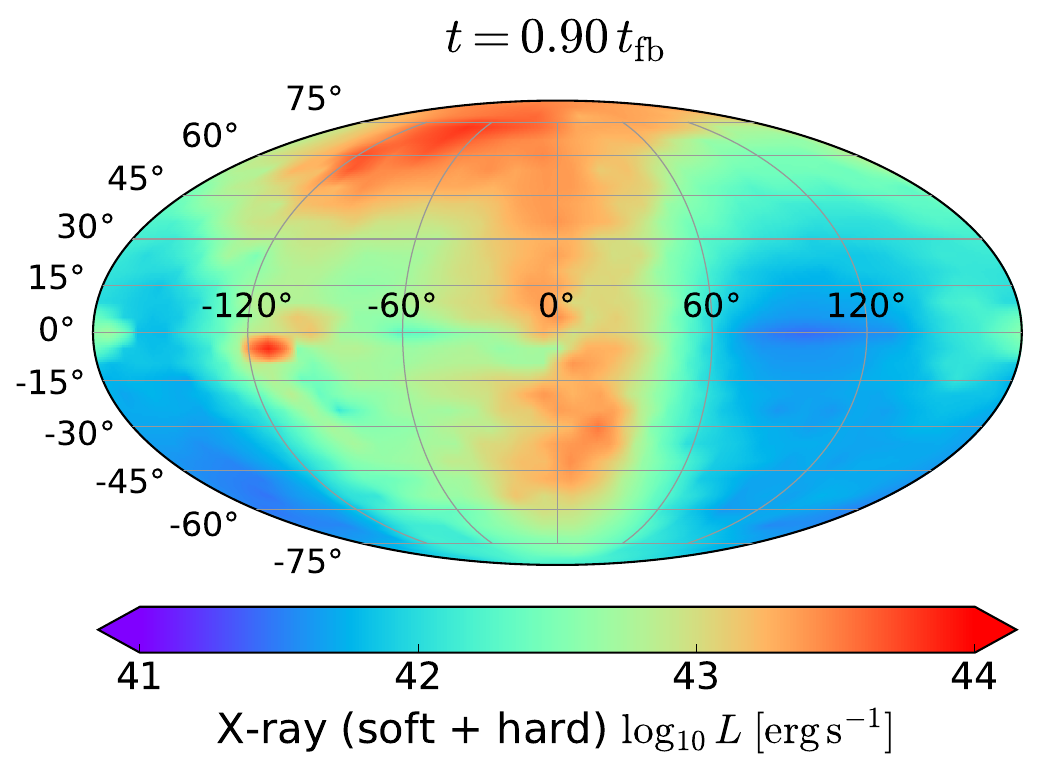}}\\
    \subfloat[Optical and UV, $1.11\,t_{\rm fb}$]{\includegraphics[width=0.46\textwidth]{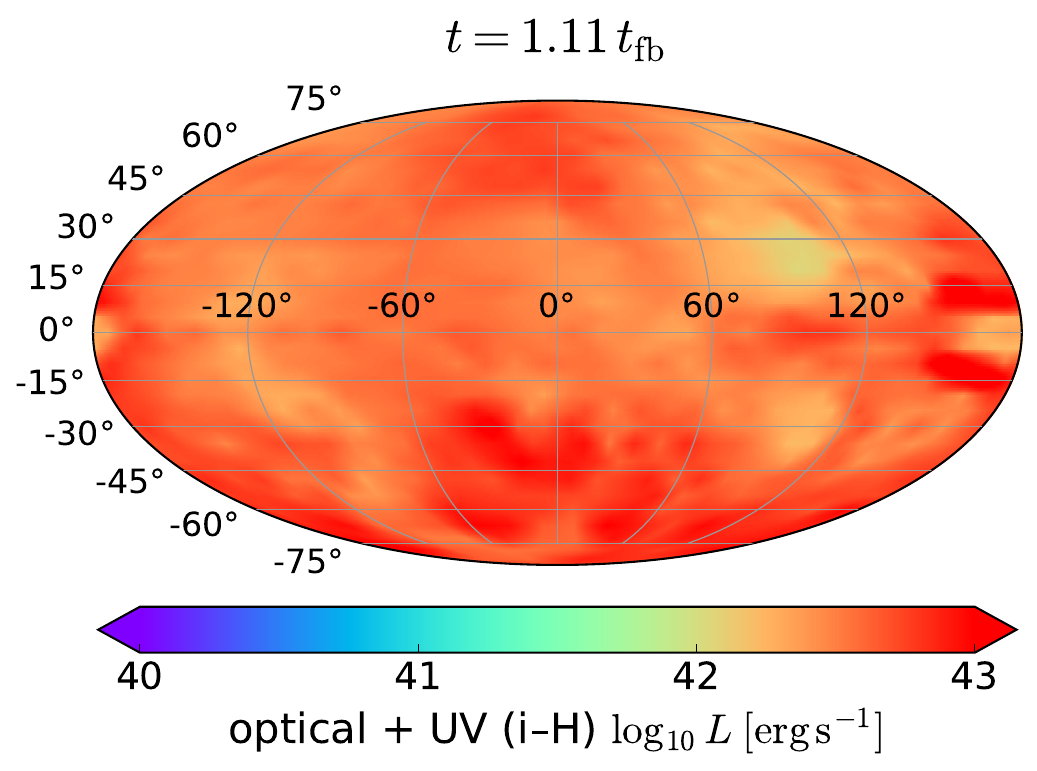}}
    \hfill
    \subfloat[X-ray, $1.11\,t_{\rm fb}$]{\includegraphics[width=0.46\textwidth]{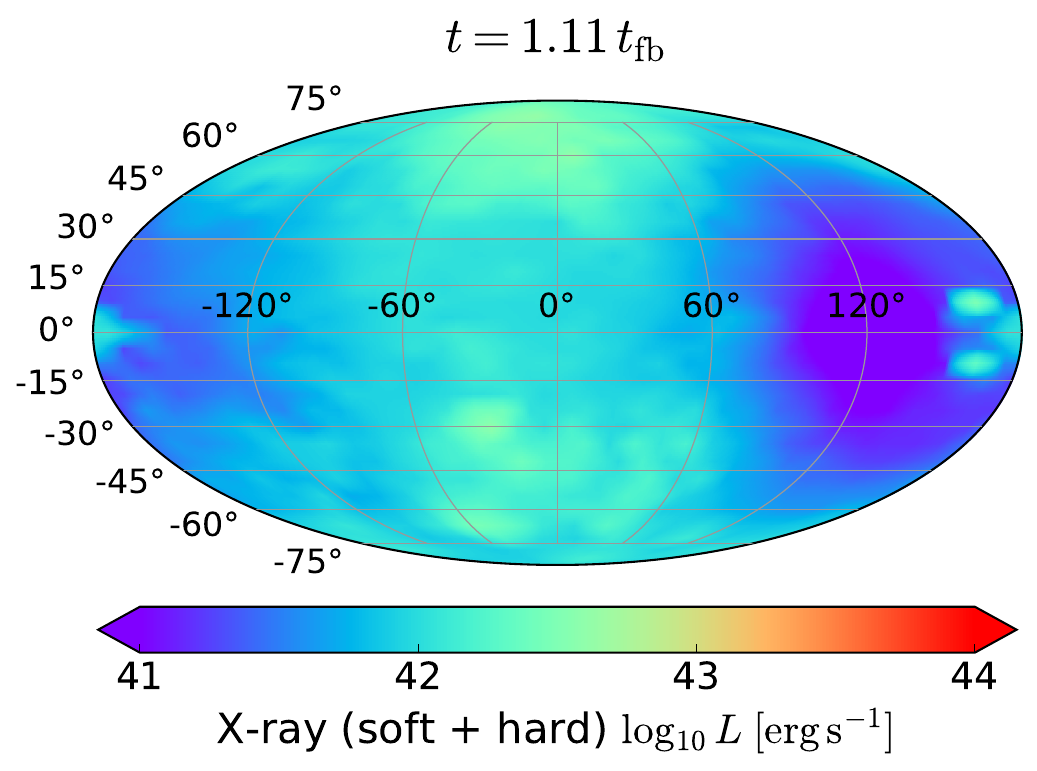}}
    \caption{Isotropic-equivalent luminosities $L_g^{\rm ph}$ of Eq.~\eqref{eq:luminosity_photosphere}, summed over the i to H groups (left) and over the soft and hard X-ray groups (right), as functions of the viewing direction in Mollweide projection at $0.66$, $0.90$, and $1.11\,t_{\rm fb}$ (top to bottom).}
    \label{fig:tde_mollweide}
\end{figure*}

Figure~\ref{fig:tde_xray} compares the rates at which Compton scattering and thermal emission add energy to the X-ray groups. Where Compton scattering dominates the coupling near the black hole (the nozzle shock region and the outflowing gas that leaves it during the flash, and the disk near $R_{\rm smooth}$ at $1.11\,t_{\rm fb}$), it adds X-ray energy tens to hundreds of times faster than thermal emission. A large ratio can also result from weak thermal emission in cool gas, so the maps show where Compton scattering is the local X-ray source, not how much X-ray energy it produces.

Figure~\ref{fig:tde_xray_power} shows that outside the photospheres, the instantaneous Compton X-ray power is a small fraction of the X-ray luminosity, except during the first half of the flash, when it is at times comparable to it. In the scattering layer it is mostly much larger than outside the photospheres. It is comparable to the X-ray luminosity at the beginning of the flash, exceeds it toward its end, at times by more than an order of magnitude, and is comparable to it after the flash. Measuring which fraction of the escaping X-rays it provides requires following the X-ray energy.

Figure~\ref{fig:tde_mollweide} shows the isotropic-equivalent optical--UV and X-ray luminosities of Eq.~\eqref{eq:luminosity_photosphere} as functions of the viewing direction at $0.66$, $0.90$, and $1.11\,t_{\rm fb}$. At $0.66\,t_{\rm fb}$ the X-rays are about as bright toward the poles as in the orbital plane, and within the plane they are faintest toward the returning stream, while the optical and UV emission is more than an order of magnitude brighter in the orbital plane than toward the poles. At $0.90$ and $1.11\,t_{\rm fb}$ the X-rays are brighter toward the poles and for $0.90\,t_{\rm fb}$ the emission from the nozzle shock region is also visible in the orbital plane, where the X-rays are bright toward the pericenter. As in \citet{Giron_2026}, the X-rays escape at higher latitudes than the optical and UV emission, consistent with X-rays escaping through the low-density polar regions \citep{Dai+18}.

\subsection{Limitations}
\label{sec:tde_limitations}

The results of this section are limited both by what the diagnostics can measure and by the simulation itself. We performed a single simulation, which includes Compton scattering, so the diagnostics show where and how strongly Compton scattering acts but not how much it changes the flow or the light curve. Measuring this requires an otherwise identical simulation without Compton scattering, which we leave to future work. The comparison of $P_{X,C}^{\rm out}$ and $P_{X,C}^{\rm sc}$ with the escaping X-ray luminosity shows that the instantaneous Compton X-ray power outside the X-ray thermalization radii is comparable to or larger than the X-ray luminosity during the flash, but not which fraction of the escaping X-rays it produces, since the diagnostics do not follow the X-ray energy from where it is produced to where it escapes. Both integrals also depend on where the photospheres and thermalization radii pass through the hot gas, and are accurate only to an order of magnitude.

The luminosities are estimated from the radiation at a single surface, the photosphere, along rays from the black hole. Before about $0.9\,t_{\rm fb}$ about half of the rays have no photosphere outside $r_p$, and on these rays the luminosity is evaluated at $r_p$, so the light curves and sky maps at these times depend on this choice.

The center region is excluded from the analysis and the gravitational potential is smoothed inside it, so the accretion flow inside $R_{\rm smooth}$, where the X-rays of TDEs are thought to originate, is not modeled. At late times the gas near the black hole in which Compton scattering dominates lies at the edge of this region, and its properties may depend on the smoothing. The flux-limited diffusion approximation is inaccurate in optically thin gas, which includes much of the distant, low-density gas in which Compton scattering dominates, and it tends to isotropize the radiation, so the sky maps likely underestimate the anisotropy of the emission. The ten energy groups resolve the X-ray band with only two groups and end at $20\,\mathrm{keV}$, which limits the spectral resolution of the Compton redistribution in hot gas. The electron density assumes fully ionized gas (Section~\ref{sec:tde_setup}), which overestimates it in cool gas. Finally, the importance of Compton scattering may differ for other black-hole masses, stellar masses, and penetration factors.

\section{Conclusions}
\label{sec:conclusions}

We have extended the multigroup FLD module of the moving-mesh RHD code \textsc{rich} \citep{Giron_2026} to include Compton scattering with the exact thermal Compton kernel, using group coefficients computed with \textsc{mace} \citep{compton_kernel_integration_paper}. We derived a semi-implicit scheme for the TRT equation, in which the temperature dependence of the Compton scattering cross section enters a generalized Fleck factor, and reduced it to the multigroup diffusion equation. The group Compton operator conserves the total photon number by construction and is corrected pairwise to satisfy detailed balance. In equilibration problems, the \textsc{rich} solutions agree with independent ODE solutions and published results and without absorption and emission the radiation relaxes to the Bose--Einstein spectrum that conserves the initial energy and photon number. In the configurations considered, the scheme remains stable for time steps that do not resolve the Compton equilibration timescale, at reduced accuracy in the temperature evolution.

In the three-dimensional TDE simulation, Compton scattering is locally important. Near the black hole, it dominates the coupling in the nozzle shock region and the outflowing gas that leaves it during the X-ray flash, and in the disk near $R_{\rm smooth}$ at later times. The gas near the nozzle shock is much hotter than the Compton temperature, so Compton scattering cools it, shifts the radiation to higher photon energies, and adds X-ray energy tens to hundreds of times faster than thermal emission. Outside the X-ray photospheres, the net Compton X-ray power is mostly a small fraction of the escaping X-ray luminosity. In the scattering layer between the X-ray thermalization radii and photospheres, however, the instantaneous rate at which Compton scattering adds X-ray energy is comparable to the escaping X-ray luminosity at the beginning of the flash and exceeds it toward its end. Compton scattering may therefore contribute substantially to the escaping X-rays during the flash, but determining this contribution requires a simulation without Compton scattering and following the X-ray energy from where it is produced.

The X-ray emission detected during the optical rise of AT~2022dsb and SRGe~J131014.7+444319 \citep{malyali_transient_2024,Zhang+25} is consistent in timing with the X-ray flash, but the observations do not determine its emission mechanism. Many more TDEs will be found, and a few thousand Rubin Observatory TDEs are expected to have pre-peak observations \citep{BricmanGomboc20}. Together with ULTRASAT discoveries \citep{Shvartzvald+24}, these observations will allow follow-up early enough to show how common such X-ray emission is. Interpreting these observations will require simulations that address the remaining limitations (Section~\ref{sec:tde_limitations}), with Monte Carlo transport, more X-ray groups, a resolved center region, and other black-hole masses, stellar masses, and penetration factors.

\section*{Acknowledgments}
The simulations were performed on the ICPL cluster at the Racah Institute of Physics.

During the preparation of this work the authors used ChatGPT \citep{openai_chatgpt} and Claude \citep{anthropic_claude} to assist with writing and refactoring the simulation and analysis code, writing code to generate the figures for the paper, and improving phrasing and wording. After using these tools, the authors reviewed and edited the content as needed and take full responsibility for the content of the published article.

\appendix
\section{Computing Group Planck Integrals}
\label{sec:computing_group_planck_integrals}

The normalized group Planck integral is evaluated in the implementation using the polylogarithmic and Taylor representations of \citet{CLARK1987311}. For group boundaries expressed through $x=\nu/(k_BT)$, define
\begin{equation}
b_g(T)=\Psi(x_{g+1/2},x_{g-1/2})
\label{eq:appendix_planck_integral}
\end{equation}
where
\begin{equation}
\Psi(x_2,x_1)=
\begin{cases}
\dfrac{15}{\pi^4}\left[\Pi_{\rm hi}(x_2)-\Pi_{\rm hi}(x_1)\right],
& x_1>2,\\[6pt]
\dfrac{15}{\pi^4}\left[\Pi_{\rm lo}(x_2)-\Pi_{\rm lo}(x_1)\right],
& x_2<2,\\[6pt]
1+\dfrac{15}{\pi^4}\left[\Pi_{\rm hi}(x_2)-\Pi_{\rm lo}(x_1)\right],
& x_1\leq2\leq x_2
\end{cases}
\label{eq:appendix_psi}
\end{equation}
The antiderivative is evaluated with a five-term exponential series at high $x$ and a Taylor expansion at low $x$,
\begin{equation}
\Pi_{\rm hi}(x)=
-x^3\sum_{l=1}^{5}\frac{e^{-l x}}{l}
-3x^2\sum_{l=1}^{5}\frac{e^{-l x}}{l^2}
\\
-6x\sum_{l=1}^{5}\frac{e^{-l x}}{l^3}
-6\sum_{l=1}^{5}\frac{e^{-l x}}{l^4}
\label{eq:appendix_pi}
\end{equation}
\begin{equation}
\Pi_{\rm lo}(x)=
\frac{x^3}{3}-\frac{x^4}{8}+\frac{x^5}{60}
-\frac{x^7}{5040}+\frac{x^9}{272160}
\\
-\frac{x^{11}}{13305600}+\frac{x^{13}}{622702080}
\label{eq:appendix_pi_lo}
\end{equation}
The additive constant in Eq.~\eqref{eq:appendix_psi} ensures continuity when a group straddles the switching point $x=2$, where the two truncated expansions agree to within $2\times10^{-6}$ of the normalized integral. The same group Planck integrals are used to initialize the equilibration radiation field, construct the equilibrium detailed-balance factors, and evaluate the group absorption and emission terms.

\section{Discrete Multigroup Equation}
\label{sec:discrete_multigroup_equation}

For completeness, we give the full form of Eq.~\eqref{eq:cell_equation} for destination group $g$ in cell $i$, including the non-Compton terms,
\begin{multline}
V_i\left(1+c\Delta t\,\kappa_{P,g,i}\right)E_{g,i}^{n+1}
-\frac{\Delta t}{2}(1-R_{2,g})\,\alpha_i E_{g,i}^{n+1}
-\Delta t\sum_{j\in\mathcal{N}(i)}
D_{g,ij}
\frac{
\vec{A}_{ij}\cdot
(\overrightarrow{CM}_j-\overrightarrow{CM}_i)
}{
\Vert\overrightarrow{CM}_j-\overrightarrow{CM}_i\Vert^2
}
\left(E_{g,j}^{n+1}-E_{g,i}^{n+1}\right)
+\Delta t\,\mathcal{D}_{g,i}
\left[\vec{E}_i^{n+1}\right]
\\
-V_ic\Delta t\sum_{g'}S_{g'g,i}^nE_{g',i}^{n+1}
-V_i\mathcal{A}_{g,i}
\left[
\sum_{g'}\kappa_{P,g',i}E_{g',i}^{n+1}
-\sum_{g'g''}S_{g'g'',i}^nE_{g',i}^{n+1}
\right]
=V_iE_{g,i}^n
+V_iU_{r,i}^n
\left(c\Delta t\,\kappa_{P,g,i}b_{g,i}-\kappa_{P,i}\mathcal{A}_{g,i}\right)
\label{eq:full_system}
\end{multline}
Here $\mathcal{N}(i)$ is the set of cells that share a face with
cell $i$. The vector $\vec{A}_{ij}$ points along the outward normal from
cell $i$ through the face shared with cell $j$ and has magnitude equal to
the face area. The center of mass of cell $i$ is denoted by
$\overrightarrow{CM}_i$, and the face-averaged gas velocity is
$\vec{v}_{ij}=(\vec{v}_i+\vec{v}_j)/2$. Define
\begin{equation}
\alpha_i=
-\sum_{j\in\mathcal{N}(i)}
\vec{A}_{ij}\cdot\vec{v}_{ij}
\label{eq:alpha_i}
\end{equation}
such that $\alpha_i\geq0$ corresponds to compression and $\alpha_i<0$
to expansion. The face diffusion coefficient $D_{g,ij}$ and the Doppler operator $\mathcal{D}_{g,i}$ are defined below. For the derivation and discretization of the non-Compton terms in
Eq.~\eqref{eq:full_system}, we refer the reader to
\citet{Giron_2026}.

The face diffusion coefficient $D_{g,ij}$ is evaluated using
face-centered quantities. We define
\begin{equation}
T_{ij}=
\left(\frac{T_i^4+T_j^4}{2}\right)^{1/4},
\quad
E_{g,ij}=\frac{E_{g,i}+E_{g,j}}{2},
\quad
\kappa_{R,g,ij}
=
\frac{\kappa_{R,g,i}+\kappa_{R,g,j}}{2}
\label{eq:face_quantities}
\end{equation}
The cell opacities entering $\kappa_{R,g,ij}$ are evaluated at the face
temperature $T_{ij}$ using the density of their respective cells. The
face-centered radiation-energy gradient is
\begin{equation}
\nabla E_{g,ij}
=
\frac{E_{g,j}-E_{g,i}}
{\Vert\overrightarrow{CM}_j-\overrightarrow{CM}_i\Vert^2}
\left(
\overrightarrow{CM}_j-\overrightarrow{CM}_i
\right)
\label{eq:face_energy_gradient}
\end{equation}
The corresponding dimensionless gradient, flux limiter, and diffusion
coefficient are
\begin{equation}
R_{g,ij}
=
\frac{|\nabla E_{g,ij}|}
{\kappa_{R,g,ij}E_{g,ij}},
\quad
\lambda_{g,ij}
=
\frac{1}{R_{g,ij}}
\left(
\coth R_{g,ij}-\frac{1}{R_{g,ij}}
\right),
\quad
D_{g,ij}
=
\frac{c\lambda_{g,ij}}{\kappa_{R,g,ij}}
\label{eq:face_diffusion_coefficient}
\end{equation}

The discrete Doppler operator
$\mathcal{D}_{g,i}[\vec{E}_i]$, where $\vec{E}_i$ denotes the vector of
radiation group energies in cell $i$, describes redistribution between
neighboring photon-energy groups. We define the spacing between adjacent
group centers as
\begin{equation}
\Delta\nu_g=\nu_{g+1}-\nu_g
\label{eq:group_center_spacing}
\end{equation}
The left and right slopes of the spectral radiation energy density in
group $g$ are
\begin{equation}
m_{L,g}
=
\frac{
E_g/\Delta_g-E_{g-1}/\Delta_{g-1}
}{
\Delta\nu_{g-1}
},
\quad
m_{R,g}
=
\frac{
E_{g+1}/\Delta_{g+1}-E_g/\Delta_g
}{
\Delta\nu_g
}
\label{eq:doppler_slopes}
\end{equation}
with slope ratio $r_g=m_{L,g}/m_{R,g}$. We use the Superbee slope limiter \citep{Roe1986},
\begin{equation}
\varphi(r)
=
\max\left[
0,
\min(2r,1),
\min(r,2)
\right]
\label{eq:superbee}
\end{equation}

For $\alpha_i\geq0$, corresponding to compression and a blueshift, the
upwind direction is from lower photon-energy groups and
\begin{multline}
\mathcal{D}_{g,i}[\vec{E}_i]=
\alpha_i\nu_{g-1/2}\frac{1-R_{2,g}}{2}
\left(
-\frac{1}{\Delta_{g-1}}
+\frac{\varphi(r_{g-1})}{2\Delta\nu_{g-1}}
\right)E_{g-1,i}
\\
+\Bigg[
\alpha_i\nu_{g+1/2}\frac{1-R_{2,g+1}}{2}
\left(
\frac{1}{\Delta_g}
-\frac{\varphi(r_g)}{2\Delta\nu_g}
\right)
-\alpha_i\nu_{g-1/2}\frac{1-R_{2,g}}{2}
\frac{\Delta_{g-1}\varphi(r_{g-1})}
{2\Delta\nu_{g-1}\Delta_g}
\Bigg]E_{g,i}
\\
+\alpha_i\nu_{g+1/2}\frac{1-R_{2,g+1}}{2}
\frac{\Delta_g\varphi(r_g)}
{2\Delta\nu_g\Delta_{g+1}}
E_{g+1,i}
\label{eq:doppler_compression}
\end{multline}

For $\alpha_i<0$, corresponding to expansion and a redshift, the upwind
direction reverses to higher photon-energy groups,
\begin{multline}
\mathcal{D}_{g,i}[\vec{E}_i]=
\alpha_i\nu_{g-1/2}\frac{1-R_{2,g}}{2}
\left(
-\frac{1}{\Delta_g}
-\frac{\varphi(r_g)}{2\Delta\nu_g}
\right)E_{g,i}
\\
+\Bigg[
\alpha_i\nu_{g+1/2}\frac{1-R_{2,g+1}}{2}
\left(
\frac{1}{\Delta_{g+1}}
+\frac{\varphi(r_{g+1})}{2\Delta\nu_{g+1}}
\right)
+\alpha_i\nu_{g-1/2}\frac{1-R_{2,g}}{2}
\frac{\Delta_g\varphi(r_g)}
{2\Delta\nu_g\Delta_{g+1}}
\Bigg]E_{g+1,i}
\\
-\alpha_i\nu_{g+1/2}\frac{1-R_{2,g+1}}{2}
\frac{\Delta_{g+1}\varphi(r_{g+1})}
{2\Delta\nu_{g+1}\Delta_{g+2}}
E_{g+2,i}
\label{eq:doppler_expansion}
\end{multline}

\bibliography{bib}{}

@article{Giron_2026,
    doi = {10.3847/1538-4365/ae4874},
    url = {https://doi.org/10.3847/1538-4365/ae4874},
    year = {2026},
    month = {mar},
    publisher = {The American Astronomical Society},
    volume = {283},
    number = {2},
    pages = {57},
    author = {Giron, Itamar and Krief, Menahem and Stone, Nicholas C. and Steinberg, Elad},
    title = {Multigroup Radiation Diffusion on a Moving Mesh: Implementation in RICH and Application to Tidal Disruption Events},
    journal = {\apjs}
}

@ARTICLE{compton_kernel_integration_paper,
       author = {Giron, Itamar and Krief, Menahem and Steinberg, Elad},
        title = "{MACE: Multigroup-Multiangle Integrated Thermal Compton Scattering Kernel}",
      journal = {arXiv e-prints},
         year = 2026,
        month = oct,
          eid = {arXiv:2610.25117},
        pages = {arXiv:2610.25117},
          doi = {10.48550/arXiv.2610.25117},
archivePrefix = {arXiv},
       eprint = {2610.25117},
 primaryClass = {astro-ph.IM}
}

@conference{Keady2017,
  author = {Keady, K. P. and Haut, T. S. and Cleveland, M. A.},
  booktitle = {International Conference on Mathematics \& Computational Methods Applied to Nuclear Science \& Engineering},
  title = {A Multigroup Interpolated Compton Scattering Treatment for the Implicit Monte Carlo Method},
  year = {2017},
  address = {Jeju, Korea},
}

@article{MCGRAW2023111980,
    title = {A new operator-split Compton scattering method},
    journal = {JCoPh},
    volume = {478},
    pages = {111980},
    year = {2023},
    issn = {0021-9991},
    doi = {10.1016/j.jcp.2023.111980},
    url = {https://www.sciencedirect.com/science/article/pii/S002199912300075X},
    author = {Carolyn McGraw and Andrew Till and James Warsa}
}

@article{TILL2020108990,
    title = {Formulations and analysis of Compton scattering for deterministic thermal radiation transport},
    journal = {JCoPh},
    volume = {400},
    pages = {108990},
    year = {2020},
    issn = {0021-9991},
    doi = {10.1016/j.jcp.2019.108990},
    url = {https://www.sciencedirect.com/science/article/pii/S0021999119306953},
    author = {Andrew T. Till and Kendra P. Long and James S. Warsa and Jim E. Morel}
}

@article{KERSHAW1986273,
    title = {A simple and fast method for computing the relativistic Compton scattering kernel for radiative transfer},
    journal = {\jqsrt},
    volume = {36},
    number = {4},
    pages = {273-282},
    year = {1986},
    issn = {0022-4073},
    doi = {10.1016/0022-4073(86)90050-6},
    url = {https://www.sciencedirect.com/science/article/pii/0022407386900506},
    author = {Kershaw, David S. and Prasad, Manoj K. and Beason, J. Douglas}
}

@book{pomraning1973equations,
  title={The Equations of Radiation Hydrodynamics},
  author={Pomraning, G. C.},
  isbn={9780080168937},
  lccn={72013242},
  year={1973},
  publisher={Pergamon Press},
  address={Oxford}
}

@article{POMRANING19721047,
    title = {Compton and inverse Compton scattering},
    journal = {\jqsrt},
    volume = {12},
    number = {6},
    pages = {1047-1061},
    year = {1972},
    issn = {0022-4073},
    doi = {10.1016/0022-4073(72)90023-4},
    url = {https://www.sciencedirect.com/science/article/pii/0022407372900234},
    author = {Pomraning, G. C.}
}

@article{WINSLOW1995262,
    title = {Multifrequency-Gray Method for Radiation Diffusion with Compton Scattering},
    journal = {JCoPh},
    volume = {117},
    number = {2},
    pages = {262-273},
    year = {1995},
    issn = {0021-9991},
    doi = {10.1006/jcph.1995.1064},
    url = {https://www.sciencedirect.com/science/article/pii/S0021999185710649},
    author = {Alan M. Winslow}
}

@book{mihalas1999foundations,
  title={Foundations of Radiation Hydrodynamics},
  author={Mihalas, D. and Weibel-Mihalas, B.},
  isbn={9780486409252},
  lccn={99042987},
  series={Dover Books on Physics},
  edition={{Dover}},
  year={1999},
  publisher={Dover Publications}
}

@inproceedings{mcgraw_till_warsa_2021,
  title     = {{BICS}: Full {Boltzmann} Implicit {Compton} Solver},
  author    = {McGraw, Carolyn and Till, Andrew and Warsa, James},
  booktitle = {Proceedings of the International Conference on Mathematics and Computational Methods Applied to Nuclear Science and Engineering (M\&C 2021)},
  address   = {Virtual Meeting},
  month     = oct,
  year      = {2021},
  publisher = {American Nuclear Society},
  pages     = {1294--1303},
  doi       = {10.13182/xyz-33718}
}

@article{Krumholz_2024,
    author = {He, Chong-Chong and Wibking, Benjamin D and Krumholz, Mark R},
    title = {A novel numerical method for mixed-frame multigroup radiation-hydrodynamics with GPU acceleration implemented in the quokka code},
    journal = {\mnras},
    volume = {535},
    number = {4},
    pages = {3059-3076},
    year = {2024},
    month = {12},
    issn = {0035-8711},
    doi = {10.1093/mnras/stae2580},
    url = {https://doi.org/10.1093/mnras/stae2580},
    archivePrefix = {arXiv},
    eprint = {2407.18304},
    primaryClass = {astro-ph.GA},
}

@inproceedings{densmore_2007,
    author={Densmore, Jeffery D.
    and Warsa, James S.
    and Morel, Jim E.},
    title={A {Fokker-Planck} Method for Modeling {Compton} Scattering in Implicit {Monte Carlo} Simulations},
    booktitle={Joint International Topical Meeting on Mathematics and Computations and Supercomputing in Nuclear Applications (M\&C + SNA 2007)},
    address={Monterey, CA},
    publisher={American Nuclear Society},
    year={2007},
    month={4}
}

@article{Pomeraning-Levermore,
       author = {{Levermore}, C.~D. and {Pomraning}, G.~C.},
        title = "{A flux-limited diffusion theory}",
      journal = {\apj},
         year = 1981,
        month = aug,
       volume = {248},
        pages = {321-334},
          doi = {10.1086/159157},
       adsurl = {https://ui.adsabs.harvard.edu/abs/1981ApJ...248..321L}
}

@article{fleck_implicit_1971,
        title = {An implicit {Monte} {Carlo} scheme for calculating time and frequency dependent nonlinear radiation transport},
        volume = {8},
        issn = {0021-9991},
        url = {https://www.sciencedirect.com/science/article/pii/0021999171900155},
        doi = {10.1016/0021-9991(71)90015-5},
        number = {3},
        journal = {JCoPh},
        author = {Fleck, J.~{\relax A., Jr} and Cummings, J.~{\relax D., Jr}},
        year = {1971},
        pages = {313--342},
}

@article{BiCGSTAB,
author = {van der Vorst, H. A.},
title = {Bi-CGSTAB: A Fast and Smoothly Converging Variant of Bi-CG for the Solution of Nonsymmetric Linear Systems},
journal = {SJSC},
volume = {13},
number = {2},
pages = {631-644},
year = {1992},
doi = {10.1137/0913035},
URL = {https://doi.org/10.1137/0913035}
}

@article{Peer2005,
    doi = {10.1086/431139},
    url = {https://doi.org/10.1086/431139},
    year = {2005},
    month = {aug},
    publisher = {},
    volume = {628},
    number = {2},
    pages = {857--866},
    author = {Pe'er, Asaf and Waxman, Eli},
    title = {Time-dependent Numerical Model for the Emission of Radiation from Relativistic Plasma},
    journal = {\apj}
}

@ARTICLE{Canfield1987,
       author = {{Canfield}, E. and {Howard}, W.~M. and {Liang}, E.~P.},
        title = "{Inverse Comptonization by one-dimensional relativistic electrons}",
      journal = {\apj},
         year = 1987,
        month = dec,
       volume = {323},
        pages = {565-574},
          doi = {10.1086/165853},
       adsurl = {https://ui.adsabs.harvard.edu/abs/1987ApJ...323..565C}
}

@article{CLARK1987311,
  title = {Computing multigroup radiation integrals using polylogarithm-based methods},
  journal = {JCoPh},
  volume = {70},
  number = {2},
  pages = {311-329},
  year = {1987},
  issn = {0021-9991},
  doi = {10.1016/0021-9991(87)90185-9},
  url = {https://www.sciencedirect.com/science/article/pii/0021999187901859},
  author = {Bradley A Clark}
}

@article{Compton1923,
  title = {A Quantum Theory of the Scattering of X-rays by Light Elements},
  author = {Compton, Arthur H.},
  journal = {Physical Review},
  volume = {21},
  issue = {5},
  pages = {483--502},
  year = {1923},
  month = {May},
  publisher = {American Physical Society},
  doi = {10.1103/PhysRev.21.483},
  url = {https://link.aps.org/doi/10.1103/PhysRev.21.483}
}

@article{Juttner1911,
  author = {Jüttner, Ferencz},
  title = {Das Maxwellsche Gesetz der Geschwindigkeitsverteilung in der Relativtheorie},
  journal = {Annalen der Physik},
  volume = {339},
  number = {5},
  pages = {856-882},
  doi = {10.1002/andp.19113390503},
  url = {https://onlinelibrary.wiley.com/doi/abs/10.1002/andp.19113390503},
  year = {1911}
}

@article{Klein1929,
  author={Klein, O.
  and Nishina, Y.},
  title={\"Uber die Streuung von Strahlung durch freie Elektronen nach der neuen relativistischen Quantendynamik von Dirac},
  journal={Zeitschrift f\"ur Physik},
  year={1929},
  month={Nov},
  day={01},
  volume={52},
  number={11-12},
  pages={853-868},
  issn={0044-3328},
  doi={10.1007/BF01366453},
  url={https://doi.org/10.1007/BF01366453}
}

@article{Madej2017,
    author = {Madej, J. and Różańska, A. and Majczyna, A. and Należyty, M.},
    title = {On the Compton scattering redistribution function in plasma},
    journal = {\mnras},
    volume = {469},
    number = {2},
    pages = {2032-2041},
    year = {2017},
    month = {08},
    issn = {0035-8711},
    doi = {10.1093/mnras/stx994},
    url = {https://doi.org/10.1093/mnras/stx994},
    archivePrefix = {arXiv},
    eprint = {1602.05088},
}

@article{Pozdnyakov1983,
       author = {{Pozdnyakov}, L.~A. and {Sobol}, I.~M. and {Syunyaev}, R.~A.},
        title = "{Comptonization and the shaping of X-ray source spectra - Monte Carlo calculations}",
      journal = {Astrophysics and Space Physics Reviews},
         year = 1983,
        month = jan,
       volume = {2},
        pages = {189-331},
       adsurl = {https://ui.adsabs.harvard.edu/abs/1983ASPRv...2..189P}
}

@article{Stone2020,
	doi = {10.3847/1538-4365/ab929b},
	url = {https://doi.org/10.3847%2F1538-4365%2Fab929b},
	year = 2020,
	month = jun,
	publisher = {American Astronomical Society},
	volume = {249},
	number = {1},
	pages = {4},
	author = {James M. Stone and Kengo Tomida and Christopher J. White and Kyle G. Felker},
	title = {The {Athena++} Adaptive Mesh Refinement Framework: Design and Magnetohydrodynamic Solvers},
	journal = {\apjs},
}

@article{Mummery2021,
    author = {Mummery, Andrew and Balbus, Steven A},
    title = {Hard X-ray emission from a Compton scattering corona in large black hole mass tidal disruption events},
    journal = {\mnras},
    volume = {504},
    number = {4},
    pages = {4730-4742},
    year = {2021},
    month = {07},
    issn = {0035-8711},
    doi = {10.1093/mnras/stab1184},
    url = {https://doi.org/10.1093/mnras/stab1184},
    archivePrefix = {arXiv},
    eprint = {2104.06195},
}

@article{BIRKINSHAW199997,
title = {The Sunyaev–Zel'dovich effect},
journal = {\physrep},
volume = {310},
number = {2-3},
pages = {97-195},
year = {1999},
issn = {0370-1573},
doi = {10.1016/S0370-1573(98)00080-5},
url = {https://www.sciencedirect.com/science/article/pii/S0370157398000805},
author = {Mark Birkinshaw}
}

@article{Zhang_2013,
  doi = {10.1088/0067-0049/204/1/7},
  url = {https://doi.org/10.1088/0067-0049/204/1/7},
  year = {2013},
  month = {jan},
  publisher = {The American Astronomical Society},
  volume = {204},
  number = {1},
  pages = {7},
  author = {Zhang, W. and Howell, L. and Almgren, A. and Burrows, A. and Dolence, J. and Bell, J.},
  title = {CASTRO: A NEW COMPRESSIBLE ASTROPHYSICAL SOLVER. III. MULTIGROUP RADIATION HYDRODYNAMICS},
  journal = {\apjs}
}

@article{Wollaeger_2013,
  doi = {10.1088/0067-0049/209/2/36},
  url = {https://doi.org/10.1088/0067-0049/209/2/36},
  year = {2013},
  month = {nov},
  publisher = {The American Astronomical Society},
  volume = {209},
  number = {2},
  pages = {36},
  author = {Wollaeger, Ryan T. and van Rossum, Daniel R. and Graziani, Carlo and Couch, Sean M. and Jordan, IV, George C. and Lamb, Donald Q. and Moses, Gregory A.},
  title = {RADIATION TRANSPORT FOR EXPLOSIVE OUTFLOWS: A MULTIGROUP HYBRID MONTE CARLO METHOD},
  journal = {\apjs}
}

@ARTICLE{Dai+18,
       author = {{Dai}, Lixin and {McKinney}, Jonathan C. and {Roth}, Nathaniel and {Ramirez-Ruiz}, Enrico and {Miller}, M. Coleman},
        title = "{A Unified Model for Tidal Disruption Events}",
      journal = {\apjl},
         year = 2018,
        month = jun,
       volume = {859},
       number = {2},
          eid = {L20},
        pages = {L20},
          doi = {10.3847/2041-8213/aab429},
archivePrefix = {arXiv},
       eprint = {1803.03265},
 primaryClass = {astro-ph.HE},
       adsurl = {https://ui.adsabs.harvard.edu/abs/2018ApJ...859L..20D}
}

@ARTICLE{LuKumar18,
       author = {{Lu}, Wenbin and {Kumar}, Pawan},
        title = "{On the Missing Energy Puzzle of Tidal Disruption Events}",
      journal = {\apj},
         year = 2018,
        month = oct,
       volume = {865},
       number = {2},
          eid = {128},
        pages = {128},
          doi = {10.3847/1538-4357/aad54a},
archivePrefix = {arXiv},
       eprint = {1802.02151},
 primaryClass = {astro-ph.HE},
       adsurl = {https://ui.adsabs.harvard.edu/abs/2018ApJ...865..128L}
}

@ARTICLE{Andalman+22,
       author = {{Andalman}, Zachary L. and {Liska}, Matthew T.~P. and {Tchekhovskoy}, Alexander and {Coughlin}, Eric R. and {Stone}, Nicholas},
        title = "{Tidal disruption discs formed and fed by stream-stream and stream-disc interactions in global GRHD simulations}",
      journal = {\mnras},
         year = 2022,
        month = feb,
       volume = {510},
       number = {2},
        pages = {1627-1648},
          doi = {10.1093/mnras/stab3444},
archivePrefix = {arXiv},
       eprint = {2008.04922},
 primaryClass = {astro-ph.HE},
       adsurl = {https://ui.adsabs.harvard.edu/abs/2022MNRAS.510.1627A}
}

@ARTICLE{Piran+15,
       author = {{Piran}, Tsvi and {Svirski}, Gilad and {Krolik}, Julian and {Cheng}, Roseanne M. and {Shiokawa}, Hotaka},
        title = "{Disk Formation Versus Disk Accretion{\textemdash}What Powers Tidal Disruption Events?}",
      journal = {\apj},
         year = 2015,
        month = jun,
       volume = {806},
       number = {2},
          eid = {164},
        pages = {164},
          doi = {10.1088/0004-637X/806/2/164},
archivePrefix = {arXiv},
       eprint = {1502.05792},
 primaryClass = {astro-ph.HE},
       adsurl = {https://ui.adsabs.harvard.edu/abs/2015ApJ...806..164P}
}

@ARTICLE{Shiokawa+15,
       author = {{Shiokawa}, Hotaka and {Krolik}, Julian H. and {Cheng}, Roseanne M. and {Piran}, Tsvi and {Noble}, Scott C.},
        title = "{General Relativistic Hydrodynamic Simulation of Accretion Flow from a Stellar Tidal Disruption}",
      journal = {\apj},
         year = 2015,
        month = may,
       volume = {804},
       number = {2},
          eid = {85},
        pages = {85},
          doi = {10.1088/0004-637X/804/2/85},
archivePrefix = {arXiv},
       eprint = {1501.04365},
 primaryClass = {astro-ph.HE},
       adsurl = {https://ui.adsabs.harvard.edu/abs/2015ApJ...804...85S}
}

@ARTICLE{BonnerotStone21,
       author = {{Bonnerot}, C. and {Stone}, N.~C.},
        title = "{Formation of an Accretion Flow}",
      journal = {\ssr},
         year = 2021,
        month = feb,
       volume = {217},
       number = {1},
          eid = {16},
        pages = {16},
          doi = {10.1007/s11214-020-00789-1},
archivePrefix = {arXiv},
       eprint = {2008.11731},
 primaryClass = {astro-ph.HE},
       adsurl = {https://ui.adsabs.harvard.edu/abs/2021SSRv..217...16B}
}

@ARTICLE{Price+24,
       author = {{Price}, Daniel J. and {Liptai}, David and {Mandel}, Ilya and {Shepherd}, Joanna and {Lodato}, Giuseppe and {Levin}, Yuri},
        title = "{Eddington Envelopes: The Fate of Stars on Parabolic Orbits Tidally Disrupted by Supermassive Black Holes}",
      journal = {\apjl},
         year = 2024,
        month = aug,
       volume = {971},
       number = {2},
          eid = {L46},
        pages = {L46},
          doi = {10.3847/2041-8213/ad6862},
archivePrefix = {arXiv},
       eprint = {2404.09381},
 primaryClass = {astro-ph.HE}
}

@ARTICLE{Chan+26,
       author = {{Chan}, Ho-Sang and {Ryu}, Taeho and {Krolik}, Julian and {Piran}, Tsvi},
        title = "{Unexpectedly Weak General Relativistic Effects in Strongly Relativistic Tidal Disruption Events}",
      journal = {\apj},
         year = 2026,
        month = jun,
       volume = {1003},
       number = {2},
          eid = {221},
        pages = {221},
          doi = {10.3847/1538-4357/ae67fa},
archivePrefix = {arXiv},
       eprint = {2603.10208},
 primaryClass = {astro-ph.HE}
}

@ARTICLE{vanVelzen+20,
       author = {{van Velzen}, Sjoert and {Holoien}, Thomas W.-S. and {Onori}, Francesca and {Hung}, Tiara and {Arcavi}, Iair},
        title = "{Optical-Ultraviolet Tidal Disruption Events}",
      journal = {\ssr},
         year = 2020,
        month = oct,
       volume = {216},
       number = {8},
          eid = {124},
        pages = {124},
          doi = {10.1007/s11214-020-00753-z},
archivePrefix = {arXiv},
       eprint = {2008.05461},
 primaryClass = {astro-ph.HE},
       adsurl = {https://ui.adsabs.harvard.edu/abs/2020SSRv..216..124V}
}

@ARTICLE{BricmanGomboc20,
       author = {{Bricman}, Katja and {Gomboc}, Andreja},
        title = "{The Prospects of Observing Tidal Disruption Events with the Large Synoptic Survey Telescope}",
      journal = {\apj},
         year = 2020,
        month = feb,
       volume = {890},
       number = {1},
          eid = {73},
        pages = {73},
          doi = {10.3847/1538-4357/ab6989},
archivePrefix = {arXiv},
       eprint = {1906.08235},
 primaryClass = {astro-ph.HE},
       adsurl = {https://ui.adsabs.harvard.edu/abs/2020ApJ...890...73B}
}

@ARTICLE{Shvartzvald+24,
       author = {{Shvartzvald}, Y. and {Waxman}, E. and {Gal-Yam}, A. and {Ofek}, E.~O. and {Ben-Ami}, S. and {Berge}, D. and {Kowalski}, M. and {B{\"u}hler}, R. and {Worm}, S. and {Rhoads}, J.~E. and {Arcavi}, I. and {Maoz}, D. and {Polishook}, D. and {Stone}, N. and {Trakhtenbrot}, B. and {Ackermann}, M. and {Aharonson}, O. and {Birnholtz}, O. and {Chelouche}, D. and {Guetta}, D. and {Hallakoun}, N. and {Horesh}, A. and {Kushnir}, D. and {Mazeh}, T. and {Nordin}, J. and {Ofir}, A. and {Ohm}, S. and {Parsons}, D. and {Pe'er}, A. and {Perets}, H.~B. and {Perdelwitz}, V. and {Poznanski}, D. and {Sadeh}, I. and {Sagiv}, I. and {Shahaf}, S. and {Soumagnac}, M. and {Tal-Or}, L. and {van Santen}, J. and {Zackay}, B. and {Guttman}, O. and {Rekhi}, P. and {Townsend}, A. and {Weinstein}, A. and {Wold}, I.},
        title = "{ULTRASAT: A Wide-field Time-domain UV Space Telescope}",
      journal = {\apj},
         year = 2024,
        month = mar,
       volume = {964},
       number = {1},
          eid = {74},
        pages = {74},
          doi = {10.3847/1538-4357/ad2704},
archivePrefix = {arXiv},
       eprint = {2304.14482},
 primaryClass = {astro-ph.IM},
       adsurl = {https://ui.adsabs.harvard.edu/abs/2024ApJ...964...74S}
}

@ARTICLE{Zhang+25,
       author = {{Zhang}, Zirui and {Yao}, Yuhan and {Gilfanov}, Marat and {Sazonov}, Sergey and {Medvedev}, Pavel and {Khorunzhev}, Georgii and {Sunyaev}, Rashid and {Ravi}, Vikram and {Kulkarni}, S.~R. and {Somalwar}, Jean and {Chornock}, Ryan and {Bikmaev}, Ilfan and {Gorbachev}, Mark A.},
        title = "{eROSITA-RU Tidal Disruption Events with Keck-I/LRIS: Sample Selection, Optical Properties, and Host Galaxy Demographics}",
      journal = {\aap},
         year = 2026,
        month = apr,
       volume = {708},
        pages = {A374},
          doi = {10.1051/0004-6361/202558567},
archivePrefix = {arXiv},
       eprint = {2512.12480},
 primaryClass = {astro-ph.HE},
       adsurl = {https://ui.adsabs.harvard.edu/abs/2026A&A...708A.374Z}
}

@article{BonnerotLu22,
    author = {Bonnerot, Cl{\'e}ment and Lu, Wenbin},
    title = {The nozzle shock in tidal disruption events},
    journal = {\mnras},
    year = {2022},
    volume = {511},
    number = {2},
    pages = {2147--2169},
    doi = {10.1093/mnras/stac146},
    eprint = {2106.01376},
    archivePrefix = {arXiv}
}

@ARTICLE{Guillochon+14,
       author = {{Guillochon}, James and {Manukian}, Haik and {Ramirez-Ruiz}, Enrico},
        title = "{PS1-10jh: The Disruption of a Main-sequence Star of Near-solar Composition}",
      journal = {\apj},
         year = 2014,
        month = mar,
       volume = {783},
       number = {1},
          eid = {23},
        pages = {23},
          doi = {10.1088/0004-637X/783/1/23},
archivePrefix = {arXiv},
       eprint = {1304.6397},
 primaryClass = {astro-ph.HE},
       adsurl = {https://ui.adsabs.harvard.edu/abs/2014ApJ...783...23G}
}

@ARTICLE{Andalman+25,
       author = {{Andalman}, Zachary L. and {Quataert}, Eliot and {Coughlin}, Eric R. and {Nixon}, C.~J.},
        title = "{Resolving the (Debate About) Nozzle Shocks in Tidal Disruption Events}",
      journal = {OJAp},
         year = 2026,
        month = jun,
       volume = {9},
          eid = {162785},
        pages = {162785},
          doi = {10.33232/001c.162785},
archivePrefix = {arXiv},
       eprint = {2512.08928},
 primaryClass = {astro-ph.HE}
}

@ARTICLE{Huang+25b,
       author = {{Huang}, Xiaoshan and {Meza}, Maria Renee and {Yun}, Sol Bin and {Mockler}, Brenna and {Davis}, Shane W. and {Jiang}, Yan-fei},
        title = "{X-ray Variability and Photosphere Evolution during Accretion Disk Formation in Tidal Disruption Events}",
      journal = {\apj},
         year = 2026,
        month = sep,
       volume = {1009},
       number = {1},
          eid = {64},
        pages = {64},
          doi = {10.3847/1538-4357/ae953e},
archivePrefix = {arXiv},
       eprint = {2512.12985},
 primaryClass = {astro-ph.HE}
}

@ARTICLE{Martire+25,
       author = {{Martire}, Paola and {Rossi}, Elena Maria and {Stone}, Nicholas C. and {Steinberg}, Elad and {Kilmetis}, Konstantinos and {Linial}, Itai},
        title = "{Wind-mediated near-Eddington emission in a $10^4\,M_\odot$ black hole tidal disruption event}",
      journal = {\mnras},
         year = 2026,
        month = jun,
       volume = {549},
       number = {3},
          eid = {stag1021},
        pages = {stag1021},
          doi = {10.1093/mnras/stag1021},
archivePrefix = {arXiv},
       eprint = {2512.10564},
 primaryClass = {astro-ph.HE}
}

@ARTICLE{Martire+26,
       author = {{Martire}, Paola and {Rossi}, Elena Maria and {Stone}, Nicholas C. and {Steinberg}, Elad and {Giron}, Itamar},
        title = "{Anisotropic wind in tidal disruption events}",
      journal = {arXiv e-prints},
         year = 2026,
        month = sep,
          eid = {arXiv:2609.10213},
        pages = {arXiv:2609.10213},
          doi = {10.48550/arXiv.2609.10213},
archivePrefix = {arXiv},
       eprint = {2609.10213},
 primaryClass = {astro-ph.HE}
}

@ARTICLE{Meza+25,
       author = {{Meza}, Maria Renee and {Huang}, Xiaoshan and {Davis}, Shane W. and {Jiang}, Yan-Fei},
        title = "{Radiation-magnetohydrodynamic Simulations of Accretion Flow Formation After a Tidal Disruption Event}",
      journal = {\apj},
         year = 2025,
        month = oct,
       volume = {993},
       number = {1},
        pages = {57},
          doi = {10.3847/1538-4357/ae044c},
archivePrefix = {arXiv},
       eprint = {2506.00109},
 primaryClass = {astro-ph.HE}
}

@ARTICLE{Steinberg+16,
       author = {{Steinberg}, Elad and {Yalinewich}, Almog and {Sari}, Re'em},
        title = "{Grid noise in moving mesh codes: fixing the volume inconsistency problem}",
      journal = {\mnras},
         year = 2016,
        month = jun,
       volume = {459},
       number = {2},
        pages = {1596-1601},
          doi = {10.1093/mnras/stw783},
archivePrefix = {arXiv},
       eprint = {1603.06753},
 primaryClass = {physics.comp-ph},
       adsurl = {https://ui.adsabs.harvard.edu/abs/2016MNRAS.459.1596S}
}

@ARTICLE{Steinberg+15,
       author = {{Steinberg}, Elad and {Yalinewich}, Almog and {Sari}, Re'em and {Duffell}, Paul},
        title = "{Balancing the Load: A Voronoi Based Scheme for Parallel Computations}",
      journal = {\apjs},
         year = 2015,
        month = jan,
       volume = {216},
       number = {1},
          eid = {14},
        pages = {14},
          doi = {10.1088/0067-0049/216/1/14},
archivePrefix = {arXiv},
       eprint = {1408.3196},
 primaryClass = {astro-ph.IM},
       adsurl = {https://ui.adsabs.harvard.edu/abs/2015ApJS..216...14S}
}

@ARTICLE{Levermore84,
       author = {{Levermore}, C.~D.},
        title = "{Relating Eddington factors to flux limiters.}",
      journal = {\jqsrt},
         year = 1984,
        month = feb,
       volume = {31},
       number = {2},
        pages = {149-160},
          doi = {10.1016/0022-4073(84)90112-2},
       adsurl = {https://ui.adsabs.harvard.edu/abs/1984JQSRT..31..149L}
}

@ARTICLE{Brutman+24,
       author = {{Brutman}, Yuval and {Steinberg}, Elad and {Balberg}, Shmuel},
        title = "{The Primary Flare Following a Stellar Collision in a Galactic Nucleus}",
      journal = {\apjl},
         year = 2024,
        month = oct,
       volume = {974},
       number = {1},
          eid = {L22},
        pages = {L22},
          doi = {10.3847/2041-8213/ad808f},
archivePrefix = {arXiv},
       eprint = {2408.16383},
 primaryClass = {astro-ph.HE},
       adsurl = {https://ui.adsabs.harvard.edu/abs/2024ApJ...974L..22B}
}

@ARTICLE{Bonnerot+21,
       author = {{Bonnerot}, Cl{\'e}ment and {Lu}, Wenbin and {Hopkins}, Philip F.},
        title = "{First light from tidal disruption events}",
      journal = {\mnras},
         year = 2021,
        month = jul,
       volume = {504},
       number = {4},
        pages = {4885-4905},
          doi = {10.1093/mnras/stab398},
archivePrefix = {arXiv},
       eprint = {2012.12271},
 primaryClass = {astro-ph.HE},
       adsurl = {https://ui.adsabs.harvard.edu/abs/2021MNRAS.504.4885B}
}

@ARTICLE{Grudic+21,
       author = {{Grudi{\'c}}, Michael Y. and {Guszejnov}, D{\'a}vid and {Hopkins}, Philip F. and {Offner}, Stella S.~R. and {Faucher-Gigu{\`e}re}, Claude-Andr{\'e}},
        title = "{STARFORGE: Towards a comprehensive numerical model of star cluster formation and feedback}",
      journal = {\mnras},
         year = 2021,
        month = sep,
       volume = {506},
       number = {2},
        pages = {2199-2231},
          doi = {10.1093/mnras/stab1347},
archivePrefix = {arXiv},
       eprint = {2010.11254},
 primaryClass = {astro-ph.IM},
       adsurl = {https://ui.adsabs.harvard.edu/abs/2021MNRAS.506.2199G}
}

@ARTICLE{Krumholz+07a,
       author = {{Krumholz}, Mark R. and {Klein}, Richard I. and {McKee}, Christopher F.},
        title = "{Radiation-Hydrodynamic Simulations of Collapse and Fragmentation in Massive Protostellar Cores}",
      journal = {\apj},
         year = {2007},
        month = feb,
       volume = {656},
       number = {2},
        pages = {959-979},
          doi = {10.1086/510664},
archivePrefix = {arXiv},
       eprint = {astro-ph/0609798},
 primaryClass = {astro-ph},
       adsurl = {https://ui.adsabs.harvard.edu/abs/2007ApJ...656..959K}
}

@ARTICLE{Jiang+19,
       author = {{Jiang}, Yan-Fei and {Stone}, James M. and {Davis}, Shane W.},
        title = "{Super-Eddington Accretion Disks around Supermassive Black Holes}",
      journal = {\apj},
         year = 2019,
        month = aug,
       volume = {880},
       number = {2},
          eid = {67},
        pages = {67},
          doi = {10.3847/1538-4357/ab29ff},
archivePrefix = {arXiv},
       eprint = {1709.02845},
 primaryClass = {astro-ph.HE},
       adsurl = {https://ui.adsabs.harvard.edu/abs/2019ApJ...880...67J}
}

@article{malyali_transient_2024,
	title = {Transient fading {X}-ray emission detected during the optical rise of a tidal disruption event},
	volume = {531},
	issn = {0035-8711},
	url = {https://doi.org/10.1093/mnras/stae927},
	doi = {10.1093/mnras/stae927},
	number = {1},
	journal = {\mnras},
	author = {Malyali, A and Rau, A and Bonnerot, C and Goodwin, A J and Liu, Z and Anderson, G E and Brink, J and Buckley, D A H and Merloni, A and Miller-Jones, J C A and Grotova, I and Kawka, A},
	month = apr,
	year = {2024},
	pages = {1256--1275},
}

@article{steinberg_streamdisk_2024,
	title = {Stream--disk shocks as the origins of peak light in tidal disruption events},
	volume = {625},
	issn = {1476-4687},
	url = {https://doi.org/10.1038/s41586-023-06875-y},
	doi = {10.1038/s41586-023-06875-y},
	number = {7995},
	journal = {\nat},
	author = {Steinberg, Elad and Stone, Nicholas C.},
	month = jan,
	year = {2024},
	pages = {463--467},
}

@book{Castor_2004,
  address={Cambridge},
  title={Radiation Hydrodynamics},
  publisher={Cambridge University Press},
  author={Castor, John I.},
  year={2004}
}

@article{yalinewich_rich_2015,
	title = {{RICH}: {OPEN}-{SOURCE} {HYDRODYNAMIC} {SIMULATION} {ON} {A} {MOVING} {VORONOI} {MESH}},
	volume = {216},
	url = {https://doi.org/10.1088/0067-0049/216/2/35},
	doi = {10.1088/0067-0049/216/2/35},
	number = {2},
	journal = {\apjs},
	author = {Yalinewich, Almog and Steinberg, Elad and Sari, Re'em},
	month = feb,
	year = {2015},
	pages = {35},
}

@article{krief2018new,
  title={A new implementation of the STA method for the calculation of opacities of local thermodynamic equilibrium plasmas},
  author={Krief, Menahem and Feigel, Alexander and Gazit, Doron},
  journal={Atoms},
  volume={6},
  number={3},
  pages={35},
  year={2018},
  publisher={MDPI},
  doi={10.3390/atoms6030035}
}

@article{krief2015variance,
  title={Variance and shift of transition arrays for electric and magnetic multipole transitions},
  author={Krief, Menahem and Feigel, Alexander},
  journal={HEDP},
  volume={17},
  pages={254--262},
  year={2015},
  publisher={Elsevier},
  doi={10.1016/j.hedp.2015.10.001}
}

@article{krief2015effect,
  title={The effect of first order superconfiguration energies on the opacity of hot dense matter},
  author={Krief, Menahem and Feigel, Alexander},
  journal={HEDP},
  volume={15},
  pages={59--66},
  year={2015},
  publisher={Elsevier},
  doi={10.1016/j.hedp.2015.04.003}
}

@article{krief2018effect,
  title={The effect of ionic correlations on radiative properties in the solar interior and terrestrial experiments},
  author={Krief, Menahem and Kurzweil, Yair and Feigel, Alexander and Gazit, Doron},
  journal={\apj},
  volume={856},
  number={2},
  pages={135},
  year={2018},
  publisher={IOP Publishing},
  doi={10.3847/1538-4357/aab353}
}

@article{Colgan2016,
  title={A New Generation of {Los Alamos} Opacity Tables},
  author={Colgan, J. and Kilcrease, D. P. and Magee, N. H. and Sherrill, M. E. and Abdallah, Jr., J. and Hakel, P. and Fontes, C. J. and Guzik, J. A. and Mussack, K. A.},
  journal={\apj},
  volume={817},
  number={2},
  pages={116},
  year={2016},
  month={feb},
  publisher={The American Astronomical Society},
  doi={10.3847/0004-637X/817/2/116}
}

@ARTICLE{HEALPix,
       author = {{G{\'o}rski}, K.~M. and {Hivon}, E. and {Banday}, A.~J. and {Wandelt}, B.~D. and {Hansen}, F.~K. and {Reinecke}, M. and {Bartelmann}, M.},
        title = "{HEALPix: A Framework for High-Resolution Discretization and Fast Analysis of Data Distributed on the Sphere}",
      journal = {\apj},
         year = 2005,
        month = apr,
       volume = {622},
       number = {2},
        pages = {759-771},
          doi = {10.1086/427976},
archivePrefix = {arXiv},
       eprint = {astro-ph/0409513},
 primaryClass = {astro-ph},
       adsurl = {https://ui.adsabs.harvard.edu/abs/2005ApJ...622..759G}
}

@ARTICLE{Kompaneets1957,
       author = {{Kompaneets}, A.~S.},
        title = "{The Establishment of Thermal Equilibrium between Quanta and Electrons}",
      journal = {Soviet Physics JETP},
         year = 1957,
       volume = {4},
       number = {5},
        pages = {730-737}
}

@ARTICLE{PaczynskiWiita1980,
       author = {{Paczy{\'n}ski}, B. and {Wiita}, P.~J.},
        title = "{Thick Accretion Disks and Supercritical Luminosities}",
      journal = {\aap},
         year = 1980,
        month = aug,
       volume = {88},
        pages = {23-31},
       adsurl = {https://ui.adsabs.harvard.edu/abs/1980A&A....88...23P}
}

@book{RybickiLightman1979,
  title={Radiative Processes in Astrophysics},
  author={Rybicki, George B. and Lightman, Alan P.},
  year={1979},
  publisher={Wiley},
  address={New York}
}

@article{Roe1986,
  title={Characteristic-Based Schemes for the {Euler} Equations},
  author={Roe, P. L.},
  journal={AnRFM},
  volume={18},
  pages={337--365},
  year={1986},
  doi={10.1146/annurev.fl.18.010186.002005}
}

@article{Densmore2009,
  title = {Stability analysis of implicit time discretizations for the Compton-scattering Fokker–Planck equation},
  journal = {JCoPh},
  volume = {228},
  number = {16},
  pages = {5933-5960},
  year = {2009},
  issn = {0021-9991},
  doi = {10.1016/j.jcp.2009.05.003},
  url = {https://www.sciencedirect.com/science/article/pii/S0021999109002459},
  author = {Jeffery D. Densmore and James S. Warsa and Robert B. Lowrie and Jim E. Morel}
}

@article{Tomida_2013,
  doi = {10.1088/0004-637X/763/1/6},
  url = {https://doi.org/10.1088/0004-637X/763/1/6},
  year = {2013},
  month = {jan},
  publisher = {The American Astronomical Society},
  volume = {763},
  number = {1},
  pages = {6},
  author = {Tomida, Kengo and Tomisaka, Kohji and Matsumoto, Tomoaki and Hori, Yasunori and Okuzumi, Satoshi and Machida, Masahiro N. and Saigo, Kazuya},
  title = {RADIATION MAGNETOHYDRODYNAMIC SIMULATIONS OF PROTOSTELLAR COLLAPSE: PROTOSTELLAR CORE FORMATION},
  journal = {\apj}
}

@ARTICLE{Tomida+15,
       author = {{Tomida}, Kengo and {Okuzumi}, Satoshi and {Machida}, Masahiro N.},
        title = "{Radiation Magnetohydrodynamic Simulations of Protostellar Collapse: Nonideal Magnetohydrodynamic Effects and Early Formation of Circumstellar Disks}",
      journal = {\apj},
         year = 2015,
        month = mar,
       volume = {801},
       number = {2},
          eid = {117},
        pages = {117},
          doi = {10.1088/0004-637X/801/2/117},
archivePrefix = {arXiv},
       eprint = {1501.04102},
 primaryClass = {astro-ph.SR},
       adsurl = {https://ui.adsabs.harvard.edu/abs/2015ApJ...801..117T}
}

@ARTICLE{Pejcha2016,
       author = {{Pejcha}, Ond{\v{r}}ej and {Metzger}, Brian D. and {Tomida}, Kengo},
        title = "{Cool and luminous transients from mass-losing binary stars}",
      journal = {\mnras},
         year = 2016,
       volume = {455},
       number = {4},
        pages = {4351--4372},
          doi = {10.1093/mnras/stv2592},
archivePrefix = {arXiv},
       eprint = {1509.02531},
 primaryClass = {astro-ph.SR}
}

@ARTICLE{Krief2016,
       author = {{Krief}, Menahem and {Feigel}, Alexander and {Gazit}, Doron},
        title = "{Solar Opacity Calculations Using the Super-transition-array Method}",
      journal = {\apj},
         year = 2016,
       volume = {821},
       number = {1},
          eid = {45},
        pages = {45},
          doi = {10.3847/0004-637X/821/1/45},
archivePrefix = {arXiv},
       eprint = {1601.01930},
 primaryClass = {astro-ph.SR}
}

@article{Wollaber23022016,
    author    = {Allan B. Wollaber},
    title     = {Four Decades of Implicit Monte Carlo},
    journal   = {JCTT},
    volume    = {45},
    number    = {1-2},
    pages     = {1--70},
    year      = {2016},
    publisher = {Taylor \& Francis},
    doi       = {10.1080/23324309.2016.1138132},
    url       = {https://doi.org/10.1080/23324309.2016.1138132}
}

@article{Thomsen2022,
    doi = {10.3847/2041-8213/ac911f},
    url = {https://doi.org/10.3847/2041-8213/ac911f},
    year = {2022},
    month = {sep},
    publisher = {The American Astronomical Society},
    volume = {937},
    number = {2},
    pages = {L28},
    author = {Thomsen, Lars L. and Kwan, Tom M. and Dai, Lixin and Wu, Samantha C. and Roth, Nathaniel and Ramirez-Ruiz, Enrico},
    title = {Dynamical Unification of Tidal Disruption Events},
    journal = {\apjl}
}

@article{Hills1975,
    author = {Hills, J. G.},
    title = {Possible power source of {Seyfert} galaxies and {QSOs}},
    journal = {\nat},
    year = {1975},
    volume = {254},
    number = {5498},
    pages = {295--298},
    doi = {10.1038/254295a0}
}

@article{Rees1988,
    author = {Rees, Martin J.},
    title = {Tidal disruption of stars by black holes of $10^6$--$10^8$ solar masses in nearby galaxies},
    journal = {\nat},
    year = {1988},
    volume = {333},
    number = {6173},
    pages = {523--528},
    doi = {10.1038/333523a0}
}

@article{Gezari21,
    author = {Gezari, Suvi},
    title = {Tidal Disruption Events},
    journal = {\araa},
    year = {2021},
    volume = {59},
    pages = {21--58},
    doi = {10.1146/annurev-astro-111720-030029},
    eprint = {2104.14580},
    archivePrefix = {arXiv}
}

@article{Saxton+20,
    author = {Saxton, R. and Komossa, S. and Auchettl, K. and Jonker, P. G.},
    title = {X-Ray Properties of {TDEs}},
    journal = {\ssr},
    year = {2020},
    volume = {216},
    number = {5},
    pages = {85},
    doi = {10.1007/s11214-020-00708-4},
    eprint = {2103.15442},
    archivePrefix = {arXiv}
}

@article{Kara+18,
    author = {Kara, E. and Dai, L. and Reynolds, C. S. and Kallman, T.},
    title = {Ultrafast outflow in tidal disruption event {ASASSN-14li}},
    journal = {\mnras},
    year = {2018},
    volume = {474},
    number = {3},
    pages = {3593--3598},
    doi = {10.1093/mnras/stx3004},
    eprint = {1711.06090},
    archivePrefix = {arXiv}
}

@article{CurdNarayan19,
    author = {Curd, Brandon and Narayan, Ramesh},
    title = {{GRRMHD} simulations of tidal disruption event accretion discs around supermassive black holes: jet formation, spectra, and detectability},
    journal = {\mnras},
    year = {2019},
    volume = {483},
    number = {1},
    pages = {565--592},
    doi = {10.1093/mnras/sty3134},
    eprint = {1811.06971},
    archivePrefix = {arXiv}
}

@article{Jiang22,
    author = {Jiang, Yan-Fei},
    title = {Multigroup Radiation Magnetohydrodynamics Based on Discrete Ordinates including {Compton} Scattering},
    journal = {\apjs},
    year = {2022},
    volume = {263},
    number = {1},
    pages = {4},
    doi = {10.3847/1538-4365/ac9231},
    eprint = {2209.06240},
    archivePrefix = {arXiv}
}

@article{SadowskiNarayan15,
    author = {S{\k{a}}dowski, Aleksander and Narayan, Ramesh},
    title = {Photon-conserving {Comptonization} in simulations of accretion discs around black holes},
    journal = {\mnras},
    year = {2015},
    volume = {454},
    number = {3},
    pages = {2372--2380},
    doi = {10.1093/mnras/stv2022},
    eprint = {1508.04980},
    archivePrefix = {arXiv}
}

@article{Ryan+15,
    author = {Ryan, B. R. and Dolence, J. C. and Gammie, C. F.},
    title = {bhlight: General Relativistic Radiation Magnetohydrodynamics with {Monte Carlo} Transport},
    journal = {\apj},
    year = {2015},
    volume = {807},
    number = {1},
    pages = {31},
    doi = {10.1088/0004-637X/807/1/31},
    eprint = {1505.05119},
    archivePrefix = {arXiv}
}

@article{Stone+20,
    author = {Stone, N. C. and Vasiliev, E. and Kesden, M. and Rossi, E. M. and Perets, H. B. and Amaro-Seoane, P.},
    title = {Rates of Stellar Tidal Disruption},
    journal = {\ssr},
    year = {2020},
    volume = {216},
    number = {3},
    pages = {35},
    doi = {10.1007/s11214-020-00651-4}
}

@article{Roth+20,
    author = {Roth, Nathaniel and Rossi, Elena Maria and Krolik, Julian and Piran, Tsvi and Mockler, Brenna and Kasen, Daniel},
    title = {Radiative Emission Mechanisms},
    journal = {\ssr},
    year = {2020},
    volume = {216},
    number = {7},
    pages = {114},
    doi = {10.1007/s11214-020-00735-1}
}

@article{Zabludoff+21,
    author = {Zabludoff, Ann and Arcavi, Iair and LaMassa, Stephanie and Perets, Hagai B. and Trakhtenbrot, Benny and Zauderer, B. Ashley and Auchettl, Katie and Dai, Jane L. and French, K. Decker and Hung, Tiara and Kara, Erin and Lodato, Giuseppe and Maksym, W. Peter and Qin, Yujing and Ramirez-Ruiz, Enrico and Roth, Nathaniel and Runnoe, Jessie C. and Wevers, Thomas},
    title = {Distinguishing Tidal Disruption Events from Impostors},
    journal = {\ssr},
    year = {2021},
    volume = {217},
    number = {4},
    pages = {54},
    doi = {10.1007/s11214-021-00829-4}
}

@article{krief_self-similar_2024,
    title = {Self-similar solutions for the non-equilibrium nonlinear supersonic {Marshak} wave problem},
    volume = {36},
    issn = {1070-6631},
    doi = {10.1063/5.0186666},
    number = {1},
    journal = {Physics of Fluids},
    author = {Krief, Menahem and McClarren, Ryan G.},
    month = jan,
    year = {2024},
    pages = {017108}
}

@article{derei_non-equilibrium_2024,
    title = {The non-equilibrium {Marshak} wave problem in non-homogeneous media},
    volume = {36},
    issn = {1070-6631},
    doi = {10.1063/5.0244247},
    number = {12},
    journal = {Physics of Fluids},
    author = {Derei, Nitay and Balberg, Shmuel and Heizler, Shay I. and Steinberg, Elad and McClarren, Ryan G. and Krief, Menahem},
    month = dec,
    year = {2024},
    pages = {127149}
}

@article{krief2021analytic,
    title = {Analytic solutions of the nonlinear radiation diffusion equation with an instantaneous point source in non-homogeneous media},
    author = {Krief, Menahem},
    journal = {Physics of Fluids},
    volume = {33},
    pages = {057105},
    number = {5},
    year = {2021},
    doi = {10.1063/5.0050422},
    publisher = {AIP Publishing}
}

@article{krief2024unified,
    title = {A unified theory of the self-similar supersonic {Marshak} wave problem},
    volume = {36},
    pages = {056129},
    issn = {1089-7666},
    doi = {10.1063/5.0208111},
    number = {5},
    journal = {Physics of Fluids},
    publisher = {AIP Publishing},
    author = {Krief, Menahem and McClarren, Ryan G.},
    year = {2024},
    month = may
}

@article{heizler2024accurate,
    title = {Accurate reaction-diffusion limit to the spherical-symmetric {Boltzmann} equation},
    author = {Heizler, Shay I. and Krief, Menahem and Assaf, Michael},
    journal = {Physical Review Research},
    volume = {6},
    number = {1},
    pages = {L012023},
    year = {2024},
    doi = {10.1103/PhysRevResearch.6.L012023},
    publisher = {APS}
}

@article{giron2021solutions,
    title = {Solutions of the imploding shock problem in a medium with varying density},
    author = {Giron, Itamar and Balberg, Shmuel and Krief, Menahem},
    journal = {Physics of Fluids},
    volume = {33},
    pages = {066105},
    number = {6},
    year = {2021},
    doi = {10.1063/5.0054525},
    publisher = {AIP Publishing}
}

@article{giron2023solutions,
    title = {Solutions of the converging and diverging shock problem in a medium with varying density},
    author = {Giron, Itamar and Balberg, Shmuel and Krief, Menahem},
    journal = {Physics of Fluids},
    volume = {35},
    pages = {066112},
    number = {6},
    year = {2023},
    doi = {10.1063/5.0151791},
    publisher = {AIP Publishing}
}

@article{krief2023piston,
    title = {Piston driven shock waves in non-homogeneous planar media},
    author = {Krief, Menahem},
    journal = {Physics of Fluids},
    volume = {35},
    pages = {046102},
    number = {4},
    year = {2023},
    doi = {10.1063/5.0145896},
    publisher = {AIP Publishing}
}

@misc{openai_chatgpt,
  author       = {{OpenAI}},
  title        = {{ChatGPT}},
  year         = {2026},
  howpublished = {Large language model, \url{https://chatgpt.com/}, accessed 2026 October 20}
}

@misc{anthropic_claude,
  author       = {{Anthropic}},
  title        = {{Claude}},
  year         = {2026},
  howpublished = {Large language model, \url{https://www.anthropic.com/claude}, accessed 2026 October 20}
}
\bibliographystyle{aasjournal}

\end{document}